\documentclass[iop,apj]{emulateapj}

\usepackage{graphicx}
\usepackage{bm}
\usepackage{subfigure}
\usepackage{mathtools}
\usepackage{color}
\usepackage{paralist}
\usepackage{amssymb,amsfonts,amsmath}%
\usepackage[dvipsnames,usenames]{xcolor}%
\usepackage{array,booktabs,units}

\graphicspath{{./fig/}{./png/}}

\newcommand{\pns}{{\rm PNS}}
\newcommand{\const}{{\rm const}}
\newcommand{\alphaem}{\ensuremath{\alpha_{\rm em}}}
\newcommand{\s}{\,{\rm s}}
\newcommand{\g}{\,{\rm g}}

\newcommand{\Mpc}{\,{\rm Mpc}}
\newcommand{\MeV}{\,{\rm MeV}}
\newcommand{\GeV}{\,{\rm GeV}}
\newcommand{\erg}{\,{\rm erg}}
\def\kB{k_{\rm B}}

\newcommand{\G}{\,{\rm G}}
\newcommand{\K}{\,{\rm K}}

\newcommand{\cm}{\,{\rm cm}}
\newcommand{\km}{\,{\rm km}}

\newcommand{\nab}{\bm{\nabla}}
\newcommand{\AAA}{\bm{A}}
\newcommand{\BB}{\bm{B}}
\newcommand{\FF}{\bm{F}}

\newcommand{\EE}{\bm{E}}

\newcommand{\UU}{\bm{U}}

\newcommand{\ff}{\bm{f}}
\newcommand{\SSSS}{\mbox{\boldmath ${\sf S}$} {}}
\newcommand{\meanrho}{\overline{\rho}}

{}
{}
{}

{}
{}
{}
{}
{}
{}
{}
{}
{}
\newcommand{\meanBB}{\overline{\mbox{\boldmath $B$}}{}}{}
{}
{}
{}
{}
{}
{}
{}
{}
\newcommand{\meanUU}{\overline{\bm{U}}}

{}
{}
{}
\newcommand{\meanA}{\overline{A}}
\newcommand{\meanB}{\overline{B}}

\newcommand{\meanmu}{\overline{\mu}}

\newcommand{\meanv}{\overline{v}}
\newcommand{\EQA}{\begin{eqnarray}}
\newcommand{\ENA}{\end{eqnarray}}

\newcommand{\Eq}[1]{Equation~(\ref{#1})}
\newcommand{\Eqs}[2]{Equations~(\ref{#1}) and~(\ref{#2})}

\newcommand{\Fig}[1]{Figure~\ref{#1}}

\newcommand{\bra}[1]{\langle #1\rangle}

\renewcommand{\max}{{\rm max}}

\def\half{{\textstyle{1\over2}}}
\def\onethird{{\textstyle{1\over3}}}

\def\Rm{{\rm Re}_{_\mathrm{M}}}
\def\Rey{{\rm Re}}
\def\Lu{\mbox{\rm Lu}}
\def\vA{v_{\rm A}}
\def\cs{c_{\rm s}}

\newcommand{\PrM}{{\rm Pr}_{_{\rm M}}}
\newcommand{\Prmu}{\mathrm{Pr}_\mu}

\usepackage{hyperref}
\newcommand\myshade{85}
\colorlet{mylinkcolor}{violet}
\colorlet{mycitecolor}{Aquamarine}
\colorlet{myurlcolor}{YellowOrange}

\hypersetup{
 linkcolor  = mylinkcolor,
 citecolor  = mycitecolor!\myshade!black,
 urlcolor   = myurlcolor!\myshade!black,
colorlinks = true,
}

\begin{document}

\title{Laminar and turbulent dynamos in chiral magnetohydrodynamics. II. Simulations}

\author{Jennifer~Schober$^{1,2}$}
\email{jennifer.schober@epfl.ch}

\author{Igor~Rogachevskii$^{3,1,4}$}
%\email{gary@bgu.ac.il}

\author{Axel~Brandenburg$^{1,4,5,6}$}
%\email{brandenb@nordita.org}

\author{Alexey~Boyarsky$^{7}$}
%\email{boyarsky@lorentz.leidenuniv.nl}

\author{J\"{u}rg~Fr\"{o}hlich$^{8}$}
%\email{juerg@phys.ethz.ch}

\author{Oleg~Ruchayskiy$^{9}$}
%\email{oleg.ruchayskiy@epfl.ch}

\author{Nathan~Kleeorin$^{3,1}$}
%\email{nat@bgu.ac.il}

\medskip
\medskip
\affiliation{
$^1$Nordita, KTH Royal Institute of Technology
 and Stockholm University, Roslagstullsbacken 23,
 10691 Stockholm, Sweden \\
$^2$Laboratoire d'Astrophysique, EPFL, CH-1290 Sauverny, Switzerland\\
$^3$Department of Mechanical Engineering,
 Ben-Gurion University of the Negev, P.O. Box 653, Beer-Sheva
 84105, Israel \\
$^4$Laboratory for Atmospheric and Space Physics,
   University of Colorado, 3665 Discovery Drive, Boulder, CO 80303, USA \\
$^5$JILA and Department of Astrophysical and Planetary Sciences,
    Box 440, University of Colorado, Boulder, CO 80303, USA\\
$^6$Department of Astronomy, AlbaNova University Center,
    Stockholm University, SE-10691 Stockholm, Sweden\\
$^7$Instituut-Lorentz for Theoretical Physics, Universiteit Leiden,
 Niels Bohrweg 2, 2333 CA Leiden, The Netherlands\\
$^8$Institute of Theoretical Physics, ETH H\"{o}nggerberg,
 CH-8093 Zurich, Switzerland \\
$^9$Discovery Center, Niels Bohr Institute, Blegdamsvej 17,
 DK-2100 Copenhagen, Denmark
}

\submitted{\today,~ $ $Revision: 1.653 $ $}
\begin{abstract}
Using direct numerical simulations (DNS),
we study laminar and turbulent dynamos in chiral
magnetohydrodynamics (MHD) with an extended set of equations that accounts
for an additional contribution to the electric current due to the chiral
magnetic effect (CME).
This quantum phenomenon originates from an asymmetry between
left- and right-handed relativistic fermions in the presence of a magnetic field
and gives rise to a chiral dynamo.
We show that the magnetic field evolution proceeds in three stages:
(1) a small-scale chiral dynamo instability;
(2) production of chiral magnetically driven turbulence
and excitation of a large-scale dynamo instability
due to a new chiral effect ($\alpha_\mu$ effect); and
(3) saturation of magnetic helicity and magnetic field growth
controlled by a conservation law for the total chirality.
The $\alpha_\mu$ effect becomes dominant at large fluid and magnetic
Reynolds numbers and is not related to kinetic helicity.
The growth rate of the large-scale magnetic field
and its characteristic scale measured in the numerical simulations
agree well with theoretical predictions based on
mean-field theory.
The previously discussed two-stage chiral magnetic scenario
did not include stage (2) during which the characteristic scale
of magnetic field variations can increase by many orders of magnitude.
Based on the findings from numerical simulations,
the relevance of the CME and the chiral effects revealed
in the relativistic plasma of the early universe
and of proto-neutron stars are discussed.
\end{abstract}


\maketitle

%%%%%%%%%%%%%
% Section 1 %
%%%%%%%%%%%%%

\section{Introduction}

Magnetic fields are observed on various spatial scales of the
universe: they are detected in planets and stars \citep{DL09,R12},
in the interstellar medium \citep{C12}, and on galactic scales \citep{B16}.
Additionally, observational lower limits on intergalactic
magnetic fields have been reported
\citep{NV10,DCRFCL11}.
Contrary to the high magnetic field strengths observed on scales below
those of galaxy clusters, which can be explained by dynamo amplification
\citep[see, e.g.,][]{BS05},
intergalactic magnetic fields, if confirmed, are most likely of primordial origin.
Because of their often large energy densities, magnetic fields can play
an important role in various astrophysical objects, a prominent
example being the $\alpha\Omega$ dynamo in solar-like stars that explains
stellar activity
\citep[see, e.g.,][]{P55,P79,M78,KR80,ZRS83,C14}.

While there is no doubt about the significant role of magnetic fields in the
dynamics of the
present-day universe, their origin and evolution over cosmic times remain a
mystery \citep{R87,GR01,W02,KZ08}.
Numerous scenarios for the generation of primordial magnetic fields have been
suggested
in the literature. The proposals span from inflation-produced magnetic fields
\citep{TW88} to field generation during cosmological phase transitions
\citep{SOJ97}.
Even though strong magnetic fields
could be generated shortly after the Big Bang, their strength subsequently
decreases in cosmic expansion unless they undergo further amplification.
Be this as it may, the presence of primordial magnetic fields can affect the
physics of the early universe.
For example, it has been shown that primordial fields could have
significant effects on the matter power spectrum by suppressing the formation
of small-scale structures \citep{KEtAl13,PEtAl15}.
This, in turn, could influence cosmological structure formation.

The theoretical framework for studying the evolution of magnetic fields is
magnetohydrodynamics (MHD).
In classical plasma physics,
the system of equations includes the induction equation,
which is derived from the Maxwell equations and Ohm's law and describes
the evolution of magnetic fields,
the continuity equation for the fluid density,
and the Navier-Stokes equation governing the evolution of the
velocity field.

At high energies, for example, in the quark-gluon plasma of the early universe,
however, an additional quantity needs to be taken into account, namely
the chiral chemical potential.
This quantity is related to an asymmetry between the number densities of
left-handed fermions (spin antiparallel to the momentum) and right-handed
fermions (spin parallel to the momentum). This leads to an additional 
contribution to the electric current along the magnetic field, known as the
{\em chiral magnetic effect} (CME).
This phenomenon was discovered by \cite{Vilenkin:80a} and
was later carefully investigated using different
theoretical approaches in a number of studies \citep{RW85,Tsokos:85,
JS97,AlekseevEtAl1998,Frohlich:2000en,Frohlich:2002fg,
Fukushima:08,Son:2009tf}.

The CME causes a small-scale dynamo instability \citep{JS97}, which has also
been revealed from a kinetic description of chiral plasmas \citep{AY13}.
The evolution equation for a nonuniform
chiral chemical potential has been derived in \citet{BFR12,BFR15} who used it
to study the inverse magnetic cascade that results in an
increase of the characteristic scale of the magnetic field.
\citet{BFR12} have shown that the chiral asymmetry can survive down to energies
of the order of $10\MeV$, due to coupling to an effective axion field.
These studies triggered various investigations related to
chiral MHD turbulence \citep{Pavlovic:2016gac,Y16}
and its role in the early universe \citep{TVV12,DvornikovSemikoz2017},
as well as in neutron stars \citep{DvornikovSemikoz2015,SiglLeite2016}.

Recently, a systematic theoretical analysis of the system of \emph{chiral MHD}
equations, including the back-reaction of the magnetic field on the chiral
chemical potential, and the coupling to the plasma velocity field has been
performed in \cite{REtAl17}, referred to here as Paper~I.
The main findings of Paper~I include a modification of MHD waves by the
CME and different kinds of laminar and turbulent dynamos.
Besides the well-studied laminar chiral dynamo caused by
the CME, a chiral--shear dynamo in the presence of a shearing velocity
was discussed there.
In addition, a mean-field theory of chiral MHD in the presence
of small-scale nonhelical turbulence was developed in Paper~I,
where a new chiral $\alpha_\mu$ effect
not related to a kinetic helicity has been found.
This effect results from an interaction of chiral magnetic fluctuations
with fluctuations of the electric current caused by the tangling magnetic
fluctuations.

In the present paper, we report on
numerical simulations that confirm and further substantiate the chiral
laminar and turbulent dynamos found in Paper~I.
To this end, we have implemented the chiral MHD equations in the
{\sc Pencil Code}\footnote{http://pencil-code.nordita.org/},
a high-order code suitable for compressible MHD turbulence.
Different situations are considered, from laminar dynamos
to chiral magnetically driven turbulence
and large-scale dynamos in externally forced turbulence.
With our direct numerical simulations (DNS), we are able to study the dynamical
evolution of a plasma that includes chiral effects in a large domain of the
parameter space.
Given that the detailed properties of relativistic astrophysical plasmas,
in particular the initial chiral asymmetry and chiral feedback mechanisms,
are not well understood at present, a broad analysis of various
scenarios is essential.
The findings from DNS can then be used to explore the possible evolution of 
astrophysical plasmas under different assumptions. 
These applications should not be regarded as realistic descriptions of 
high-energy plasmas; they aim to find out under what conditions the
CME plays a significant role in the evolution of a
plasma of relativistic charged fermions (electrons) and to test the
importance of chirality flips changing the handedness of the fermions.
We are not pretending that the regimes accessible to our simulations
are truly realistic in the context of the physics of the early universe or in
neutron stars.

The outline of the present paper is as follows.
In Section~\ref{sec:simulations} we review the governing equations and
the numerical setup,
and we discuss the physics related to the two main nonlinear effects
in chiral MHD, which lead to different scenarios of the magnetic
field evolution.
In Section~\ref{sec:lamdyn} we present numerical results on laminar chiral
dynamos.
In Section~\ref{sec_turbdynamo1} we show how magnetic fields, amplified by
the CME, produce turbulence (chiral magnetically driven turbulence).
We discuss how this gives rise to the chiral $\alpha_\mu$ effect.
We also study this effect in Section~\ref{sec_turbdynamo} for a system
where external forcing is employed to produce turbulence.
After a discussion of chiral MHD in astrophysical and cosmological processes in
Section~\ref{sec_astro}, we draw conclusions in Section~\ref{sec_concl}.

%%%%%%%%%%%%%
% Section 2 %
%%%%%%%%%%%%%

\section{Chiral MHD in numerical simulations}
\label{sec:simulations}

\subsection{Equations of chiral MHD}
\label{sec:eqs-chiral MHD}

The system of chiral MHD equations
includes the induction equation for the magnetic field $\BB$,
the Navier-Stokes equation for the velocity field $\UU$ of the plasma,
the continuity equation for the plasma density $\rho$,
and an evolution equation for the normalized chiral chemical potential
$\mu$:
\begin{eqnarray}
\frac{\partial \BB}{\partial t} &=&
\nab   \times   \left[{\UU}  \times   {\BB}
- \eta \, \left(\nab   \times   {\BB}
- \mu {\BB} \right) \right] ,
\label{ind-DNS}\\
\rho{D \UU \over D t}&=& (\nab   \times   {\BB})  \times   \BB
-\nab  p + \nab  {\bm \cdot} (2\nu \rho \SSSS)
+\rho \ff ,
\label{UU-DNS}\\
\frac{D \rho}{D t} &=& - \rho \, \nab  \cdot \UU ,
\label{rho-DNS}\\
\frac{D \mu}{D t} &=& D_5 \, \Delta \mu
+ \lambda \, \eta \, \left[{\BB} {\bm \cdot} (\nab   \times   {\BB})
- \mu {\BB}^2\right]
 -\Gamma_{\rm\!f}\mu,
\label{mu-DNS}
\end{eqnarray}
where $\BB$ is normalized such that the magnetic energy
density is $\BB^2/2$ (without the $4\pi$ factor), and
$D/D t = \partial/\partial t + \UU \cdot \nab$ is the
advective derivative.
Further, $\eta$ is the microscopic magnetic diffusivity,
$p$ is the fluid pressure,
${\sf S}_{ij}=\half(U_{i,j}+U_{j,i})-\onethird\delta_{ij} {\bm \nabla}
{\bm \cdot} \UU$
are the components of the trace-free strain tensor $\SSSS$ (commas denote
partial spatial derivatives),
$\nu$ is the kinematic viscosity,
and $\ff$ is the turbulent forcing function.

Equation~(\ref{mu-DNS}) describes the evolution of the chiral
chemical potential $\mu_5 \equiv\mu_L - \mu_R$, with $\mu_L$ and $\mu_R$ being
the chemical potentials of left- and right-handed chiral fermions, which is normalized such that
$\mu = (4 \alphaem /\hbar c) \mu_5$ has the dimension of an inverse length.
Here $D_5$ is the diffusion constant of the chiral chemical potential $\mu$, and
the parameter $\lambda$, referred to in Paper~I as
the \textit{chiral feedback parameter},
characterizes the strength of the
coupling of the electromagnetic field to $\mu$.
The expression of the feedback term in Equation~(\ref{mu-DNS}) was
derived in Paper~I and is valid for the limit of small magnetic
diffusivities.
For hot plasmas, when $\kB T \gg \max(|\mu_L|,|\mu_R|)$,
the parameter $\lambda$ is given
by\footnote{The definition of $\lambda$ in the case of a degenerate Fermi
gas will be given in Section~\ref{sec:PNS}.}
\begin{eqnarray}
  \lambda=3 \hbar c \left({8 \alphaem \over \kB T} \right)^2,
\label{eq_lambda}
\end{eqnarray}
where $\alphaem \approx 1/137$ is the fine structure constant,
$T$ is the temperature, $\kB$ is the Boltzmann constant,
$c$ is the speed of light, and $\hbar$ is the reduced Planck constant.
We note that
$\lambda^{-1}$ has the dimension of energy per unit length.
The last term in Equation~(\ref{mu-DNS}), proportional to $\Gamma_{\rm\!f} \mu$,
characterizes chirality flipping processes due to finite fermion masses.
This term is included in a phenomenological way.
The detailed
dependence of $\Gamma_{\rm\!f}$ on the plasma parameters in realistic systems
is still not fully understood.
In most of the runs,
the chirality flipping effect is neglected because we concentrate in this paper
on the high-temperature regime, where the other terms in
Equation~(\ref{mu-DNS}) dominate.
However, we study its effect on the nonlinear evolution of
$\mu$ in Section~\ref{sec_flip}.

We stress that the effects related to the chiral anomaly cannot be
separated from the rest of the equations. This is one of the
essential features of the chiral MHD equations that we are studying.
The equations interconnect the chiral chemical potential to the
electromagnetic field.
However, the chiral anomaly couples the electromagnetic field
not directly to the chiral chemical potential but to
the chiral charge density, a conjugate variable in the sense of statistical
mechanics. 
The parameter $\lambda$ is nothing but a susceptibility, that is,
a (inverse) proportionality coefficient
quantifying the response of the
axial charge to a change in the chiral chemical potential; see Paper~I.

The system of equations~(\ref{ind-DNS})--(\ref{mu-DNS}) and their range of
validity have been discussed in detail in Paper~I.
Below we present a short summary of the assumptions made in deriving these
equations.
We focus our attention on an isothermal plasma, $T=\mathrm{const}$.
The equilibration rate of the temperature gradients is related to the shortest
timescales of the plasma (of the order of the plasma frequency or below) and is
much shorter than the time-scales that we consider in the present study.
For an isothermal equation of state, the pressure $p$ is related to
the density $\rho$ via $p=c_{\rm s}^2\rho$, where $c_{\rm s}$ is the sound speed.
We apply a one-fluid MHD model that
follows from a two-fluid plasma model
\citep{Artsimovich-Sagdeev,Melrose-QuantumPlasmadynamicsMagnetized,Biskamp:97}.
This implies that we do not consider here kinetic effects and effects related to
the two-fluid plasma model.
We note that the MHD formalism is valid for scales above the mean free path that can be 
approximated as \citep{ArnoldEtAl2000}
\begin{equation}
  \ell_\mathrm{mfp} \approx \frac{1}{(4 \pi \alpha_\mathrm{em})^2 \ln{((4 \pi \alpha_\mathrm{em})^{-1/2})}} \frac{\hbar c}{\kB T}.
\label{eq_mfp}
\end{equation}
Further, we study the nonrelativistic bulk motion of a highly relativistic plasma.
The latter leads to a term in the Maxwell equations that destabilizes
the nonmagnetic equilibrium and causes an exponential growth of
the magnetic field.
Such plasmas arise in the description of certain astrophysical systems,
where, for example,
a nonrelativistic plasma interacts with cosmic rays consisting of
relativistic particles with small number density;
see, for example, \cite{Schlickeiser-book}.
We study the case of small magnetic diffusivity
typical of many astrophysical systems with large magnetic Reynolds numbers, so
we neglect terms of the order of $\sim O(\eta^2)$ in the electric field; see Paper I.

A key difference in the induction equations of chiral
and classical MHD is the last term
$\propto \nab \times (\eta \, \mu \, {\BB})$ in Equation~(\ref{ind-DNS}).
This is reminiscent of mean-field dynamo theory, where a mean magnetic field $\meanBB$ is
amplified by an $\alpha$ effect due to a term $\propto \nab \times (\alpha \, {\meanBB})$
in the mean-field induction equation, which results in an $\alpha^2$ dynamo.
In analogy with mean-field dynamo theory, we use the name
\textit{$v_\mu^2$ dynamo}, introduced in Paper~I, where
\begin{equation}
  v_\mu\equiv\eta\mu_0
  \label{eq:16}
\end{equation}
plays the role of $\alpha$ (see Equation~(\ref{ind-DNS})),
and $\mu_0$ is the initial value of the normalized chiral chemical potential.
These different notions are motivated by the fact that the $v_\mu$ effect
is not related to any turbulence effects; that is, it is not determined by
the mean electromotive force, but originates from the
CME; see Paper~I for details.
We will discuss the differences between chiral and classical MHD in more
detail in Section~\ref{sec:difference}.

The system of Equations~(\ref{ind-DNS})--(\ref{mu-DNS}) implies a conservation law:
\begin{equation}
\frac{\partial }{\partial t} \left({\lambda \over 2}  {\bm A} {\bm \cdot} \BB
+ \mu \right) + \nab  {\bm \cdot} \FF_{\rm tot} = 0,
\label{CL}
\end{equation}
where
\begin{equation}
\FF_{\rm tot} =
{\lambda \over 2} \left({\bm \EE} \times   {\bm A} + \BB \, \Phi\right)
-  D_5 \nab  \mu
\end{equation}
is the flux of total chirality and
$\BB = {\bm \nabla} {\bm \times} {\bm A}$,
where ${\bm A}$ is the vector potential, ${\bm \EE}=
- c^{-1} \, \{ {\bm \UU} {\bm \times} {\BB} + \eta \, \big(\mu {\BB}  -
{\bm \nabla} {\bm \times} {\BB} \big) \}$ is the electric field, $\Phi$ is the
electrostatic potential, $\lambda$ is assumed to be constant,
and the chiral flipping term, $-\Gamma_{\rm\!f}\mu$,
in Equation~(\ref{mu-DNS}) is assumed to be negligibly small.
This implies that the total chirality is a conserved quantity:
\begin{equation}
  \label{cons_law}
\frac\lambda 2 \langle \AAA \cdot \BB\rangle + \langle \mu \rangle = \mu_0 = \mathrm{const},
\end{equation}
where $\langle \mu \rangle$
is the value of the chemical potential and
$ \langle \AAA \cdot \BB\rangle \equiv V^{-1}\int \AAA \cdot \BB \, dV$
is the mean magnetic helicity density over the volume $V$.

\subsection{Chiral MHD equations in dimensionless form}

We study the system of chiral MHD equations~(\ref{ind-DNS})--(\ref{mu-DNS})
in numerical simulations to analyze various laminar and turbulent dynamos,
as well as the production of turbulence by the CME.
It is, therefore, useful to rewrite this system of equations in
dimensionless form, where velocity is measured in units of the sound speed
$c_\mathrm{s}$, length is measured in units of $\ell_\mu \equiv \mu_0^{-1}$, so time
is measured in units of $\ell_\mu/c_\mathrm{s}$,
the magnetic field is measured in units of $\sqrt{\meanrho} \, c_\mathrm{s}$,
fluid density is measured in units of $\meanrho$
and the chiral chemical potential is measured in units of $\ell_\mu^{-1}$,
where $\meanrho$ is the volume-averaged density.
Thus, we introduce the following dimensionless functions, indicated by a
tilde:
${\bm \BB}=\sqrt{\meanrho} \, c_\mathrm{s}\tilde{\bm \BB}$,
${\bm \UU}=c_\mathrm{s}\tilde{\bm \UU}$,
$\mu=\ell_\mu^{-1} \tilde\mu$
and $\rho= \meanrho \, \tilde\rho$.
The chiral MHD equations in dimensionless form are given by
\begin{eqnarray}
\frac{\partial \tilde\BB}{\partial \tilde t}  &=&   \tilde{\bm \nabla}
{\bm \times} \biggl[\tilde{\bm \UU} {\bm \times} \tilde{\BB}  + {\rm Ma}_\mu \,
\Big(\tilde\mu \tilde{\BB}  - \tilde{\bm \nabla} {\bm \times} {\tilde\BB} \Big)
\biggr],
\label{ind-NS}\\
\tilde \rho {D \tilde \UU \over D \tilde t}&=& (\tilde \nab \times \tilde{\BB})  \times
\tilde \BB
-\tilde \nab \tilde \rho + {\rm Re}_\mu^{-1} \tilde\nab {\bm \cdot} (2\nu \tilde
\rho \SSSS)
+ \tilde \rho \ff ,
\nonumber\\
\label{UU-NS}\\
\frac{D \tilde\rho}{D \tilde t} &=& - \tilde\rho \, \tilde\nab  \cdot \tilde\UU ,
\label{rho-NS}\\
\frac{D \tilde\mu}{D \tilde t}  &=& D_\mu \, \tilde\Delta \tilde\mu
+ \Lambda_\mu \, \Big[{\tilde\BB} {\bm \cdot} (\tilde\nab {\bm \times}
{\tilde\BB})
- \tilde\mu {\tilde\BB}^2 \Big]
- \tilde \Gamma_\mathrm{f} \tilde \mu ,
\label{mu-NS}
\end{eqnarray}
where we introduce the following dimensionless parameters:
\begin{itemize}
\item{Chiral Mach number:
\begin{eqnarray}
{\rm Ma}_\mu = \frac{\eta\mu_0}{c_\mathrm{s}} \equiv \frac{v_{\mu}}
{c_\mathrm{s}},
\label{Ma_mu_def}
\end{eqnarray}
}
\item{Magnetic Prandtl number:
\begin{eqnarray}
\PrM = \frac{\nu}{\eta},
\end{eqnarray}
}
\item{Chiral Prandtl number:
\begin{eqnarray}
{\rm Pr}_\mu = \frac{\nu}{D_5} ,
\end{eqnarray}
}
\item{Chiral nonlinearity parameter:
\begin{eqnarray}
  \lambda_\mu = \lambda \eta^2 \meanrho ,
\label{eq_lambamu}
\end{eqnarray}
}
\item{Chiral flipping parameter:
\begin{eqnarray}
  \tilde\Gamma_\mathrm{f} = \frac{\Gamma_\mathrm{f}}{\mu_0 c_\mathrm{s}} .
\end{eqnarray}
}
\end{itemize}
Then, $D_\mu={\rm Ma}_\mu \, \PrM / {\rm Pr}_\mu$,
$\, \Lambda_\mu =\lambda_\mu/{\rm Ma}_\mu$
and ${\rm Re}_\mu = \left({\rm Ma}_\mu \, \PrM\right)^{-1}$.

\subsection{Physics of different regimes of magnetic field evolution}

There are two key nonlinear effects that determine the dynamics
of the magnetic field in chiral MHD.
The first nonlinear effect is determined by
the conservation law~(\ref{CL}) for the total chirality,
which follows from the induction equation
and the equation for the chiral magnetic potential.
The second nonlinear effect is determined by the
Lorentz force in the Navier-Stokes equation.

If the evolution of the magnetic field starts from a very small
force-free magnetic field, the
second nonlinear effect, due to the
Lorentz force, vanishes if we assume that
the magnetic field remains force-free.
The magnetic field is generated by the chiral
magnetic dynamo instability
with a maximum growth rate $\gamma^\mathrm{max}_\mu=v_\mu^2/4 \eta$
attained at the wavenumber $k_\mu=\mu_0/2$ \citep{JS97}.

Since the total chirality is conserved, the increase of
the magnetic field in the nonlinear regime results in a decrease of the
chiral chemical potential, so that the characteristic scale
at which the growth rate is maximum increases in time.
This nonlinear effect has been interpreted in terms of
an inverse magnetic cascade \citep{BFR12}.
The maximum saturated level of the magnetic field can be
estimated from the conservation law~(\ref{CL}):
$B_{\rm sat} \sim (\mu_0 k_{\rm M}/\lambda)^{1/2} < \mu_0/\lambda^{1/2}$.
Here, $\mu_{\rm sat} \ll \mu_0$ is the chiral chemical potential at saturation with
the characteristic wavenumber $k_{\rm M} < \mu_0$, corresponding
to the maximum of the magnetic energy.

However, the growing force-free magnetic field cannot stay
force-free in the nonlinear regime of the magnetic field evolution.
If the Lorentz force does not vanish, it generates small-scale velocity
fluctuations.
This nonlinear stage begins when the nonlinear term ${\UU}\times{\BB}$ in
Equation~(\ref{ind-DNS}) is of the order of the dynamo generating
term $v_\mu {\BB}$, that is, when the velocity reaches the level
of $U \sim v_\mu$.
The effect described here results in the production of
chiral magnetically driven turbulence,
with the level of turbulent kinetic energy being determined
by the balance of the nonlinear term, $(\UU \cdot \nab)\UU$, in
Equation~(\ref{UU-DNS}) and the Lorentz force, $(\nab \times {\BB}) \times \BB$,
so that the turbulent velocity can reach the Alfv\'en speed
$\vA=({\BB^2}/\meanrho)^{1/2}$.

The chiral magnetically driven turbulence causes
complicated dynamics:
it produces the mean electromotive force that includes
the turbulent magnetic diffusion
and the chiral $\alpha_\mu$ effect that
generates large-scale magnetic fields; see Paper~I.
The resulting large-scale magnetic fields are concentrated
at the wavenumber $k_\alpha = 2 k_\mu (\ln \Rm)/(3\Rm)$,
for $\Rm \gg 1$; see Paper~I.
The saturated value of the large-scale
magnetic field controlled by the conservation law~(\ref{CL}), is
$B_{\rm sat} \sim (\mu_0 k_\alpha/\lambda)^{1/2}$.
Here, $\Rm$ is the magnetic Reynolds number based on the integral
scale of turbulence and the turbulent velocity at this scale.

Depending on the chiral nonlinearity parameter $\lambda_\mu$
(see Equation~\ref{eq_lambamu}), there
are either two or three stages of magnetic field evolution.
In particular, when $\lambda_\mu$ is very small,
there is sufficient time to produce turbulence and excite the large-scale dynamo,
so that the magnetic field evolution includes three stages:
\begin{asparaenum}[\it (1)]
\item{
the small-scale chiral dynamo instability,}
\item{
the production of chiral magnetically driven MHD turbulence
and the excitation of a large-scale dynamo instability, and}
\item{
the saturation of magnetic helicity and magnetic field growth
controlled by the conservation law~(\ref{CL}).}
\end{asparaenum}

If $\lambda_\mu$ is not very small, such that the saturated value
of the magnetic field is not large, there is not enough time to excite
the large-scale dynamo instability. In this case,
the magnetic field dynamics includes two stages:
\begin{asparaenum}[\it (1)]
\item{
the chiral dynamo instability, and}
\item{
the saturation of magnetic helicity and magnetic field growth
controlled by the conservation law~(\ref{CL}) for the total chirality.}
\end{asparaenum}

\subsection{Characteristic scales of chiral magnetically driven turbulence}
\label{sec_chiScales}

In the nonlinear regime, once turbulence is fully developed,
small-scale magnetic fields can be excited over a broad range of wavenumbers
up to the diffusion cutoff wavenumber.
Using dimensional arguments and numerical simulations,
\cite{BSRKBFRK17} found that, for chiral magnetically driven turbulence,
the magnetic energy spectrum $E_{\rm M}(k,t)$ obeys
\begin{equation}
   E_{\rm M}(k,t)=C_\mu\,\meanrho\mu_0^3\eta^2k^{-2},
\label{Cmu}
\end{equation}
where $C_\mu\approx16$ is a chiral magnetic Kolmogorov-type constant.
Here, $E_{\rm M}(k,t)$ is normalized such that
$\mathcal{E}_{\rm M} = \int E_{\rm M}(k)~\mathrm{d}k=\bra{\BB^2}/2$
is the mean magnetic energy density.
It was also confirmed numerically in \cite{BSRKBFRK17}
that the magnetic energy spectrum $E_{\rm M}(k)$ is bound from above
by $C_\lambda \mu_0/\lambda$,
where $C_\lambda\approx1$ is another empirical constant.
This yields a critical minimum wavenumber,
\begin{equation}
   k_\lambda=\sqrt{\meanrho\lambda \frac{C_\mu}{C_\lambda}}\,\mu_0\eta,
\label{klambda}
\end{equation}
below which the spectrum will no longer be proportional to $k^{-2}$.

The spectrum extends to larger wavenumbers up to a diffusive cutoff wavenumber
$k_{\rm diff}$.
The diffusion scale for magnetically produced turbulence
is determined by the condition $\Lu(k_{\rm diff})=1$,
where $\Lu(k)=\vA(k)/\eta k$ is the scale-dependent Lundquist number,
$\vA(k)=(\bra{\BB^2}_k/\meanrho)^{1/2}$
is the scale-dependent Alfv\'en speed,
and $\bra{\BB^2}_k=2\int_{k_\lambda}^{k} E_{\rm M}(k)~\mathrm{d}k$.
To determine the Alfv\'en speed, $\vA(k)$, we integrate Equation~(\ref{Cmu}) over $k$
and obtain
\begin{equation}
\vA(k)=\eta \mu_0\left({2\,C_\mu\,\mu_0\over k_\lambda}\right)^{1/2}
\left(1 - {k_\lambda \over k}\right)^{1/2}.
\end{equation}
The conditions $\Lu(k_{\rm diff})=1$ and $k_{\rm diff} \gg k_\lambda$ yield
\begin{equation}
   k_{\rm diff}=\sqrt{2}\left(\frac{C_\mu C_\lambda}{\lambda_\mu}\right)^{1/4}\mu_0
              \approx2.8\,\lambda_\mu^{-1/4}\,\mu_0.
\label{eq_kdiffchi}
\end{equation}
Numerical simulations reported in \cite{BSRKBFRK17} have been performed for
$0.75 \leq k_{\rm diff}/\mu_0 \leq 75$.
In the present DNS, we use values in the range from 4.5 to 503.

\subsection{Differences between chiral and standard MHD}
\label{sec:difference}

The system of equations~(\ref{ind-DNS})--(\ref{mu-DNS}) describing chiral
MHD exhibits the following key differences from standard MHD:
\begin{itemize}
 \item{
     The presence of the term
     ${\bm \nabla} \times (\eta \, \mu \, {\BB})$
     in Equation~(\ref{ind-DNS}) causes a chiral dynamo
     instability and results in chiral magnetically driven turbulence.}
 \item{
      Because of the finite value of $\lambda$, the presence of a helical magnetic
      field affects the evolution of $\mu$; see Equation~(\ref{mu-DNS}).}
 \item{
      For $\Gamma_\mathrm{f}=0$, the total chirality,
      $\int (\half \lambda {\bm A} {\bm \cdot} \BB + \mu) \, dV$,
      is strictly conserved, and not just in the limit $\eta\to 0$.
      This conservation law determines the level of the saturated magnetic field.}
  \item{
      The excitation of a large-scale magnetic field is caused by
      (i)  the combined action of the chiral dynamo instability
           and the inverse magnetic cascade due to the conservation
           of total chirality, as well as by
      (ii) the chiral $\alpha_\mu$ effect resulting in
           chiral magnetically driven turbulence.
           This effect is not related to kinetic helicity
           and becomes dominant at large fluid and magnetic
           Reynolds numbers; see Paper~I.}
\end{itemize}

The chiral term in Equation~(\ref{ind-DNS}) and the evolution
of $\mu$ governed by Equation~(\ref{mu-DNS}) are responsible for
different behaviors in chiral and standard MHD.
In particular, in standard MHD,
the following phenomena and a conservation law are established:
\begin{itemize}
 \item{
     The magnetic helicity $\int {\bm A} {\bm \cdot} \BB \, dV$
     is only conserved in the limit of $\eta \to 0$.}
  \item{
      Turbulence does not have an intrinsic source.
      Instead, it can be produced externally by a stirring force,
      or due to large-scale shear at large fluid Reynolds numbers,
      the Bell instability in the presence of a cosmic-ray current,
      \citep{RNBE2012,BL14},
      the magnetorotational instability \citep{HGB95,BNST95},
      or just an initial irregular magnetic field \citep{BKT15}.
      }
  \item{
      A large-scale magnetic field can be generated by:
      (i) helical turbulence with nonzero mean kinetic helicity
           that is produced either by external helical forcing or by
           rotating, density-stratified, or inhomogeneous turbulence
           (so-called mean-field $\alpha^2$ dynamo);
      (ii) helical turbulence with large-scale shear, which
           results in an additional mechanism of large-scale dynamo action
           referred to as an $\alpha \Omega$ or $\alpha^2 \Omega$ dynamo
           \citep{M78,P79,KR80,ZRS83};
     (iii) nonhelical turbulence with large-scale shear, which causes
           a large-scale shear dynamo \citep{vishniac1997,RK03,RK04,sridhar2010,sridhar2014};
     and
     (iv) in different nonhelical deterministic flows due to negative
     effective magnetic diffusivity \citep[in Roberts flow IV, see][]{Devlen+13},
     or time delay of an effective pumping velocity of the magnetic field
     associated with the off-diagonal components of the $\alpha$
     tensor that are either antisymmetric (known as the $\gamma$
     effect) in Roberts flow III or symmetric in Roberts flow II;
     see \cite{Rheinhardt+14}.
     All effects in items (i)--(iv) can work in chiral MHD as well.
     However, which one of these effects is dominant depends on the flow
     properties and the
     governing parameters.
}
  \end{itemize}

\subsection{DNS with the {\sc Pencil Code}}
\label{sec:Penci-Code}

We solve Equations~(\ref{ind-NS})--(\ref{mu-NS})
numerically using the {\sc Pencil Code}.
This code uses sixth-order explicit finite differences in space
and a third-order accurate time-stepping method \citep{BD02,Bra03}.
The boundary conditions are periodic in all three directions.
All simulations presented in Sections~\ref{sec:lamdyn} and~\ref{sec_turbdynamo1}
are performed without external forcing of turbulence.
In Section~\ref{sec_turbdynamo} we apply a turbulent forcing function
$\ff$ in the Navier-Stokes equation, which consists of random plane
transverse white-in-time, unpolarized waves.
In the following, when we discuss numerical simulations,
all quantities are considered as dimensionless quantities, and we drop the
``tildes'' in Equations~(\ref{ind-NS})--(\ref{mu-NS}) from now on.
The wavenumber $k_1=2\pi/L$ is based on the size of the box $L=2\pi$.
In all runs, we set $k_1=1$, $c_\mathrm{s}=1$ and the mean fluid density $\meanrho=1$.

\begin{table}
\centering
\caption{
Overview of Runs for the Laminar $v_\mu^2$ Dynamos
(Reference Run in Bold)}
     \begin{tabular}{l|lllll}
      \hline
      \hline
      \\
	simulation 	& $\PrM$        & $\lambda_\mu$  & $\dfrac{{\rm Ma}_\mu}{10^{-3}}$ & $\dfrac{k_\lambda}{10^{-4}\mu_0}$ & $\dfrac{k_\mathrm{diff}}{\mu_0}$	\\	                        		\\
      \hline
      	La2-1B  	& $1.0$		& $1\times10^{-8}$	& $2$	& $4.0$ & $283$ \\
 	La2-2B  	& $0.5$		& $4\times10^{-8}$	& $4$	& $8.0$ & $200$ \\
     	La2-3B  	& $0.2$		& $2.5\times10^{-7}$	& $10$	& $20$ & $126$ \\
       	La2-4B  	& $2.0$		& $2.5\times10^{-9}$	& $1$	& $2.0$ & $400$ \\
      	La2-5B  	& $1.0$		& $1\times10^{-9}$	& $1.5$	& $1.3$ & $503$ \\
        La2-5G  	& $1.0$		& $1\times10^{-8}$	& $1.5$	& $4.0$ & $283$ \\
        La2-6G  	& $1.0$		& $1\times10^{-5}$	& $2$	& $130$ & $50$  \\
     	La2-7B   	& $1.0$		& $1\times10^{-9}$	& $3$	& $4.0$ & $283$ \\
        La2-7G   	& $1.0$		& $1\times10^{-9}$	& $3$	& $4.0$ & $283$ \\
    	La2-8B   	& $1.0$		& $1\times10^{-9}$	& $5$	& $4.0$ & $283$ \\
        La2-8G   	& $1.0$		& $1\times10^{-9}$	& $5$	& $4.0$ & $283$ \\
    	La2-9B   	& $1.0$		& $1\times10^{-9}$	& $10$	& $4.0$ & $283$ \\
        La2-9G   	& $1.0$		& $1\times10^{-9}$	& $10$	& $4.0$ & $283$ \\
     	La2-10B  	& $1.0$		& $1\times10^{-5}$	& $20$	& $130$ & $50$ \\
      	La2-10Bkmax  	& $1.0$		& $1\times10^{-5}$	& $20$	& $130$ & $50$ \\
     	La2-10G  	& $1.0$		& $1\times10^{-5}$	& $20$	& $4.0$ & $283$ \\
    	La2-11B   	& $1.0$		& $1\times10^{-9}$	& $50$	& $1.3$ & $503$ \\
    	La2-11G   	& $1.0$		& $1\times10^{-8}$	& $50$	& $4.0$ & $283$ \\
      	La2-12B  	& $1.0$		& $1\times10^{-9}$	& $2$	& $1.3$ & $503$ \\
      	La2-13B  	& $1.0$		& $1\times10^{-7}$	& $2$	& $13$ & $159$ \\
      	La2-14B  	& $1.0$		& $3\times10^{-9}$	& $2$	& $2.2$ & $382$ \\
      	\textbf{La2-15B}  & $\mathbf{1.0}$& $\mathbf{1\times10^{-5}}$& $\mathbf{2}$ & $\mathbf{130}$ & $\mathbf{50}$\\
      	La2-16B  	& $1.0$		& $3\times10^{-8}$	& $2$	& $6.9$ & $215$ \\
      \hline
      \hline
    \end{tabular}
  \label{table_simulations_vmu2}
\end{table}
\medskip

%%%%%%%%%%%%%
% Section 3 %
%%%%%%%%%%%%%

\section{Laminar chiral dynamos}
\label{sec:lamdyn}

In this section, we study numerically laminar chiral dynamos
in the absence of any turbulence (externally or chiral magnetically driven).

\subsection{Numerical setup}

Parameters and initial conditions for all
laminar dynamo simulations are listed in
Tables~\ref{table_simulations_vmu2} and~\ref{table_simulations_vmushear}.
All of these simulations are two-dimensional and have a resolution of $256^2$.
Runs with names ending with `B' are with the initial conditions
for the magnetic field in the form of a Beltrami magnetic field:
$B(t=0)=10^{-4}$(0, sin$\,x$, cos$\,x$),
while runs with names ending with `G' are initiated with Gaussian noise.
The initial conditions for the velocity field for the laminar $v_\mu^2$ dynamo
are $U(t=0)=(0,0,0)$, and for the laminar chiral--shear dynamos
(the $v_\mu^2$--shear or $v_\mu$--shear dynamos) are
$U(t=0)=(0, S_0~\mathrm{cos}\,x, 0)$, with
the dimensionless shear rate $S_0$ given for all runs.

We set the chiral Prandtl number $\Prmu=1$ in all runs.
In many runs the magnetic Prandtl number $\PrM=1$ (except in several

runs for the laminar $v_\mu^2$ dynamo, see Table~\ref{table_simulations_vmu2}).
The reference runs for the laminar $v_\mu^2$ dynamo (La2-15B)
and the chiral--shear dynamos (LaU-4G) are shown in bold in
Tables~\ref{table_simulations_vmu2} and~\ref{table_simulations_vmushear}.
The results of numerical simulations are compared with
theoretical predictions.

\subsection{Laminar $v_\mu^2$ dynamo}
\label{sec_laminara2}

We start with the situation without an imposed fluid flow,
where the chiral laminar $v_\mu^2$ dynamo can be excited.

\subsubsection{Theoretical aspects}

In this section, we outline the theoretical predictions
for a laminar chiral dynamo; for details see Paper~I.
To determine the chiral dynamo growth rate, we seek a solution of
the linearized Equation~(\ref{ind-DNS})
for small perturbations of the following form:
$\BB(t,x,z)=B_y(t,x,z) {\bm e}_y + \nab   \times  [A(t,x,z) {\bm e}_y]$,
where ${\bm e}_y$ is the unit vector in the $y$ direction.

We consider the equilibrium configuration:
$\mu=\mu_0=\const$ and ${\UU}_0=0$.
The functions $B_y(t,x,z)$ and $A(t,x,z)$ are determined by the equations
\begin{eqnarray}
 \frac{\partial A(t,x,z)}{\partial t} &=& v_\mu \, B_y + \eta \Delta A,
\label{A-eq} \\
\frac{\partial B_y(t,x,z)}{\partial t}&=& -v_\mu \, \Delta A + \eta \Delta B_y ,
\label{By-eq}
\end{eqnarray}
where $v_\mu=\eta \, \mu_0$,  $\Delta=\nabla_x^2 + \nabla_z^2$,
and the remaining components of the magnetic field are given by
$B_x=-\nabla_z A$ and $B_z=\nabla_x A$.
We seek a solution to Equations~(\ref{A-eq}) and~(\ref{By-eq}) of the form
$A, B_y \propto \exp[\gamma t + i (k_x x + k_z z)]$.
The growth rate of the dynamo instability is given by
\begin{eqnarray}
\gamma = |v_\mu \, k| - \eta k^2 ,
\label{gamma}
\end{eqnarray}
where $k^2=k_x^2 + k_z^2$.
The dynamo instability is excited (i.e., $\gamma> 0)$ for $k < |\mu_0|$.
The maximum growth rate of the dynamo instability,
\begin{eqnarray}
\gamma^{\rm max}_\mu = \frac{v_\mu^2}{4 \eta},
\label{gamma-max}
\end{eqnarray}
is attained at
\begin{equation}
k_\mu =\frac12|\mu_0|.
\label{eq_kmax}
\end{equation}

\subsubsection{Time evolution}

%%%%%%
\begin{figure}[t]
\begin{center}
   \includegraphics[width=\columnwidth]{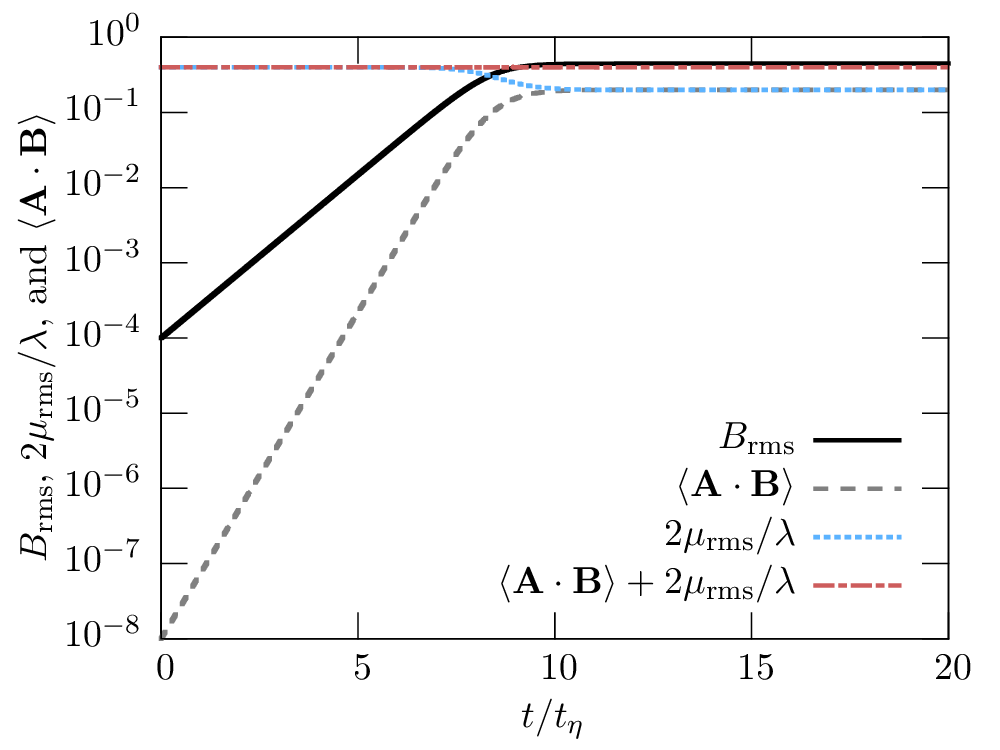}
\end{center}
\caption[]{\textbf{Laminar $v_\mu^2$ dynamo:}
time evolution of $B_\mathrm{rms}$ (solid black line),
$\langle \mathbf{A} \cdot \mathbf{B} \rangle$ (dashed gray line),
$\mu_\mathrm{rms}$ (multiplied by
$2/\lambda$, dotted blue line), and
$\langle \mathbf{A} \cdot \mathbf{B} \rangle +2\mu_\mathrm{rms}/\lambda$
(dash-dotted red line) for reference run La2-15B
(see table \ref{table_simulations_vmu2}).
}
\label{fig__La2_15B__ts}
\end{figure}
%%%%%%

In Figure~\ref{fig__La2_15B__ts} we show the time evolution
of the rms magnetic field $B_\mathrm{rms}$,
the magnetic helicity $\langle \mathbf{A} \cdot \mathbf{B} \rangle$,
the chemical potential $\mu_\mathrm{rms}$ (multiplied by a factor of
$2/\lambda$), and $\langle \mathbf{A} \cdot \mathbf{B} \rangle
+2\mu_\mathrm{rms}/\lambda$ for reference run La2-15B.
In simulations, the time is measured in units of diffusion time
$t_\eta=(\eta k_1^2)^{-1}$.
The initial conditions for the magnetic field are chosen in the form of
a Beltrami field on $k=k_1=1$.

The magnetic field is amplified
exponentially over more than four orders
of magnitude until it saturates after roughly eight diffusive times.
Within the same time, the magnetic helicity
$\langle \mathbf{A} \cdot \mathbf{B} \rangle$ increases
over more than eight orders of magnitude.
Since the sum of magnetic helicity and $2 \mu/\lambda$ is conserved,
the chemical potential $\mu$ decreases, in a nonlinear era
of evolution, from the initial value $\mu_0=2$ to $\mu=1$,
resulting in a saturation of the laminar $v_\mu^2$ dynamo.

%%%%%%
\begin{figure}[t!]
\begin{center}
\includegraphics[width=\columnwidth]{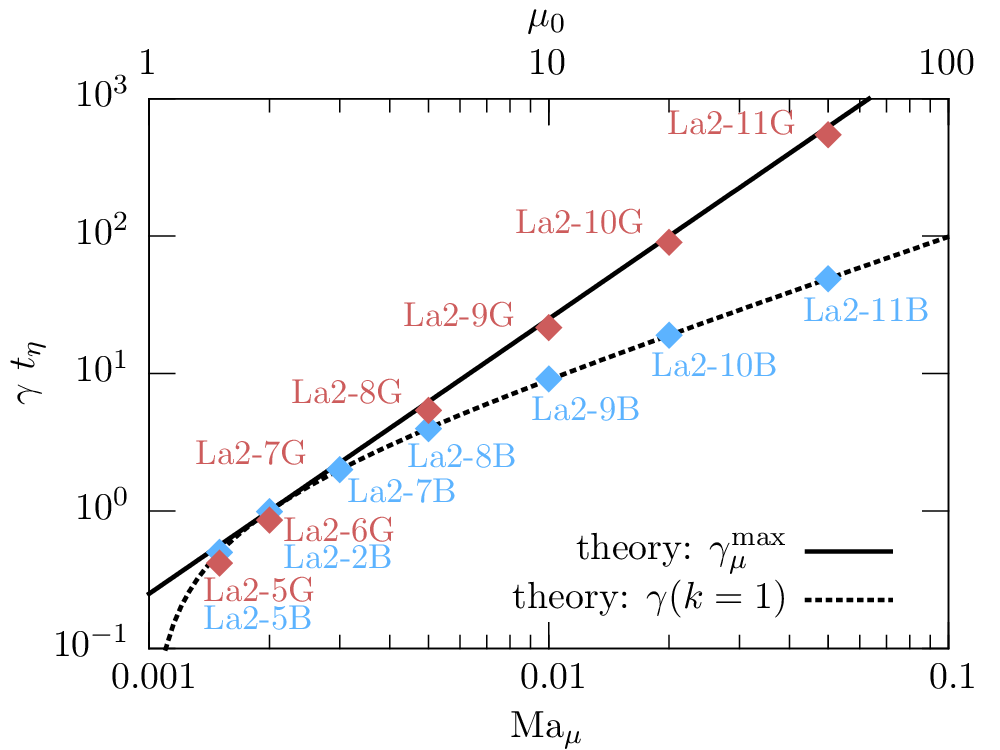}
\end{center}
\caption[]{\textbf{Laminar $v_\mu^2$ dynamo:}
growth rates as a function of ${\rm Ma}_\mu$,
for simulations with $\mu_0=2$.
The black line is the theoretical prediction for the maximum growth rate
$\gamma^\mathrm{max}_\mu$
(see Equation~\ref{gamma-max}) that is attained at
$k_\mu=\mu_0/2=1$
(see Equation~\ref{eq_kmax}).
The runs with Gaussian initial fields, shown as red diamonds,
lie on the theoretically predicted
$\gamma^\mathrm{max}_\mu$.
The dotted line corresponds to the theoretical prediction for the growth rate
$\gamma(k=1)$ at the scale of the box.
The runs with an initial magnetic Beltrami field on $k=1$,
shown as blue diamonds, lie on the theoretically predicted dotted
curve $\gamma(k=1)$.
}
\label{fig__Gamma_eta}
\end{figure}
%%%%%%

\subsubsection{Dynamo growth rate}

In Figure~\ref{fig__Gamma_eta}, we show the growth rate of the magnetic field
as a function of the chiral Mach number, ${\rm Ma}_\mu$.
The black solid line
in this figure shows the theoretical prediction for the maximum
growth rate $\gamma^\mathrm{max}_\mu$ that is attained at $k_\mu=\mu_0/2=1$;
see Equations~(\ref{gamma-max}) and (\ref{eq_kmax}).
When the initial magnetic field is distributed over all spatial scales,
like in the case of initial magnetic Gaussian noise, in which
there is a nonvanishing magnetic field at
$k_\mu$; that is inside the computational domain,
the initial magnetic field is excited with the maximum growth rate
as observed in the simulations.
Consequently, the runs with Gaussian initial fields shown as red diamonds in
Figure~\ref{fig__Gamma_eta}, lie on the theoretical curve
$\gamma^\mathrm{max}_\mu$.
The dotted line in Figure~\ref{fig__Gamma_eta} corresponds to the theoretical
prediction for the growth rate $\gamma$ at the scale of the box $(k=1)$.
The excitation of the magnetic field from an initial Beltrami field on $k=1$
occurs with growth rates in agreement with the theoretical dotted curve;
see blue diamonds in Figure~\ref{fig__Gamma_eta}.

\subsubsection{Dependence on initial conditions}

%%%%%%
\begin{figure}[t]\begin{center}
\includegraphics[width=\columnwidth]{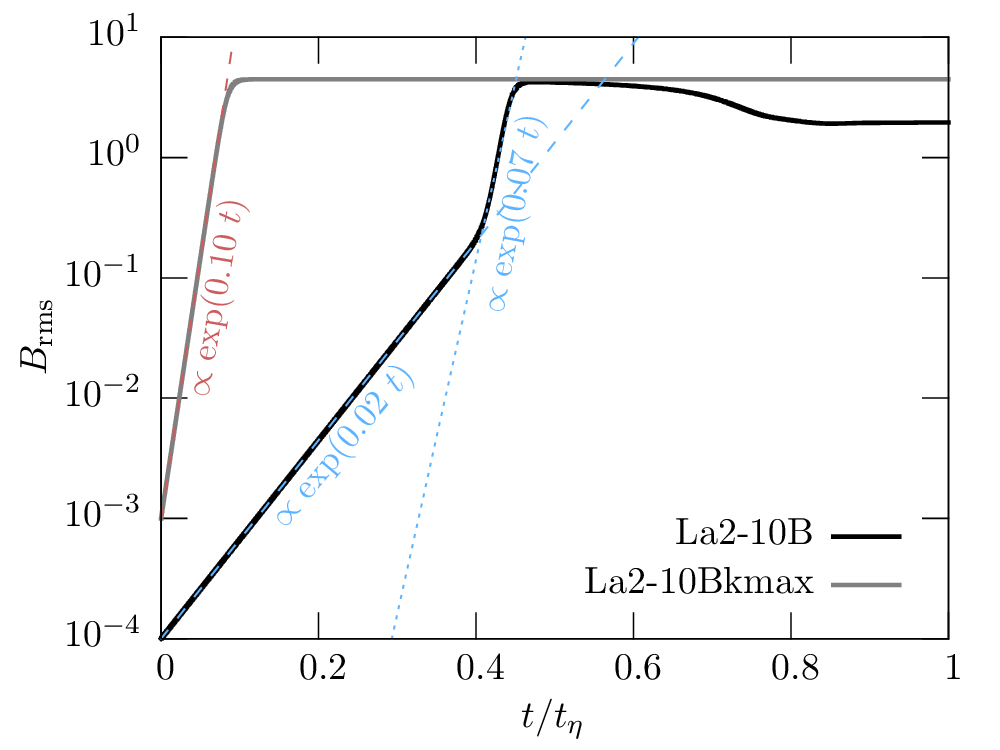}
\end{center}
\caption[]{\textbf{Laminar $v_\mu^2$ dynamo:}
Time evolution of $B_\mathrm{rms}$ for two different initial conditions.
The black line is for the dynamo instability started from an initial
Beltrami field at $k=1$ (run La2-10B), while the blue line
is for an initial Beltrami field with
$k=10$ (run La2-10Bkmax).
Fits in different regimes are indicated by thin lines.
Both runs are for the initial value $\mu_0=20$, so that $k_\mu=10$;
and $\gamma^\mathrm{max}_\mu=0.1$ (see~Equation~\ref{gamma-max}).
}
\label{fig__La2_B_t}
\end{figure}
%%%%%%

The initial conditions for the magnetic field are important mostly at early times.
If the magnetic field is initially concentrated on the box scale, we expect to
observe a growth rate $\gamma(k=1)$ as given by Equation~(\ref{gamma}).
At later times, the spectrum of the magnetic field can, however, be changed,
due to mode coupling, and be amplified with a larger growth rate.
This behavior is observed in Figure~\ref{fig__La2_B_t}, where
an initial Beltrami field with $k=10$ is excited with maximum growth rate,
since $\mu_0=20$.
In Figure~\ref{fig__La2_B_t} we also consider another situation
where the dynamo is started from an initial Beltrami field with $k=1$ (La2-10B).
In this case, the dynamo starts with a growth rate $\gamma=0.019$,
which is consistent with the theoretical prediction for $\gamma(k=1)$.
Later, after approximately $0.4\,t_\eta$, the dynamo growth
rate increases up to the value $\gamma=0.07$, which is close to
the maximum growth rate $\gamma^\mathrm{max}_\mu=0.1$.

\subsubsection{Saturation}

\begin{figure}[t]\begin{center}
\includegraphics[width=\columnwidth]{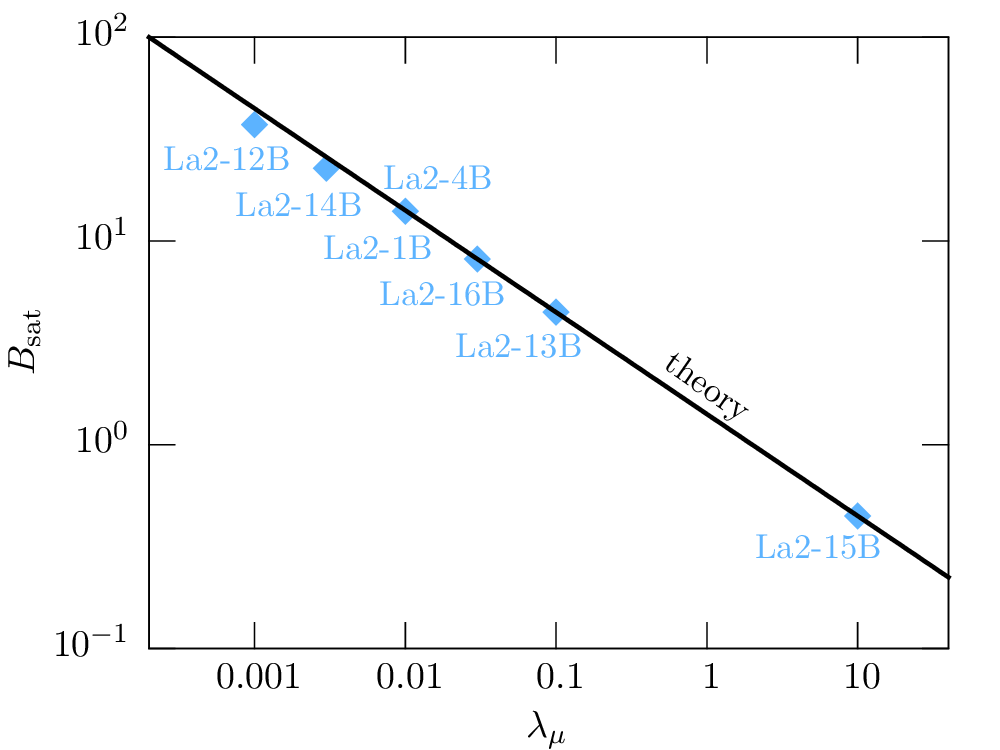}
\end{center}
\caption{{\bf Laminar $v_\mu^2$ dynamo:}
the saturation magnetic field strength for simulations
with different $\lambda_\mu$.
Details for the different runs, given by labeled blue diamonds, can be found in
Table~\ref{table_simulations_vmu2}.}
\label{fig_Bsat_lambda}
\end{figure}

The parameter $\lambda$ in the evolution Equation~(\ref{mu-DNS}),
or the corresponding dimensionless parameter $\lambda_\mu$
in Equation~(\ref{mu-NS}), for the chiral chemical potential determines
the nonlinear saturation of the chiral dynamo.
We determine the saturation value of the
magnetic field $B_\mathrm{sat}$ numerically for different values of $\lambda_\mu$;
see Figure~\ref{fig_Bsat_lambda}.
We find that the saturation value of the magnetic field
increases with decreasing $\lambda_\mu$.
This can be expected from the conservation law (\ref{CL}).
If the initial magnetic energy is very small,
we find from Equation~(\ref{CL}) the following estimate
for the saturated magnetic field during laminar chiral dynamo action:
\begin{eqnarray}
   B_\mathrm{sat} \sim \left[\frac{\mu_0(\mu_0 -\mu_\mathrm{sat})}{\lambda}\right]^{1/2},
\label{eq_Bsat}
\end{eqnarray}
where $\mu_\mathrm{sat}$ is the chiral chemical potential at saturation, and
we use the estimate $A$ by $2 B /\mu_0$.
Inspection of Figure~\ref{fig_Bsat_lambda} demonstrates
a good agreement between
theoretical (solid line) and numerical results (blue diamonds).

%%%%%%
%%%%%%

\subsubsection{Effect of a nonvanishing flipping rate}
\label{sec_flip}

In this section, we consider the influence of a nonvanishing chiral
flipping rate on the $v_\mu^2$ dynamo.
A large flipping rate $\Gamma_\mathrm{f}$ decreases the chiral chemical
potential $\mu$; see Equation~(\ref{mu-DNS}). It can stop
the growth of the magnetic field caused by the chiral dynamo instability.

\begin{figure}[t]\begin{center}
\includegraphics[width=\columnwidth]{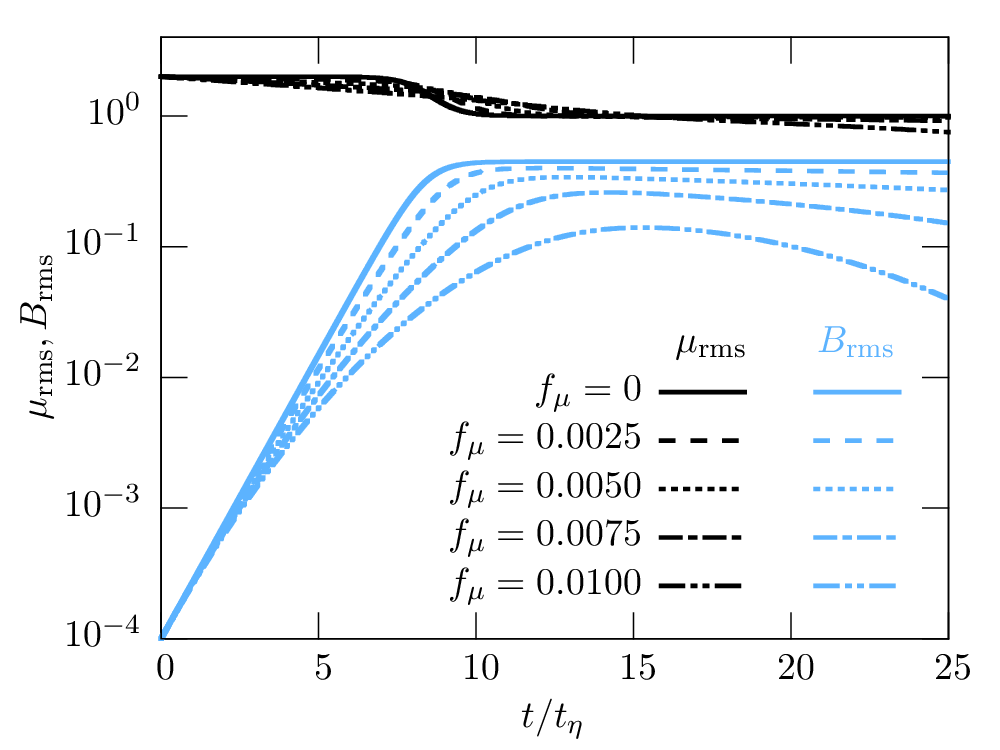}
\end{center}
\caption{{\bf Laminar $v_\mu^2$ dynamo:}
time evolution of the chiral chemical potential $\mu_{\rm rms}$ (black lines) and
the magnetic field $B_{\rm rms}$ (blue lines)
for $f_\mu=0$ (solid), $f_\mu = 0.0025$ (dashed), and $f_\mu = 0.01$ (dotted).
}
\label{fig_ts_flip}
\end{figure}

Quantitatively, the influence of the flipping term can be estimated by
comparing the last two terms of Equation~(\ref{mu-DNS}).
The ratio of these terms is
\begin{eqnarray}
   f_\mu \equiv \frac{\Gamma_\mathrm{f}\mu_0}{\lambda\eta\mu_0 B_\mathrm{sat}^2}
   = \frac{\Gamma_\mathrm{f}}{\eta\mu_0^2},
\label{eq_fmu}
\end{eqnarray}
where we have used Equation~(\ref{eq_Bsat}) with $\mu_\mathrm{sat} \ll \mu_0$
for the saturation value of the magnetic field strength.
In Figure~\ref{fig_ts_flip} we present the time evolution of
$B_\mathrm{rms}$ and $\mu_\mathrm{rms}$
for different values of $f_\mu$.
The reference run La2-15B, with zero flipping rate ($f_\mu=0$),
has been repeated with a finite flipping term.
As a result, the magnetic field grows more
slowly in the nonlinear era, due to the flipping effect, and it
decreases the saturation level of the magnetic field; see Figure~\ref{fig_ts_flip}.
For larger values of $f_\mu$, the chiral chemical potential
$\mu$ decreases quickly, leading to strong quenching of the $v_\mu^2$ dynamo;
see the blue lines in Figure~\ref{fig_ts_flip}.

%%%%%%

\subsection{Laminar chiral--shear dynamos}
\label{sec_laminarashear}

In this section, we consider laminar chiral dynamos
in the presence of an imposed shearing velocity.
Such a nonuniform velocity profile can be created
in different astrophysical flows.

\subsubsection{Theoretical aspects}

We start by outlining the theoretical predictions
for laminar chiral dynamos in the presence
of an imposed shearing velocity; for details see Paper~I.
We consider the equilibrium configuration specified by the shear velocity
${\UU}_{\rm eq}=(0,S\, x,0)$,
and $\mu=\mu_0=$ const.
This implies that the fluid has nonzero vorticity
${\bm W} = (0,0,S)$ similar to differential (nonuniform) rotation.
The functions $B_y(t,x,z)$ and $A(t,x,z)$ are determined by
\begin{eqnarray}
&&\frac{\partial A(t,x,z)}{\partial t} = v_\mu \,  B_y + \eta \Delta A,
\label{A1-eq}\\
&&\frac{\partial B_y(t,x,z)}{\partial t} = - S\nabla_z A - v_\mu \,  \Delta A
+ \eta \Delta B_y .
\label{By1-eq}
\end{eqnarray}
We look for a solution to Equations~(\ref{A1-eq})
and~(\ref{By1-eq}) of the form $A, B_y \propto \exp[\gamma t + i (k_x x + k_z z-
\omega t)]$.
The growth rate of the dynamo instability and the frequency of the dynamo waves
are given by
\begin{eqnarray}
 \gamma = {|v_\mu \, k| \over \sqrt{2}} \,
\left\{1 + \left[1 + \left({S k_z \over v_\mu \, k^2}\right)^2 \right]^{1\over 2}
\right\}^{1\over 2} - \eta k^2
\label{eq_gamma_aS}
\end{eqnarray}
and
\begin{eqnarray}
 \omega=  {\rm sgn} \left(\mu_0  k_z\right) \, {S k_z \over \sqrt{2} k}
\,
\left\{1 + \left[1 + \left({S k_z \over v_\mu \, k^2}\right)^2\right]^{1\over 2}
\right\}^{-{1\over 2}} ,
\nonumber\\
\label{omega10}
\end{eqnarray}
respectively.
This solution describes a laminar $v_\mu^2$--shear dynamo
for arbitrary values of the shear rate $S$.

Next, we consider a situation where the shear term on the
right side of Equation~(\ref{By1-eq}) dominates,
that is, where $|S \nabla_z A| \gg |v_\mu \, \Delta A|$.
The growth rate of the dynamo instability
and the frequency of the dynamo waves are then given by
\begin{eqnarray}
&& \gamma = \left({ |v_\mu  \, S \, k_z| \over 2}\right)^{1/2} - \eta k^2 ,
\label{gamma1}\\
&& \omega= {\rm sgn} \left(\mu_0  k_z\right) \, \left({|v_\mu  \, S \,
k_z| \over 2}\right)^{1/2} .
\label{omega}
\end{eqnarray}
The dynamo is excited for $k < |v_\mu \, S \, k_z /2\eta^2|^{1/4}$.
The maximum growth rate of the dynamo instability
and the frequency $\omega=\omega (k=k_z^\mu)$
of the dynamo waves are attained at
\begin{equation}
  k_z^\mu ={1 \over 4} \left({2|S \, v_\mu| \over \eta^2} \right)^{1/3},
\label{kz-max}
\end{equation}
and are given by
\begin{eqnarray}
&& \gamma^{\rm max}_\mu
  = {3 \over 8} \left({S^2 \, v_\mu^2 \over 2\eta}\right)^{1/3}
  - \eta k_x^2,
\label{gam-max}\\
&& \omega(k=k_z^\mu)
  = {{\rm sgn} \left(v_\mu \, k_z\right) \over 2\eta}  \,
  \left({S^2 \, v_\mu^2 \over 2 \eta}\right)^{1/3} .
\label{omega-max}
\end{eqnarray}
This solution describes the laminar $v_\mu$--shear dynamo.

\subsubsection{Simulations of the laminar $v_\mu$--shear dynamo}

\begin{table}
\centering
\caption{
Overview of Runs for the Chiral--Shear Dynamos
(Reference Run in Bold)}
     \begin{tabular}{l|llllll}
      \hline
      \hline
      \\
	simulation 	& $\lambda_\mu$  	& $\dfrac{{\rm Ma}_\mu}{10^{-3}}$ & $u_S$	& $\dfrac{k_\lambda}{10^{-4}\mu_0}$ & $\dfrac{k_\mathrm{diff}}{\mu_0}$	\\\\	                        		\\
      \hline
   	LaU-1B	   	& $1\times10^{-9}$	& $2.0$	& 0.01	& 1.3 & 503 \\
   	LaU-1G	   	& $1\times10^{-9}$	& $2.0$	& 0.01	& 1.3 & 503 \\
   	LaU-2B	   	& $1\times10^{-9}$	& $2.0$	& 0.02	& 1.3 & 503 \\
   	LaU-2G	   	& $1\times10^{-9}$	& $2.0$	& 0.02	& 1.3 & 503 \\
   	LaU-3B	   	& $1\times10^{-9}$	& $2.0$	& 0.05	& 1.3 & 503 \\
   	LaU-3G	   	& $1\times10^{-9}$	& $2.0$	& 0.05	& 1.3 & 503 \\
   	LaU-4B	   	& $1\times10^{-9}$	& $2.0$	& 0.10	& 1.3 & 503 \\
   	\textbf{LaU-4G}   & $\mathbf{1\times10^{-5}}$	& $\mathbf{2.0}$ & $\mathbf{0.10}$ & $\mathbf{126}$ &  $\mathbf{50}$ \\
   	LaU-5B	   	& $1\times10^{-9}$	& $2.0$	& 0.20 	& 1.3 & 503 \\
   	LaU-5G	   	& $1\times10^{-9}$	& $2.0$	& 0.20  & 1.3 & 503 \\
   	LaU-6B	   	& $1\times10^{-9}$	& $2.0$ & 0.50	& 1.3 & 503 \\
   	LaU-6G	   	& $1\times10^{-9}$	& $2.0$ & 0.50	& 1.3 & 503 \\
   	LaU-7G	   	& $1\times10^{-8}$	& $10$  & 0.01	& 4.0 & 283 \\
   	LaU-8G	   	& $1\times10^{-8}$	& $10$  & 0.05	& 4.0 & 283 \\
   	LaU-9G	   	& $1\times10^{-8}$	& $10$  & 0.10	& 4.0 & 283 \\
   	LaU-10G		& $1\times10^{-8}$	& $10$	& 0.50	& 4.0 & 283 \\
      \hline
      \hline
    \end{tabular}
  \label{table_simulations_vmushear}
\end{table}

%%%%%%
\begin{figure}[t]
\begin{center}
   \includegraphics[width=\columnwidth]{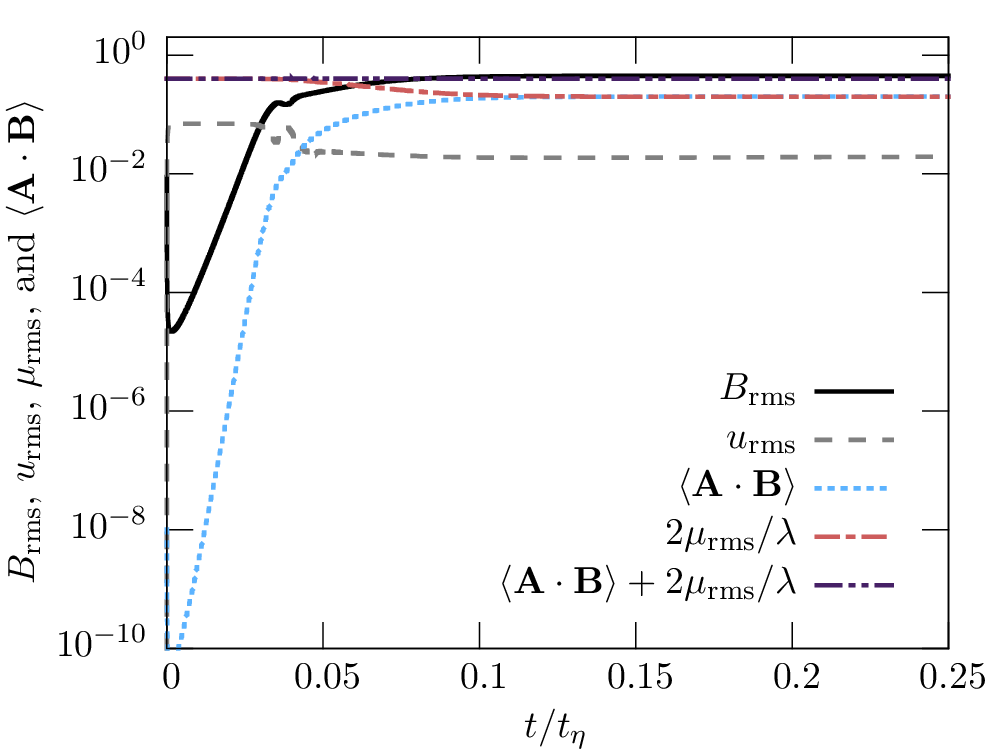}
\end{center}
\caption[]{\textbf{Laminar $v_\mu$--shear dynamo:}
time evolution of the magnetic field
$B_\mathrm{rms}$, the velocity $u_\mathrm{rms}$, the magnetic helicity
$\langle \mathbf{A} \cdot \mathbf{B} \rangle$,
the chemical potential $\mu_\mathrm{rms}$ (multiplied by a factor
of $2/\lambda$), and $\langle \mathbf{A} \cdot \mathbf{B}
\rangle +2\mu_\mathrm{rms}/\lambda$
(run LaU-4G).}
\label{fig_AlphaShear_t}
\end{figure}
%%%%%%

Since our simulations have periodic boundary conditions, we model shear
velocities as $U_S=(0, u_S \cos x, 0)$.
The mean shear velocity $\overline{u}_S$ over half the box is
$\overline{u}_S = (2/\pi) u_S$.
In Figure~\ref{fig_AlphaShear_t} we show the time evolution of the magnetic field
(which starts to be excited from a Gaussian initial field),
the velocity $u_\mathrm{rms}$, the magnetic helicity
$\langle \mathbf{A} \cdot \mathbf{B} \rangle$,
the chemical potential $\mu_\mathrm{rms}$ (multiplied by a factor
of $2/\lambda$), and $\langle \mathbf{A} \cdot \mathbf{B}
\rangle +2\mu_\mathrm{rms}/\lambda$ for run LaU-4G.
The growth rate for the chiral--shear dynamo (the $v_\mu^2$--shear dynamo)
is larger than that for the laminar chiral dynamo (the $v_\mu^2$--dynamo).
After a time of roughly $0.03~t_\eta$, the system enters a nonlinear
phase, in which the velocity field is affected by the magnetic field,
but the magnetic field can still increase slowly.
Saturation of the dynamo occurs after approximately $0.1~t_\eta$.

For Gaussian initial fields, we have observed a short delay in the
growth of the magnetic field.
In both cases, the dynamo growth rate increases with increasing shear.
As for the chiral $v_\mu^2$ dynamo, we observe perfect
conservation of the quantity $\langle \mathbf{A} \cdot \mathbf{B} \rangle +
2\mu_\mathrm{rms}/\lambda$ in the simulations of the laminar $v_\mu$--shear
dynamo.

%%%%%%
\begin{figure}[t]
  \begin{center}
    \includegraphics[width=\columnwidth]{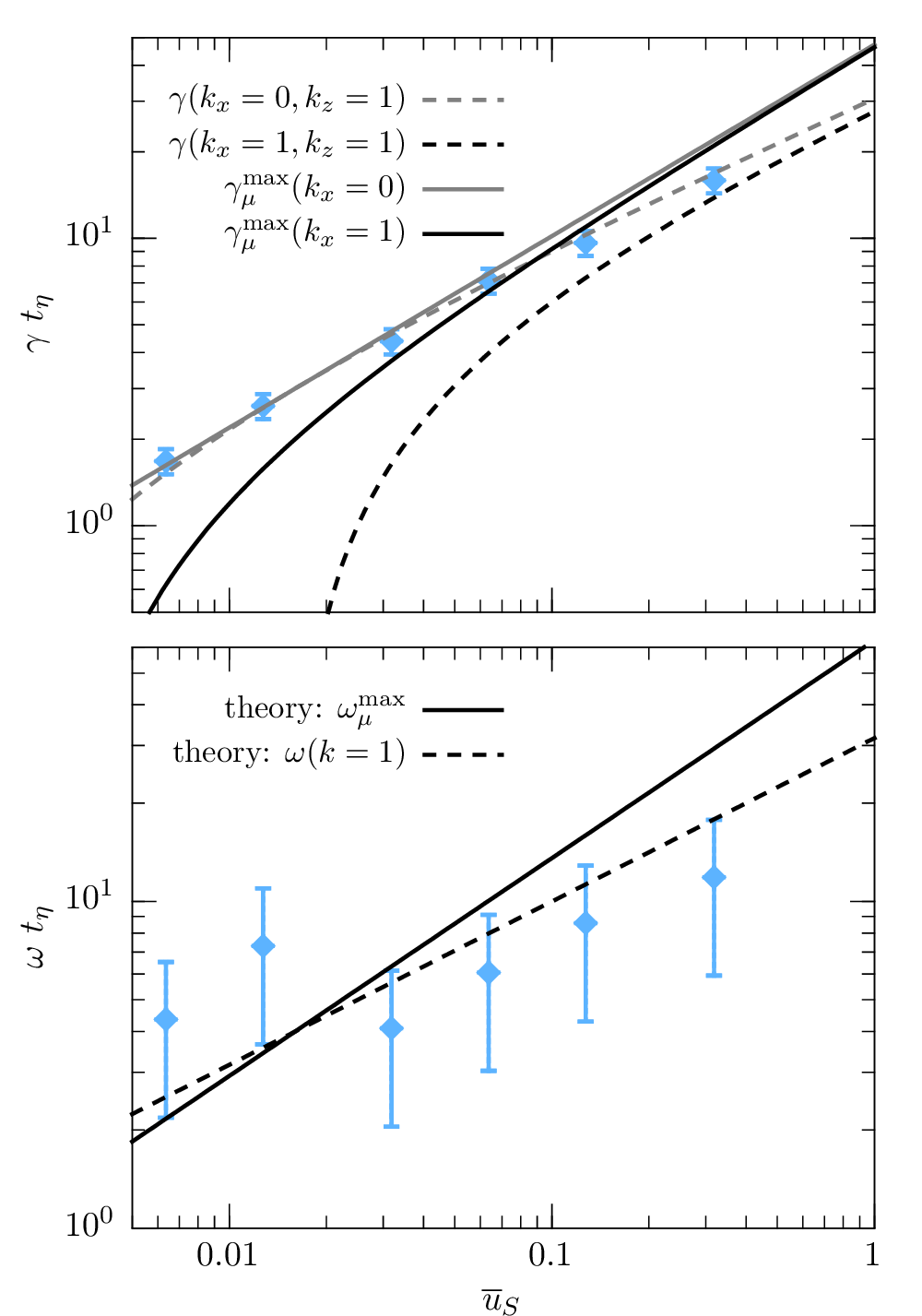}
\end{center}
\caption[]{\textbf{Laminar $v_\mu$--shear dynamo:}
growth rate (top panel) and dynamo frequency (bottom panel) as a function
of the mean shear $\overline{u}_S$ for the Beltrami initial field
(runs LaU-$n$B with $n=1$--$6$; see Table~\ref{table_simulations_vmushear}).}
\label{Gamma_Us_Beltrami}
\end{figure}
%%%%%%

In Figure~\ref{Gamma_Us_Beltrami} we show
the theoretical dependence of the growth rate $\gamma$ and the dynamo frequency
$\omega$ on the shear velocity $\overline{u}_S$
for Beltrami initial conditions at different wavenumbers;
see Equations~(\ref{gamma1}) and~(\ref{gam-max}).
The dynamo growth rate is estimated from an exponential fit.
The result of the fit depends slightly on the fitting regime, leading
to an error of the order of 10\%.
The dynamo frequency is determined afterward by dividing the magnetic field
strength by $\mathrm{exp}(\gamma t)$ and fitting a sine function.
Due to the small amplitude and a limited number of periods of dynamo waves,
the result is sensitive to the fit regime considered.
Hence we assume a conservative error of 50\% for the dynamo frequency.
The blue diamonds correspond to the numerical results.
Within the error bars, the theoretical and numerical results are in agreement.

\subsubsection{Simulations of the laminar $v_\mu^2$--shear dynamo}

\begin{figure}[t]
\begin{center}
   \includegraphics[width=\columnwidth]{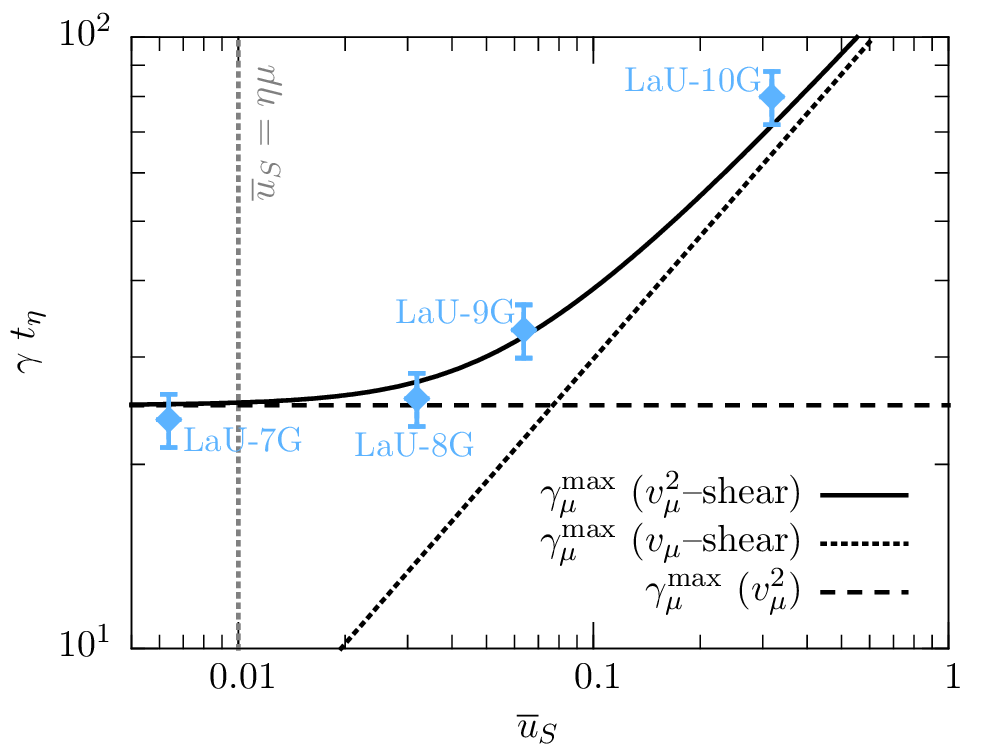}
\end{center}
\caption[]{\textbf{Laminar $v_\mu^2$--shear dynamo:}
growth rate $\gamma$ as a function of
mean shear $\overline{u}_S$.
For comparison, we plot the maximum growth rate of $v_\mu^2$ dynamo
(\ref{gamma-max}) and of the $v_\mu$--shear dynamo (\ref{gam-max}).
The solid black line is the theoretically predicted maximum growth rate
(see Equation~\ref{eq_gamma_aS}).
}\label{fig_Gamma_Us_aShear}
\end{figure}
%%%%%%

The growth rate of chiral--shear dynamos versus mean shear in
the range between $u_S=0.01$
and $0.5$ is shown in Figure~\ref{fig_Gamma_Us_aShear}.
We choose a large initial value of the chemical potential, i.e.\
$\mu_0=10$, to ensure that $k_\mathrm{max}$ is
inside the box for all values of $\overline{u}_S$.
We overplot the growth rates found from the simulations with the maximum growth
rate given by Equation (\ref{eq_gamma_aS}).
In addition, we show the theoretical predictions for the limiting cases
of the $v_\mu^2$ and
$v_\mu$--shear dynamos; see Equations~(\ref{gamma-max}) and (\ref{gam-max}).
Inspection of Figure~\ref{fig_Gamma_Us_aShear} shows that the results
obtained from the simulations agree with theoretical predictions.

%%%%%%%%%%%%%
% Section 4 %
%%%%%%%%%%%%%

\section{Chiral magnetically driven turbulence}
\label{sec_turbdynamo1}

%%%%%%%%%%%%%%%%%%%%%%%%%%%%%%%%
% link to simulation:
% La2-17: /pencil-code/jenny/chiral_fluids/laminar_dynamos/alpha2/576_3d_eta1e-4_nu1e-4_l2e1_mu020_Gaussian
%%%%%%%%%%%%%%%%%%%%%%%%%%%%%%%%

In this section we show that the CME
can drive turbulence via the Lorentz force in the Navier-Stokes equation.
When the magnetic field increases exponentially, due to the
small-scale chiral magnetic dynamo with growth rate $\gamma$,
the Lorentz force,
$({\bm \nabla} {\bm \times} {\BB}) {\bm \times} \BB$,
increases at the rate $2\gamma$.
The laminar dynamo occurs only up to the first nonlinear phase,
when the Lorentz force starts to produce turbulence
(referred to as chiral magnetically driven turbulence).
We will also demonstrate here that, during the second nonlinear phase,
a large-scale dynamo is excited by the chiral $\alpha_\mu$ effect
arising in chiral magnetically driven turbulence.
The chiral $\alpha_\mu$ effect was studied using different
analytical approaches in Paper~I.
This effect is caused by an interaction of the CME
and fluctuations of the small-scale
current produced by tangling magnetic fluctuations.
These fluctuations are generated by tangling of the large-scale
magnetic field through sheared velocity fluctuations.
Once the large-scale magnetic field becomes strong enough,
the chiral chemical potential decreases, resulting in the saturation
of the large-scale dynamo instability.

This situation is similar to that of
driving small-scale turbulence via the Bell instability
in a system with an external cosmic-ray current
\citep{B04,BL14}, and the generation of a
large-scale magnetic field by the Bell turbulence; see \citet{RNBE2012}
for details.

\subsection{Mean-field theory for large-scale dynamos}
\label{sec_meanfieldMHD}

In this section, we outline the theoretical predictions
for large-scale dynamos based on mean-field theory;
see Paper~I for details.
The mean induction equation is given by
\begin{eqnarray}
\frac{\partial \meanBB}{\partial t} &=&
\nab   \times   \left[\meanUU  \times   \meanBB
+ (\meanv_\mu + \alpha_\mu) \meanBB
- (\eta+ \, \eta_{_{T}})\nab   \times   \meanBB\right], \nonumber\\
\label{ind4-eq}
\end{eqnarray}
where $\meanv_\mu = \eta \meanmu_{0}$, and we consider the following equilibrium
state: $\meanmu_{\rm eq}=\meanmu_{0}=\const$ and ${\bm \meanUU}_{\rm eq}=0$.
This mean-field equation contains additional terms that are
related to the chiral $\alpha_\mu$ effect and the turbulent magnetic diffusivity
$\eta_{_{T}}$.
In the mean-field equation, the chiral $v_\mu$ effect is replaced
by the mean chiral $\meanv_\mu$ effect.
Note, however, that at large fluid and magnetic Reynolds numbers, the $\alpha_\mu$ effect
dominates the $\meanv_\mu$ effect.

To study the large-scale dynamo,
we seek a solution to Equation~(\ref{ind4-eq}), for small perturbations in
the form
$\meanBB(t,x,z)=\meanB_y(t,x,z) {\bm e}_y + \nab   \times
[\meanA(t,x,z) {\bm e}_y]$,
where ${\bm e}_y$ is the unit vector directed along the $y$ axis.
The functions $\meanB_y(t,x,z)$ and $\meanA(t,x,z)$ are determined by
\begin{multline}
  \frac{\partial \meanA(t,x,z)}{\partial t}
  =(\meanv_\mu + \alpha_\mu)\, \meanB_y
 + (\eta+ \, \eta_{_{T}}) \, \Delta \meanA,
 \label{me-A-eq}
\end{multline}
\begin{multline}
  \frac{\partial \meanB_y(t,x,z)}{\partial t}
  =-(\meanv_\mu + \alpha_\mu) \, \Delta \meanA
 + (\eta+ \, \eta_{_{T}}) \, \Delta \meanB_y ,
\label{me-By-eq}
\end{multline}
where $\Delta=\nabla_x^2 + \nabla_z^2$, and the other components of the magnetic
field are $\meanB_x=-\nabla_z \meanA$ and $\meanB_z=\nabla_x \meanA$.

We look for a solution of the mean-field equations~(\ref{me-A-eq})
and~(\ref{me-By-eq}) in the form
\begin{eqnarray}
  \meanA, \meanB_y \propto \exp[\gamma t + i (k_x x + k_z z)],
\end{eqnarray}
where the growth rate of the large-scale dynamo instability is given by
\begin{eqnarray}
\gamma = |(\meanv_\mu + \alpha_\mu)\, k| - (\eta+ \, \eta_{_{T}}) \, k^2,
\label{gamma_turb}
\end{eqnarray}
with $k^2=k_x^2 + k_z^2$.
The maximum growth rate of the large-scale dynamo instability, attained at
the wavenumber
\begin{equation}
   k \equiv k_\alpha
   ={|\meanv_\mu + \alpha_\mu|
   \over 2(\eta+ \, \eta_{_{T}})},
\label{kmax_turb}
\end{equation}
is given by
\begin{eqnarray}
\gamma^{\rm max}_\alpha
= {(\meanv_\mu + \alpha_\mu)^2\over 4 (\eta+ \, \eta_{_{T}})}
= {(\meanv_\mu + \alpha_\mu)^2\over 4 \eta \, (1 + \, \Rm/3)}.
\label{gammamax_turb}
\end{eqnarray}
For small magnetic Reynolds numbers,
$\Rm=u_0 \ell_0/\eta = 3 \eta_{_{T}}/\eta$, this equation
yields the correct result for the laminar $v_\mu^2$ dynamo;
see Equation~(\ref{gamma-max}).

As was shown in Paper~I, the CME
in the presence of turbulence gives rise to the chiral $\alpha_\mu$ effect.
The expression for $\alpha_\mu$
found for large Reynolds numbers and a weak
mean magnetic field is
\begin{eqnarray}
  \alpha_\mu = - {2 \over 3} \meanv_\mu \ln \Rm.
\label{alphamu}
\end{eqnarray}
Since the $\alpha_\mu$ effect in homogeneous turbulence
is always negative, while the $\meanv_\mu$ effect is positive,
the chiral $\alpha_\mu$ effect decreases the $\meanv_\mu$ effect.
Both effects compensate each other at $\Rm=4.5$ (see Paper~I).
However, for large fluid and magnetic Reynolds numbers, $\meanv_\mu \ll
|\alpha_\mu|$, and we can neglect $\meanv_\mu$ in these equations.
This regime corresponds to the large-scale $\alpha_\mu^2$ dynamo.

\subsection{DNS of chiral magnetically driven turbulence}

We have performed a higher resolution $(576^3)$ three-dimensional
numerical simulation to study chiral magnetically driven turbulence.
The chiral Mach number of this simulation is ${\rm Ma}_\mu=2\times10^{-3}$,
the chiral nonlinearity parameter is $\lambda_\mu=2\times10^{-7}$, and the
magnetic and the chiral Prandtl numbers are unity.
The velocity field is initially zero, and the magnetic field is Gaussian noise,
with $B=10^{-6}$.

%%%%%%
\begin{figure}[t]
\centering
  \subfigure{\includegraphics[width=\columnwidth]{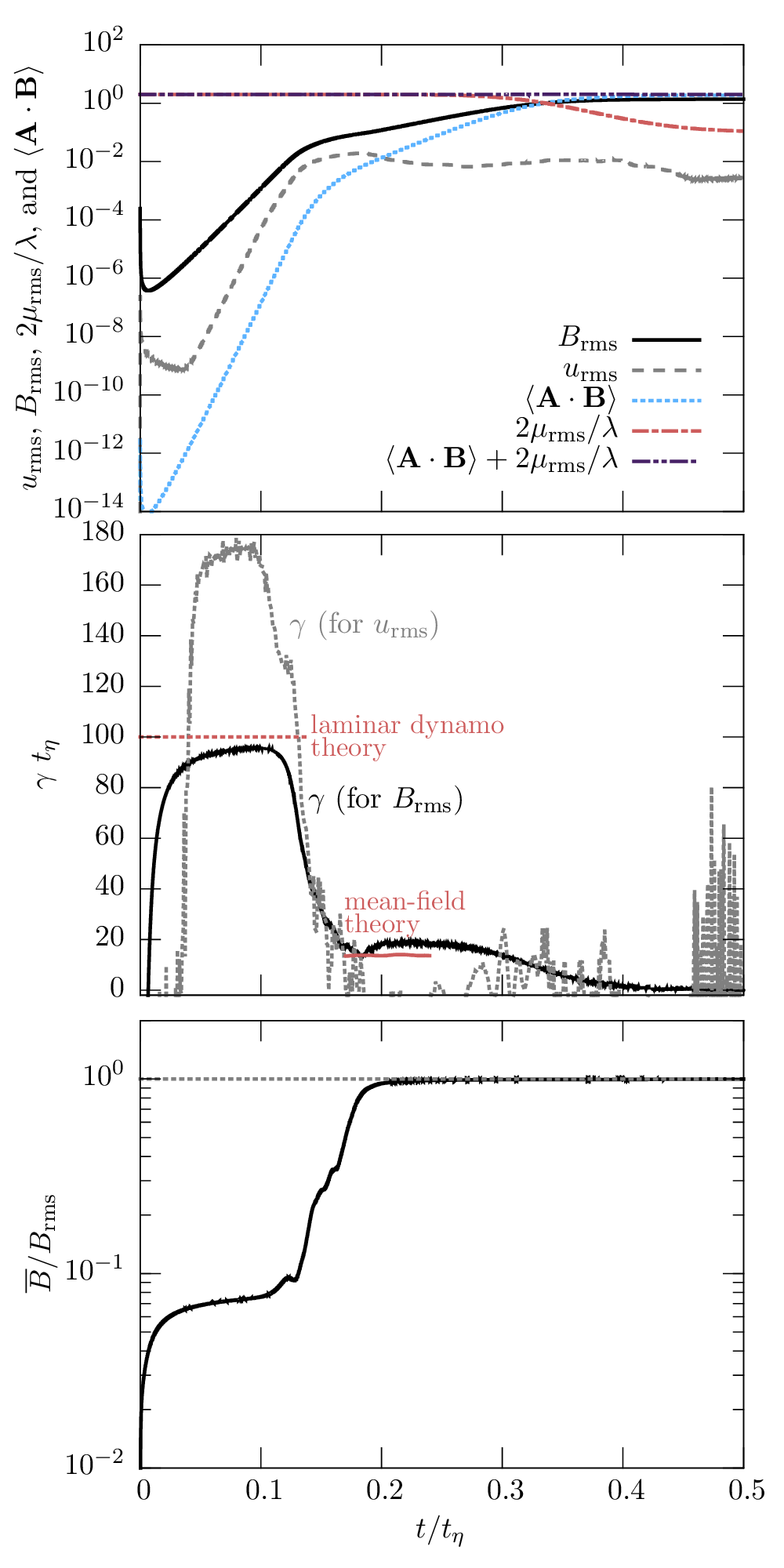}}
\caption{
{\bf Chiral magnetically driven turbulence.}
Time evolution for different quantities.
}
  \label{fig_LTts}
\end{figure}
%%%%%%

The time evolution of $B_\mathrm{rms}$, $u_\mathrm{rms}$,
$\langle \mathbf{A} \cdot \mathbf{B} \rangle$,
$\mu_\mathrm{rms}$ (multiplied by
$2/\lambda$), and $\langle \mathbf{A} \cdot \mathbf{B}
\rangle +2\mu_\mathrm{rms}/\lambda$
of chiral magnetically driven turbulence
is shown in the top panel of Figure~\ref{fig_LTts}.
Four phases can be distinguished:
\begin{asparaenum}[\it (1)]
\item{The kinematic phase of small-scale chiral dynamo instability
resulting in exponential growth of small-scale magnetic field
due to the CME.
This phase ends approximately at $t=0.05 t_\eta$.
}
\item{The first nonlinear phase resulting in production
of chiral magnetically driven turbulence.
In this phase, $u_\mathrm{rms}$ grows from very weak noise
over seven orders of magnitude up to nearly the equipartition value
between turbulent kinetic and magnetic energies,
due to the Lorentz force in the Navier-Stokes equation.
}
\item{The second nonlinear phase resulting in large-scale dynamos.
In particular, the evolution of $B_\mathrm{rms}$ for $t > 0.12 t_\eta$
is affected by the velocity field.
During this phase, the velocity stays approximately constant,
while the magnetic field continues to increase
at a reduced growth rate in comparison with that of the
small-scale chiral dynamo instability.
In this phase, we also observe the formation of inverse energy transfer
with a $k^{-2}$ magnetic energy spectrum that was previously found
and comprehensively analyzed by \cite{BSRKBFRK17} in DNS of chiral MHD
with different parameters.
}
\item{The third nonlinear phase resulting in saturation of the large-scale
dynamos, which ends at $\approx 0.45 t_\eta$ when the
large-scale magnetic field reaches the maximum value.
The conserved quantity $\langle \mathbf{A} \cdot \mathbf{B} \rangle
+2\mu_\mathrm{rms}/\lambda$ stays constant over all four phases.
Saturation is caused by the $\lambda$ term in the evolution equation of the chiral
chemical potential, which leads to a decrease of $\mu$ from its initial value to 1.
}
\end{asparaenum}

The middle panel of Figure~\ref{fig_LTts} shows the measured growth rate of
$B_\mathrm{rms}$ as a function of time.
In the kinematic phase, $\gamma$ agrees
with the theoretical prediction for the
laminar chiral dynamo instability; see Equation~(\ref{gamma-max}),
which is indicated by the dashed red horizontal line in the middle panel of
Figure~\ref{fig_LTts}.
During this phase, the growth rate
of the velocity field, given by the dotted gray line in Figure~\ref{fig_LTts}, is
larger by roughly a factor of two than that of the magnetic field.
This is expected when turbulence is driven via the Lorentz force, which is
quadratic in the magnetic field.

Once the kinetic energy is of the same order
as the magnetic energy, the growth rate of
the magnetic field decreases abruptly by a factor of more than five.
This is expected in the presence of turbulence, because
the energy dissipation of the magnetic field is increased by turbulence due to
turbulent magnetic diffusion.
Additionally, however, a positive contribution to the
growth rate comes from the chiral $\alpha_\mu$ effect
that causes large-scale dynamo instability.

The time evolution of the ratio of the mean magnetic field to the total
field, $\overline{B}/B_\mathrm{rms}$, is presented in the bottom panel
of Figure~\ref{fig_LTts}.
The mean magnetic field grows faster than the rms of the total magnetic
field in the time interval between 0.14 and 0.2 $t_\eta$.
During this time, the large-scale (mean-field) dynamo operates,
so magnetic energy is transferred to larger spatial scales.
We now determine, directly from DNS,
the growth rate of the large-scale dynamo using
Equation~(\ref{gamma_turb}).
To this end, we determine the Reynolds number and the strength of the $\alpha_\mu$
effect using the data from our DNS.
Whereas the rms velocity is a direct output of the simulation, the turbulent
forcing scale can be found from analysis of the energy spectra.
The theoretical value based on these estimates at the time $0.2\,t_\eta$ is
indicated as the solid red horizontal line in the middle
panel of Figure~\ref{fig_LTts}.

%%%%%%
\begin{figure}[t]
\centering
  \subfigure{\includegraphics[width=\columnwidth]{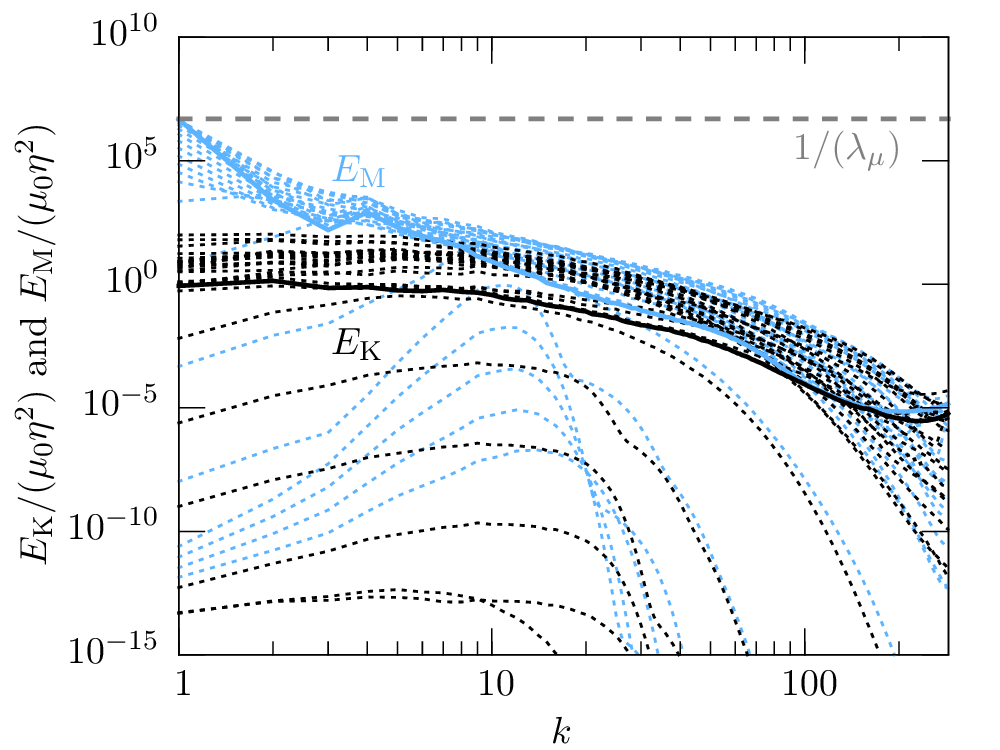}}
\caption{
{\bf Chiral magnetically driven turbulence.}
Magnetic (blue lines) and kinetic (black lines) energy
spectra are calculated at equal time differences,
and the very last spectra are shown as solid lines.}
  \label{fig_LTspec}
\end{figure}
%%%%%%

The evolution of kinetic and magnetic energy spectra is
shown in Figure~\ref{fig_LTspec}.
We use equal time steps between the different spectra, covering the whole
simulation time.
The magnetic energy, indicated by blue
lines, increases initially at $k=\mu_0/2=10$, which agrees with the theoretical
prediction for the chiral laminar dynamo.
The magnetic field drives a turbulent spectrum of the kinetic energy, as can
clearly be seen in
Figure~\ref{fig_LTspec} (indicated by black lines in Figure~\ref{fig_LTspec}).
The final spectral slope of the kinetic energy is roughly $-5/3$.
The magnetic field continues to grow at small wavenumbers,
producing a peak at $k=1$ in the final stage of the time evolution.

%%%%%%
\begin{figure}[t]
\centering
   \subfigure{\includegraphics[width=\columnwidth]{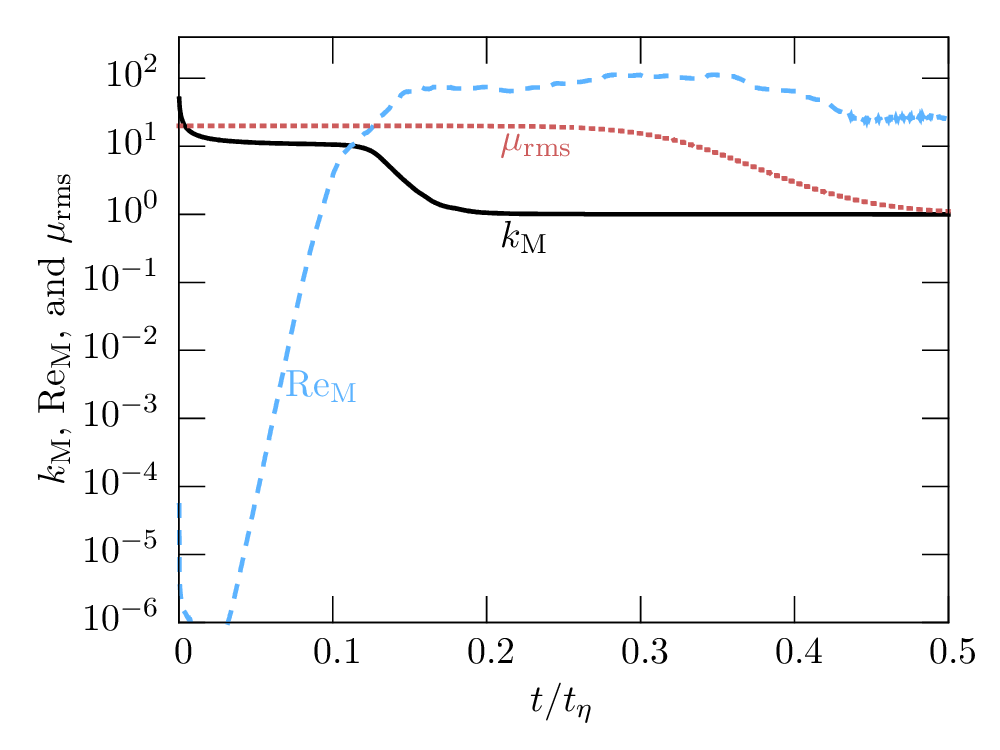}}
\caption{
{\bf Chiral magnetically driven turbulence.}
The black solid line shows the inverse correlation length, $k_{\rm M}$,
of the magnetic energy, defined by Equation (\ref{eq_kcorr}), as a
function of time $t$.
Using this wavenumber and the rms velocity, the fluid and magnetic
Reynolds numbers are estimated (see Equation~\ref{eq_Rmkcorr}), that is
shown by the dashed blue line.}
  \label{fig_LTkmax}
\end{figure}
%%%%%%

We determine the correlation length of the magnetic field from
the magnetic energy spectrum via
\begin{equation}
\xi_{\rm M}(t)\equiv
  k_{\rm M}^{-1}(t) = \frac{1}{\mathcal{E}_{\rm M}(t)} \int k^{-1}
  E_{\rm M}(k,t)~\mathrm{d}k .
\label{eq_kcorr}
\end{equation}
The wavenumber $k_{\rm M}$ so defined coincides (up to a numerical factor of order unity) with
the so-called tracking solution, $\Delta \mu_{\rm tr}$ in \citet{BFR12}.
There it was demonstrated that, in the course of evolution, the chiral
chemical potential follows $k_{\rm M}(t)$.
And, indeed, the evolution of $k_{\rm M}$, shown in Figure~\ref{fig_LTkmax},
starts at around 10 (the value of $\mu_0/2$ in this simulation) and then decreases to
$k_{\rm M} = k_1$ (corresponding to the simulation box size) at $t\approx0.18~t_\eta$.
Interestingly, the chemical potential is affected by
magnetic helicity only at much later times, as can be seen in
Figure~\ref{fig_LTkmax}.
Based on the wavenumber, $k_{\rm M}$, we estimate
the Reynolds numbers as
\begin{equation}
  \Rm = \mathrm{Re} = \frac{u_\mathrm{rms}}{\nu k_{\rm M}} .
\label{eq_Rmkcorr}
\end{equation}
Figure~\ref{fig_LTkmax} shows that the Reynolds number increases exponentially,
mostly due to the fast increase of $u_\mathrm{rms}$, and saturates
later at $\Rm\approx 10^2$.
Similarly, the turbulent diffusivity can be estimated as
\begin{equation}
  \eta_{_{\rm T}} = \frac{u_\mathrm{rms}}{3~k_{\rm M}} .
\label{eq_etaTkcorr}
\end{equation}
During the operation of the mean-field large-scale dynamo, we find
that $\eta_{_{\rm T}}\approx2.4\times10^{-3}$,
which is about 24 times larger than the molecular diffusivity $\eta$.
Using these estimates, we determine
the chiral magnetic $\alpha_\mu$ effect from Equation~(\ref{alphamu}).
The large-scale dynamo growth rate (\ref{gamma_turb}) is
shown as the solid red horizontal line in the middle panel of Figure~\ref{fig_LTts}
and is in agreement with the DNS results shown as the black solid line.

%%%%%%
\begin{figure}[t]
\centering
  \subfigure{\includegraphics[width=\columnwidth]{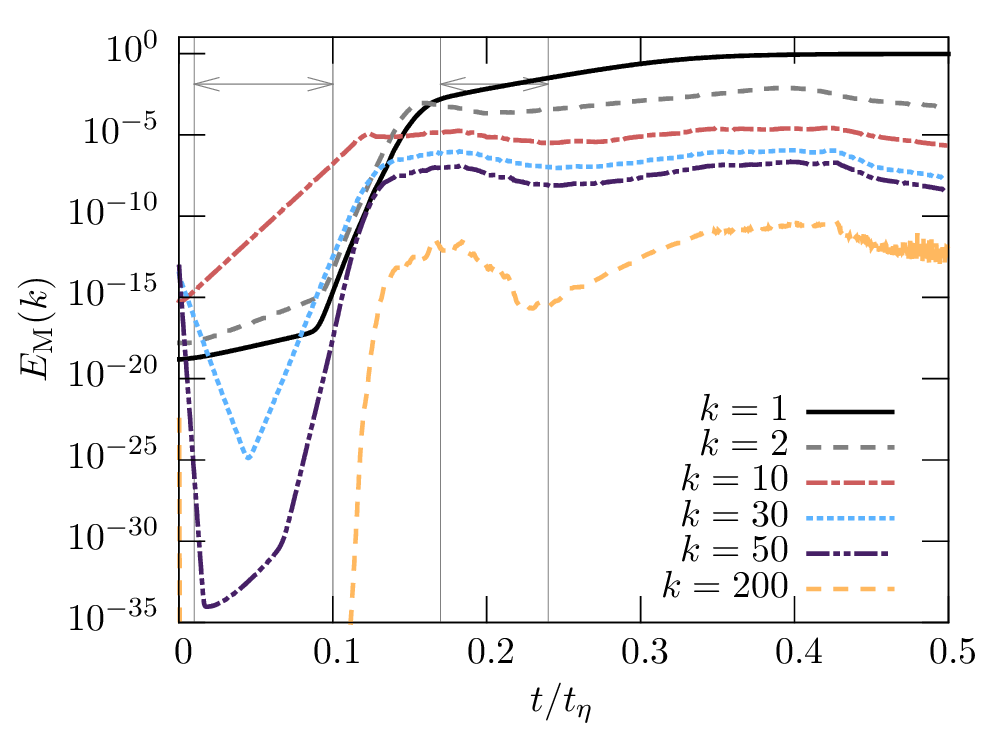}}
  \subfigure{\includegraphics[width=\columnwidth]{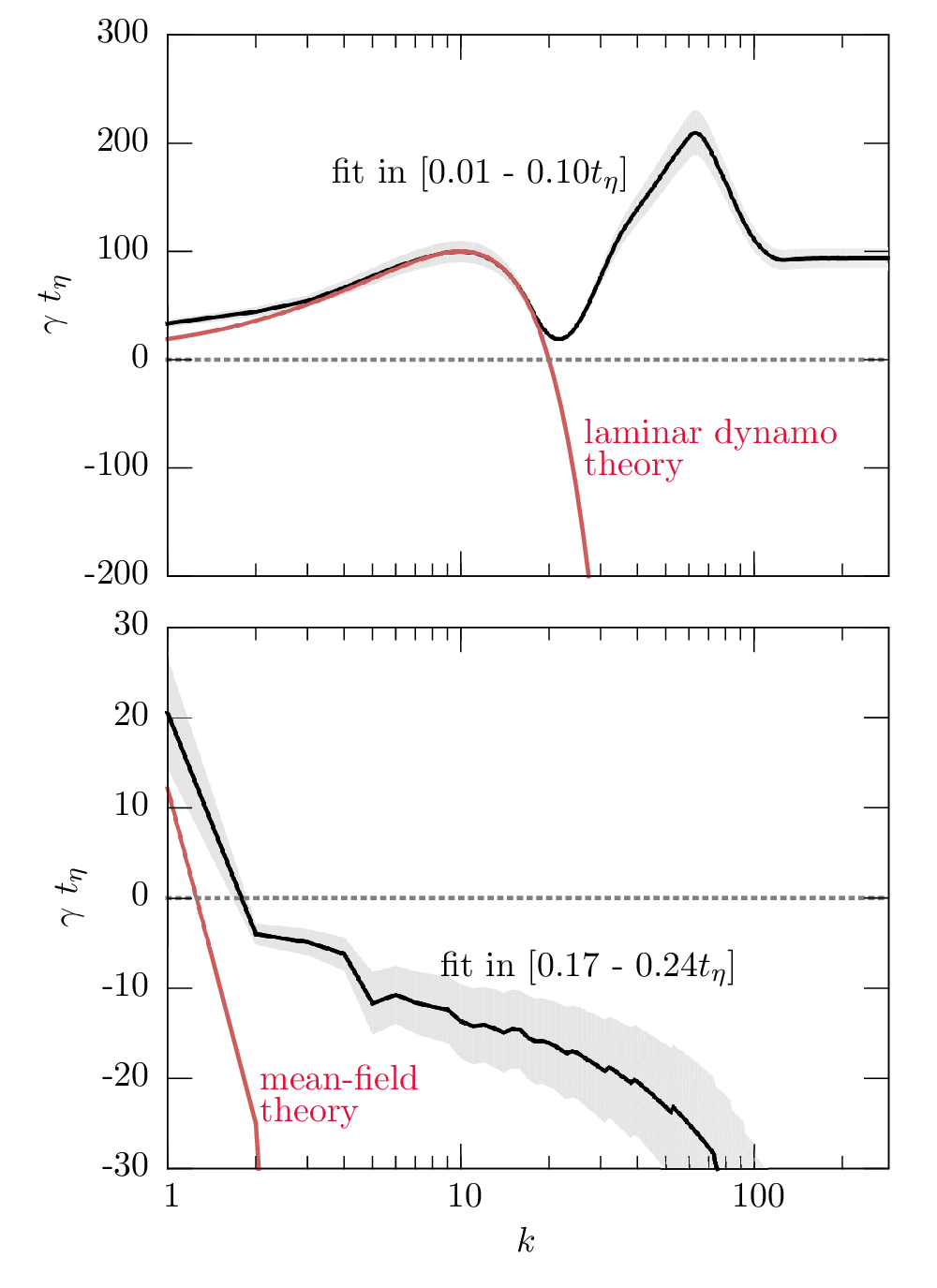}}
\caption{
{\bf Chiral magnetically driven turbulence.}
The evolution of the magnetic energy $E_{\rm M}$ on different wavenumbers $k$
(top panel).
The growth rate as a function of $k$ in different time intervals as given in the
plot legend. The black line corresponds to a fit, while the theoretical
expectations are given as a red
line.}
  \label{fig_LT_gammak}
\end{figure}
%%%%%%

Further analysis of the evolution of the magnetic field at different wavenumbers
is presented in Figure~\ref{fig_LT_gammak}.
In the top panel, we display the magnetic energy at various
wavenumbers as a function of time.
In the kinematic phase, for $t<0.1~t_\eta$, the fastest amplification
occurs at $k=10$, as can also be seen in the energy spectra.
At wavenumbers $k<k_\mu$ there is an initial phase of magnetic dissipation,
followed by an exponential increase of the field.
The rapid transition between the two phases, which occurs at
$t=0.05\,t_\eta$ for $k=30$ in our example, may lead to the impression of an
interpolation between long time steps.
In reality, however, the range $t=0$--$0.1\,t_\eta$ is
resolved by approximately 500 time steps.
At $t\approx0.18\,t_\eta$, the magnetic field grows only at $k=1$.
This confirms the idea that a large-scale (mean-field) dynamo operates.
In the next two panels, we compare the observed growth rates as a function
of wavenumber at different time intervals.
The middle panel of Figure~\ref{fig_LT_gammak}
shows the growth rate in the laminar phase, where we find
good agreement with the theoretical predictions below $k\approx20$.
The resulting value for the growth rate
depends on the accuracy of the fitting, and
a typical error of 10\% is shown by
a gray uncertainty band in the middle panel of Figure~\ref{fig_LT_gammak}.
Also, the observed growth rate of the mean-field dynamo, which we find from
fitting growth rates in the time interval $0.17$--$0.24\,t_\eta$, is comparable to
the prediction from mean-field theory, using our estimates for the Reynolds
number (\ref{eq_Rmkcorr}) and the turbulent diffusivity (\ref{eq_etaTkcorr}).
As the mean-field dynamo phase is followed by the nonlinear phase, the growth
rate is more sensitive to the fitting regime.
Hence we indicate a 30\% uncertainty band in this phase.
The time intervals for the two different fitting regimes are indicated by gray
arrows in the top panel.

\subsection{The effect of a strong initial magnetic field}

The effect of changing the chiral nonlinearity parameter
$\lambda_\mu$ is explored in \cite{BSRKBFRK17},
who considered values between $2\times10^{-6}$ and $200$.
Using dimensional analysis and simulations, they showed that the extension
of the inertial range of the turbulence
is approximately $\lambda_\mu^{-1/2}/4$.
The ratio $\mu/k_\lambda$ is approximately $660$ in our reference
run for chiral magnetically driven turbulence, which was presented in the last section.

\begin{figure}[t]
\begin{center}
\includegraphics[width=\columnwidth]{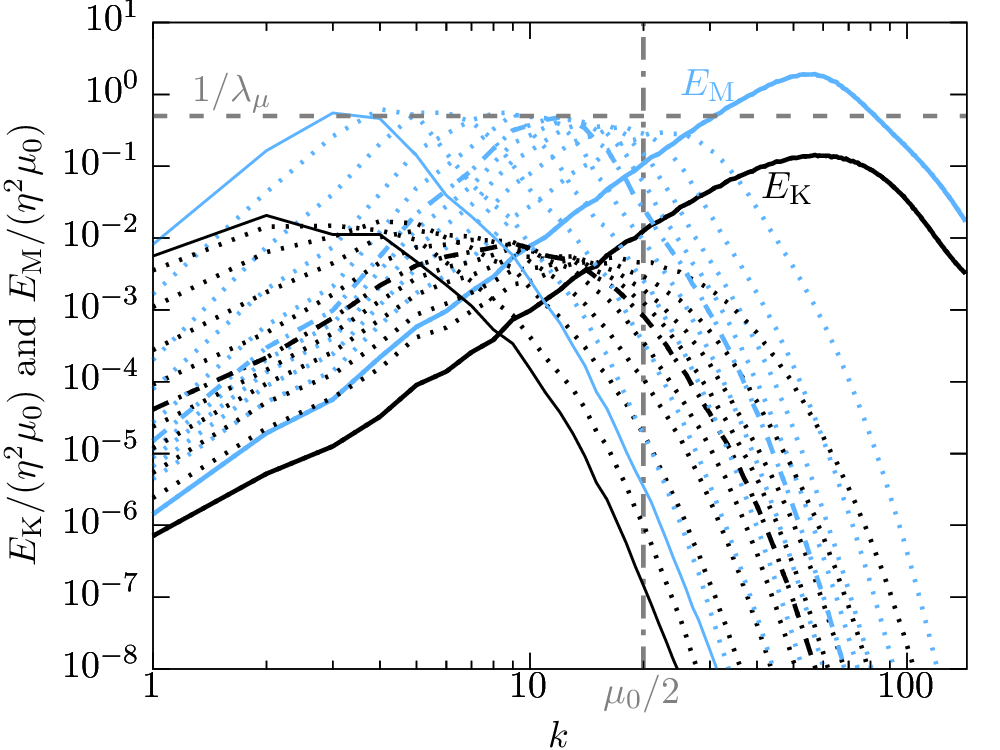}
\end{center}\caption[]{
{\bf Chiral magnetically driven turbulence.}
Evolution of the magnetic (blue lines) and kinetic (black lines) energy spectrum
for a run with large initial
magnetic field on a small spatial scale.
The initial spectra are shown as thick solid lines; later spectra have
equal time intervals up to 0.025 $t_\eta$ (shown in thick dashed lines).
Above $t = 0.025~t_\eta$, the time intervals increase by a factor of two,
until the final spectra are reached, presented here as thin solid lines.
The horizontal dashed gray line shows $1/\lambda_\mu$, the upper limit
predicted by the chiral conservation law, and the vertical gray line shows the
scale where
the growth rate of the small-scale chiral instability
reaches its maximum.
}
\label{pspec_select_288_3D_kf60_mu040_lambda1e9d_B3em3}
\end{figure}

\cite{BSRKBFRK17} found that $E_{\rm M}(k,t)$ is bound from above
by the value of $\mu/\lambda$.
It is interesting to note that this also applies when the initial
magnetic field strength exceeds this limit.
To demonstrate this, we now present a simulation with an
initial magnetic energy spectrum $\propto k^4$ for $k/k_1<60$
and exponential decrease for larger $k$ with $\vA/\cs=0.089$.
We use $\mu_0=40$, $\eta=5\times10^{-5}$,
$\meanrho\lambda=8\times10^8$,
${\rm Ma}_\mu=0.0014$ and $\lambda_\mu=2$.
The result is shown in \Fig{pspec_select_288_3D_kf60_mu040_lambda1e9d_B3em3}.

%%%%%
%%%%%

At early times, $E_{\rm M}/(\eta^2 \mu_0)$ overshoots the value of
$(\mu/\lambda)/(\eta^2 \mu_0)=1/\lambda_\mu$,
but after a short time it follows this limit almost precisely.
This shows that the bound on $E_{\rm M}(k,t)$ is obeyed even when
the initial field strength exceeds this value.

\subsection{Stages of chiral magnetically driven turbulence}
\label{sec:stages}

This DNS demonstrates that the
magnetic field evolution proceeds in the following
distinct stages:
\begin{asparaenum}[\it (1)]
\item{Small-scale chiral dynamo instability.}
\item{First nonlinear stage when the Lorentz force drives
small-scale turbulence.}
\item{Formation of inverse energy transfer
with a $k^{-2}$ magnetic energy spectrum;
see \cite{BSRKBFRK17} for details.}
\item{Generation of large-scale magnetic fields
by chiral magnetically driven turbulence.}
\item{Decrease of the chemical potential and saturation
of the large-scale chiral dynamo.}
\end{asparaenum}
Although the magnetic field cannot grow any further, the spectrum
continues to move to smaller wavenumbers in a shape-invariant fashion
\citep[see][]{BK17}.
This implies that the magnetic integral scale
$\xi_{\rm M}\equiv k_{\rm M}^{-1}$ continues to grow and
the magnetic energy continues to decrease proportional to $t^{-2/3}$
with $\langle \BB^2 \rangle \xi_{\rm M} \approx \const$.

%%%%%%%%%%%%%
% Section 5 %
%%%%%%%%%%%%%

\section{DNS of large-scale dynamos in forced, nonhelical, and homogeneous
turbulence}
\label{sec_turbdynamo}

In this section, we study the evolution of the magnetic field in the presence
of forced, nonhelical, and homogeneous turbulence in order
to control the turbulence parameters in the chiral MHD simulations.
Chiral dynamos in forced turbulence can be described by the mean-field chiral
MHD equations. The theoretical results related to
the mean-field chiral dynamos obtained in Paper~I
have been outlined in Section~\ref{sec_meanfieldMHD}.

\subsection{DNS setup for externally forced turbulence}

To study chiral large-scale dynamos, we perform three-dimensional DNS
with externally forced turbulence and a spatial resolution of $200^3$.
In run Ta2-10, the resolution is $280^3$ (see Table~\ref{table_simulations_forced}).
Turbulence is driven via the forcing term $\ff(\mathbf{x},t)$ in
Equation~(\ref{UU-DNS}).
The forcing function is nonhelical and localized around the wavenumber
$k_\mathrm{f}$; see \cite{HBD2004} for details.
For the runs presented in the following,
we choose $k_\mathrm{f}= 4$ and 10.
These values are small enough for the fluid and magnetic Reynolds numbers,
$\Rey = u_\mathrm{rms}/(\nu k_\mathrm{f})$ and
$\Rm = u_\mathrm{rms}/(\eta k_\mathrm{f})$, respectively, to be sufficiently
large for turbulence to develop.
At the same time, $k_\mathrm{f}$ is large enough for a clear separation between the
box scale and the forcing scale, allowing to study
of mean-field (large-scale) dynamos.
In the numerical simulations, we vary ${\rm Ma}_\mu$, $\lambda_\mu$, and $\Rey$
(see Table~\ref{table_simulations_forced}).

For comparison with the results from mean-field theory, the simulations need to
fulfill the following criteria:\\
\begin{itemize}
\item{
To capture the maximum amplification inside the numerical domain with
$k_\mathrm{box}=2\pi/L_\mathrm{box}=1$, the condition $k_\mathrm{max}>1$ needs
to be fulfilled.
As shown in Equation (\ref{kmax_turb}), $k_\mathrm{max}$ is proportional
to $\eta/\eta_{_{\rm T}}$, which is inversely proportional to the magnetic
Reynolds number $\Rm$.
As a result, the chemical potential needs to be sufficiently large for
$k_\mathrm{max}>1$.}
\item{
Due to nonlocal effects, the turbulent diffusivity $\eta_{_{\rm T}}$
is generally scale-dependent and
decreases above $k_\mathrm{f}$ \citep{BRS08}.
For comparison with mean-field theory, the chiral dynamo
instability has to occur on scales $k < k_\mathrm{f}$, where
$\eta_{_{\rm T}} \approx u_\mathrm{rms}/(3k_\mathrm{f})$.
Note, however, that
the presence of a mean kinetic helicity in the system
caused by the CME (see Paper~I)
can increase the turbulent diffusivity $\eta_{_{\rm T}}$
for moderate magnetic Reynolds numbers by up to 50\% \citep{BSR17}.}
\item{
To simplify the system, we
avoid classical small-scale dynamo action, which
occurs at magnetic Reynolds numbers larger than
${\rm Re}_{_{\mathrm{M,crit}}}\approx50$.}
\end{itemize}

% links to simulations:
% reference run:
% Ta2-1: /pencil-code/jenny/chiral_fluids/turbulent_dynamos/homogeneous/alpha2/200_3D_eta4e-4_nu4e-4_kf10_force17e-2_mu20_lambda1e2
% Ta2-2: /pencil-code/jenny/chiral_fluids/turbulent_dynamos/homogeneous/alpha2/200_3D_eta2e-4_nu2e-4_kf10_force1e-2_mu20_lambda1e2
% Ta2-3: /pencil-code/jenny/chiral_fluids/turbulent_dynamos/homogeneous/alpha2/200_3D_eta4e-4_nu4e-4_kf10_force1e-2_mu20_lambda1e2
% Ta2-4: /pencil-code/jenny/chiral_fluids/turbulent_dynamos/homogeneous/alpha2/200_3D_eta4e-4_nu4e-4_kf10_force1e-2_mu20_lambda1e2
% Ta2-5: /pencil-code/jenny/chiral_fluids/turbulent_dynamos/homogeneous/alpha2/200_3D_eta4e-4_nu4e-4_kf10_force17e-2_mu20_lambda1e1
% Ta2-6: /pencil-code/jenny/chiral_fluids/turbulent_dynamos/homogeneous/alpha2/200_3D_eta4e-4_nu4e-4_kf10_force1e-2_mu20_lambda1e1
% Ta2-7: /pencil-code/jenny/chiral_fluids/turbulent_dynamos/homogeneous/alpha2/200_3D_eta4e-4_nu4e-4_kf4_force1e-2_mu30_lambda2e2
% Ta2-8: /pencil-code/jenny/chiral_fluids/turbulent_dynamos/homogeneous/alpha2/200_3D_eta3e-4_nu3e-4_kf4_force1e-2_mu30_lambda2e2
% Ta2-9: /pencil-code/jenny/chiral_fluids/turbulent_dynamos/homogeneous/alpha2/200_3D_eta3e-4_nu3e-4_kf4_force15e-2_mu30_lambda15e2
% Ta2-10: /pencil-code/jenny/chiral_fluids/turbulent_dynamos/homogeneous/alpha2/280_3D_eta2e-4_nu2e-4_kf4_force14e-2_mu40_lambda4e2

\begin{table}[t]
\caption{
Overview of Runs With Externally Forced Turbulence
(Reference Run in Bold)}
\centering
     \begin{tabular}{l|lllllll}
      \hline
      \hline
      \\
	~ 	        & $\mu_0$ 	& $\dfrac{{\rm Ma}_\mu}{10^{-3}}$ 	& $\dfrac{\lambda_\mu}{10^{-6}}$  	& $\dfrac{k_\lambda}{10^{-3} \mu_0}$ & $\dfrac{k_\mathrm{diff}}{\mu_0}$ & $k_\mathrm{f}$ 	& $\Rm$		\\	
	~ 	        & 		& ~		 	& ~		  	& ~		& ~		& ~	 	& (early $\rightarrow$ late)		\\	                \hline
      	Ta2-1  	& $20$		& $8$	& $16$ 	& 160		  	& 4.5	& $10$ 		& $24 \rightarrow 19$   \\
      	Ta2-2  	& $20$		& $4$	& $4.0$ & 80		  	& 63	& $10$ 		& $36 \rightarrow 28$ 	\\
      	Ta2-3  	& $20$		& $8$	& $16$ 	& 160		  	& 45	& $10$ 		& $16 \rightarrow   14$ 	\\
      	Ta2-4  	& $20$		& $4$	& $4.0$ & 80		  	& 63	& $10$ 		& $4 \rightarrow  13$ 	\\
      	\textbf{Ta2-5} & $\mathbf{20}$	& $\mathbf{8}$ & $\mathbf{160}$ & $\mathbf{51}$ 	& $\mathbf{80}$	& $\mathbf{10}$ 	& $\mathbf{24} \rightarrow\mathbf{18}$ 	\\
      	Ta2-6  	& $20$		& $8$	& $1.6$ & 51		  	& 80	& $10$ 		& $16 \rightarrow 14$ 	\\
      	Ta2-7  	& $30$		& $12$	& $32$ 	& 230		  	& 38	& $4$ 		& $42 \rightarrow  58$ 	\\
      	Ta2-8  	& $30$		& $9$	& $18$ 	& 160		  	& 43	& $4$ 		& $58 \rightarrow  65$ 	\\
      	Ta2-9  	& $30$		& $9$	& $13.5$ & 150		  	& 47	& $4$ 		& $82 \rightarrow  74$ 	\\
      	Ta2-10  & $40$		& $8$	& $16$ 	& 160		  	& 45	& $4$ 		& $119 \rightarrow 107$ \\
      \hline
      \hline
    \end{tabular}
  \label{table_simulations_forced}
\end{table}

\subsection{DNS of chiral dynamos in forced turbulence}
\label{sec:forced-chiral-dynamos}

\begin{figure}[t]
\centering
  \includegraphics[width=\columnwidth]{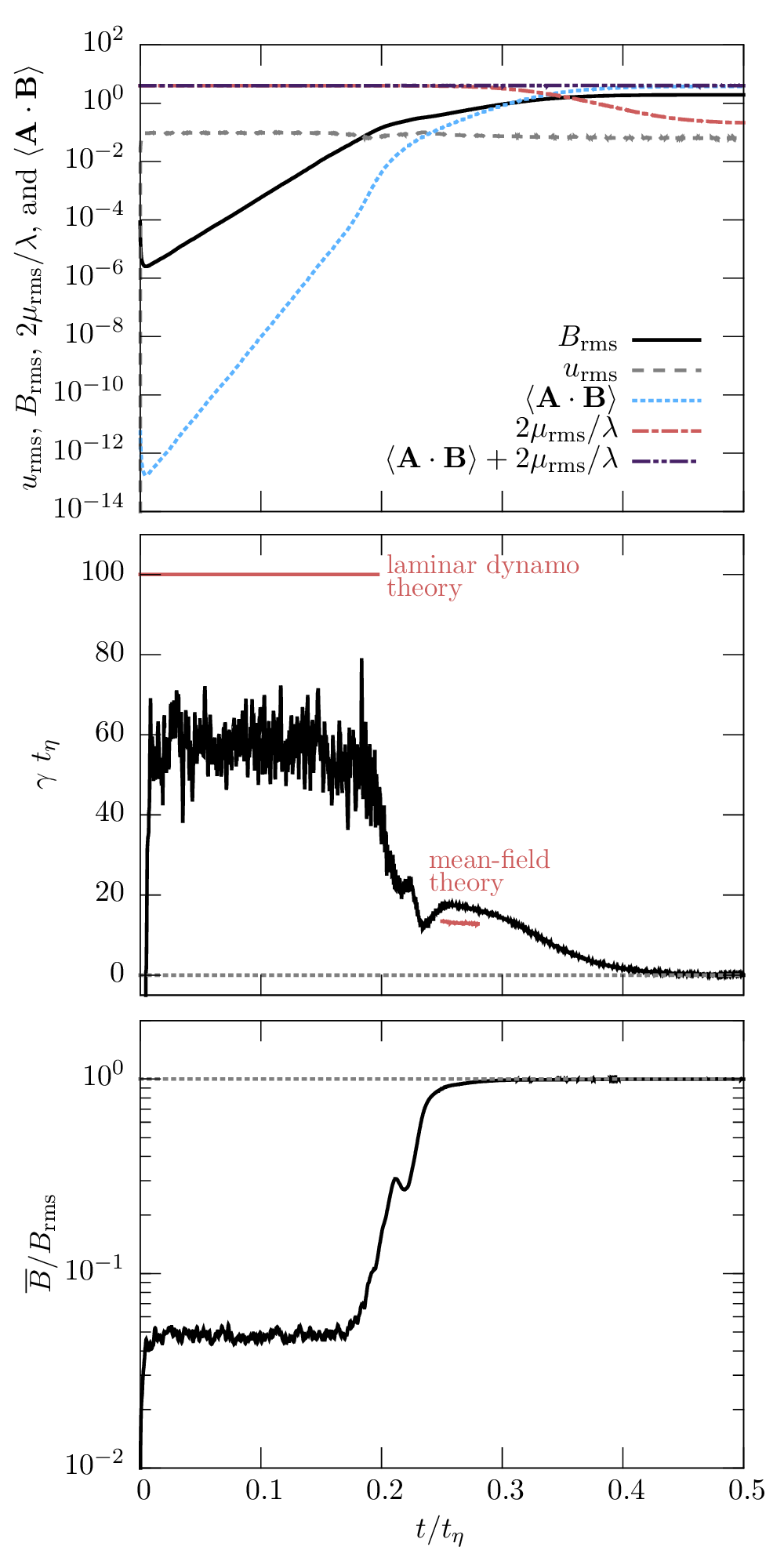}
\caption{
{\bf Externally forced turbulence.}
Time evolution of the magnetic field, the velocity field, and
the chemical potential, as well as the mean value of the magnetic
helicity (top panel).
The middle panel shows the growth rate of $B_\mathrm{rms}$ as a function of
time (solid black line).
The red lines are theoretical expectations in different dynamo phases.
In the bottom panel, the ratio of the mean magnetic field to the total field
$B_\mathrm{rms}$ is presented.}
\label{fig_alphamu_ts}
\end{figure}

The time evolution of different quantities in our reference run is presented in
Figure~\ref{fig_alphamu_ts}.
The magnetic field first increases first exponentially,
with a growth rate $\gamma \approx 60~t_\eta^{-1}$, which is
about a factor of 1.6 lower than that expected for the laminar $v_\mu^2$ dynamo;
see the middle panel of Figure~\ref{fig_alphamu_ts}.
This difference seems to be caused by the presence of random forcing;
see discussion below.
At approximately 0.2 $t_\eta$, the growth rate decreases to a value
of $\gamma\approx 15~t_\eta^{-1}$ consistent with that of the mean-field
chiral $\alpha_\mu^2$ dynamo, before saturation occurs at $0.4\,t_\eta$.
The evolution of $B_\mathrm{rms}$ is comparable qualitatively
in chiral magnetically produced turbulence; see Figure~\ref{fig_LTts}.
An additional difference from the latter
is the value of $u_\mathrm{rms} \approx 0.1$ for
externally forced turbulence, which is controlled by the intensity
of the forcing function.
An indication of the presence of a mean-field dynamo is the evolution of
$\overline{B}/B_\mathrm{rms}$ in the bottom panel of
Figure~\ref{fig_alphamu_ts}, which reaches a value of unity at $0.3\,t_\eta$.

\begin{figure}[t]
\centering
   \subfigure{\includegraphics[width=\columnwidth]{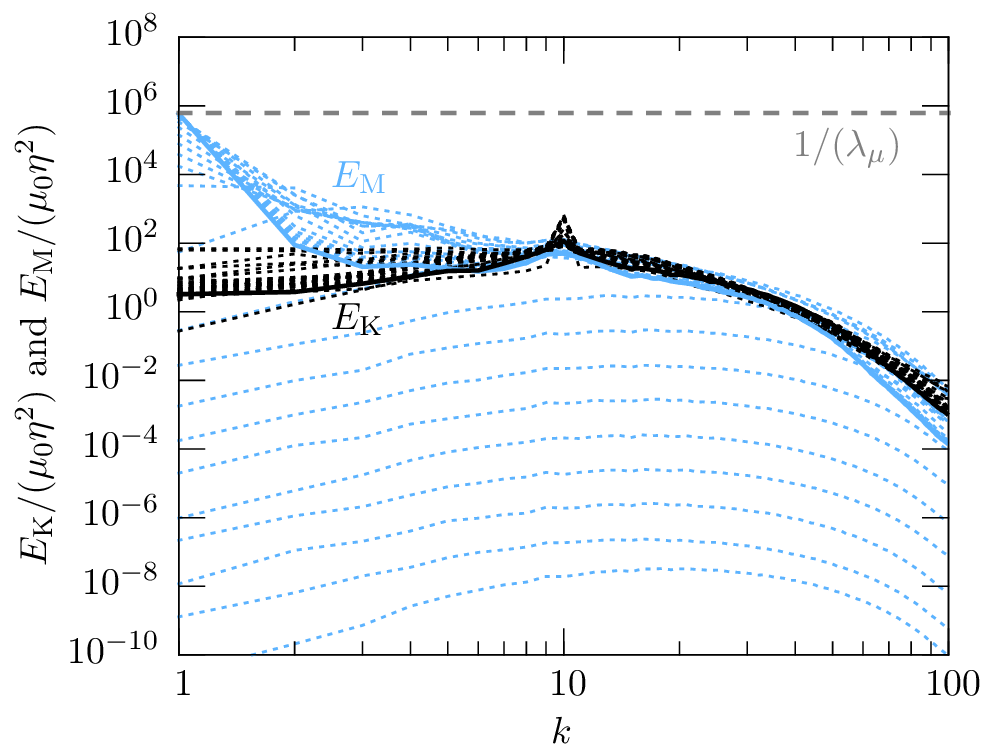}}
\caption{
{\bf Externally forced turbulence.}
Evolution of kinetic (black lines) and magnetic energy spectra (blue lines)
for the reference run Ta2-5.
The ratio $\mu_0/\lambda$ is indicated by the horizontal dashed line.}
\label{fig_alphamu_spec}
\end{figure}

The energy spectra presented in Figure~\ref{fig_alphamu_spec} support the
large-scale dynamo scenario.
First, the magnetic energy increases at all scales,
and, at later times, the maximum of the magnetic energy
is shifted to smaller wavenumbers, finally producing a
peak at $k=1$, i.e., the smallest possible wavenumber in our periodic domain.

\begin{figure}[t]
\centering
   \subfigure{\includegraphics[width=\columnwidth]{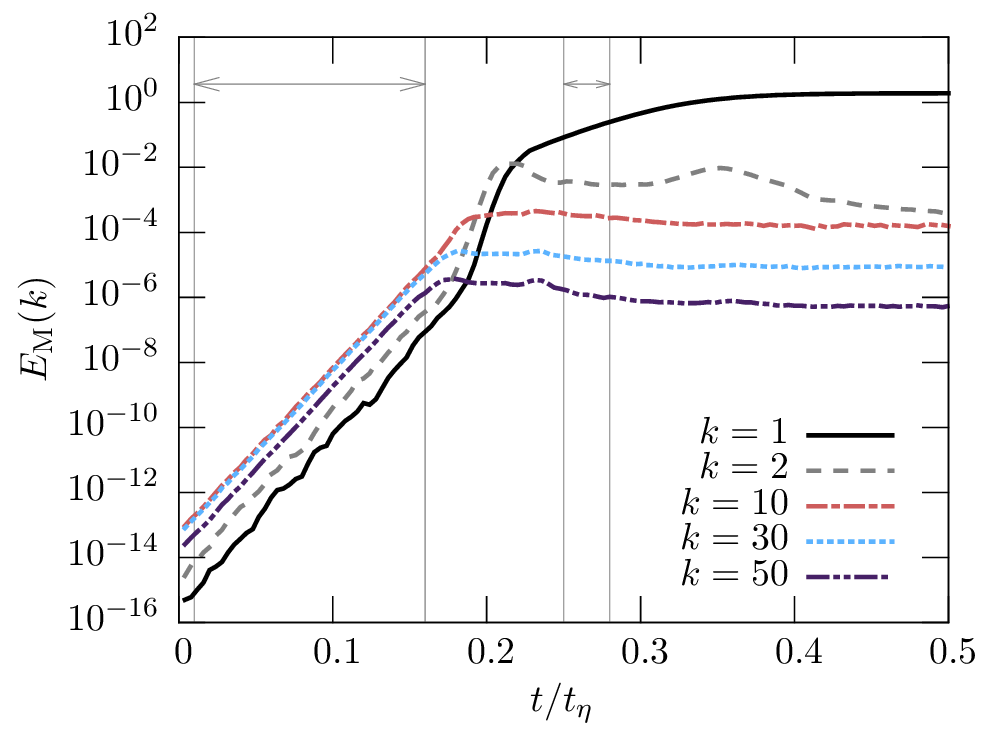}}
   \subfigure{\includegraphics[width=\columnwidth]{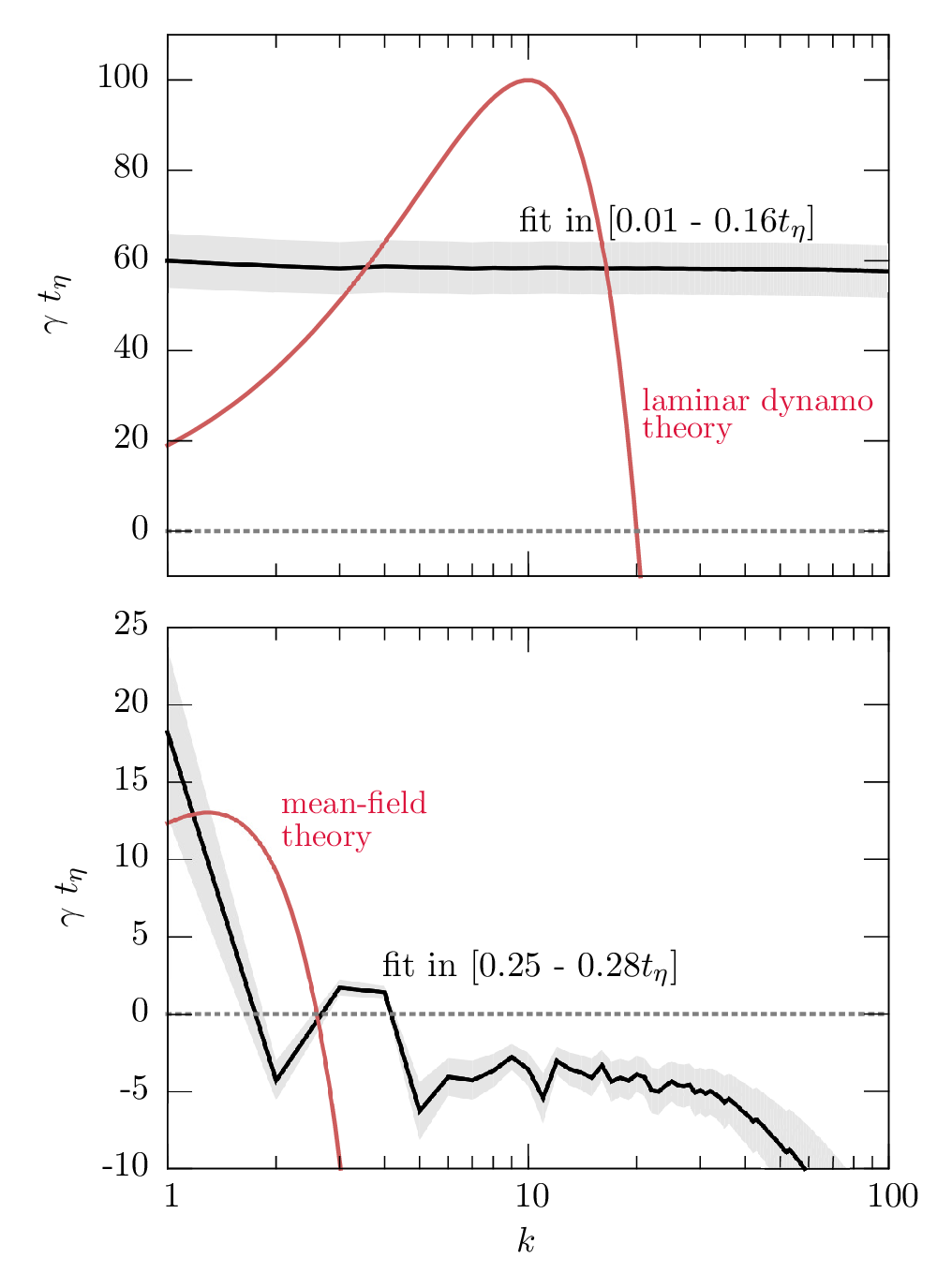}}
\caption{
{\bf Externally forced turbulence.}
Time evolution of the magnetic energy at different wavenumbers $k$
(top panel).
The remaining panels show the growth rates as a function of scale in
different fit intervals.}
\label{fig_alphamu_specana}
\end{figure}

A detailed analysis of the growth of magnetic energy is presented in
Figure~\ref{fig_alphamu_specana}.
In the first phase, the growth rate of the magnetic field is
independent of the wavenumber $k$ (see top panel), due to a coupling
between different modes.
The growth rate measured in this phase is less than that
in the laminar case (see middle panel), due to a scale-dependent
turbulent diffusion caused by the random forcing.

Within the time interval $(0.22$--$0.28)\,t_\eta$, only the magnetic field at
$k=1$ increases.
This is clearly seen in the bottom panel of Figure~\ref{fig_alphamu_specana},
where we show the evolution of the magnetic energy at different wavenumbers $k$.
The growth rate of the mean-field dynamo, which is determined
at $k=1$, agrees with the result from
mean-field theory, given by Equation~(\ref{gammamax_turbRm}).
There is a small dependence of the resulting mean-field growth rate on the
exact fitting regime.
If the phase of the
mean-field dynamo is very short, changing the fitting range can affect the
result by a factor up to 30 \%.
We use the latter value as an estimate of the uncertainty in the growth rate, and,
in addition, indicate an error of 20 \% in determining the Reynolds number,
which is caused by the temporal variations of $u_\mathrm{rms}$.

\subsection{Dependence on the magnetic Reynolds number}

Based on the mean-field theory developed in Paper~I, we expect
the following.
Using the expression for the $\alpha_\mu$ effect
given by
Equation (\ref{alphamu}), the maximum growth rate (\ref{gammamax_turb})
for the mean-field dynamo can be rewritten as a function of the magnetic
Reynolds number:
\begin{eqnarray}
\gamma_{\rm max} (\Rm) = \frac{\meanv_\mu^2 (1 - 2/3~\ln\Rm )^2}{4 \eta \,
(1 + \, \Rm/3)} ,
\label{gammamax_turbRm}
\end{eqnarray}
where the ratio $\eta_{_{\rm T}}/\eta = \Rm/3$.

We perform DNS with different Reynolds numbers to
test the scaling of $\gamma_{\rm max}(\Rm)$ given by
Equation~(\ref{gammamax_turbRm}).
The parameters of the runs with externally forced
turbulence are summarized in Table \ref{table_simulations_forced}.
We vary $\nu \, (=\eta$), the forcing wavenumber $k_\mathrm{f}$, as well as the
amplitude of the forcing,
to determine the function $\gamma_{\rm max}(\Rm)$.
In the initial phase, $u_\mathrm{rms}$ is constant in time.
Once large-scale turbulent dynamo action occurs,
there are additional minor
variations in $u_\mathrm{rms}$, because the system is already
in the nonlinear phase.
The nonlinear terms in the Navier-Stokes equation lead to a modification of
the velocity field at small spatial scales, which affects the value of
$u_\mathrm{rms}$ and results in
the small difference between
the initial and final values of the Reynolds numbers
(see Table \ref{table_simulations_forced}).

According to Equation~(\ref{kmax_turb}),
the wavenumber associated with the maximum growth rate
of the large-scale turbulent
dynamo instability decreases with increasing $\Rm$.
In order to keep this mode inside the computational domain and hence to
compare the measured growth rate with the maximum one given by
Equation~(\ref{gammamax_turbRm}),
we vary the value of $\mu_0$ in our simulations.
The variation of $\mu_0$ and the additional variation of $\eta$ for scanning
through the $\Rm$ parameter space, implies that ${\rm Ma}_\mu$ changes
correspondingly.

The values of the nonlinear parameter $\lambda$ should be within a certain range.
Indeed, the saturation value of the magnetic field, given by
Equation~(\ref{eq_Bsat}), is proportional to $\lambda^{-1/2}$.
In order for the Alfv\'en velocity not to exceed the sound speed,
which would result in a very small time step in DNS,
$\lambda$ should not be below a certain value.
On the other hand, $\lambda$ should not be too large,
as in this case the dynamo would saturate quickly, and there is only a very short
time interval of the large-scale dynamo.
In this case, determining the growth rate of the mean-field dynamo,
and hence comparing with the mean-field theory, is difficult.

%%%%%%
\begin{figure}[t]
\centering
  \includegraphics[width=\columnwidth]{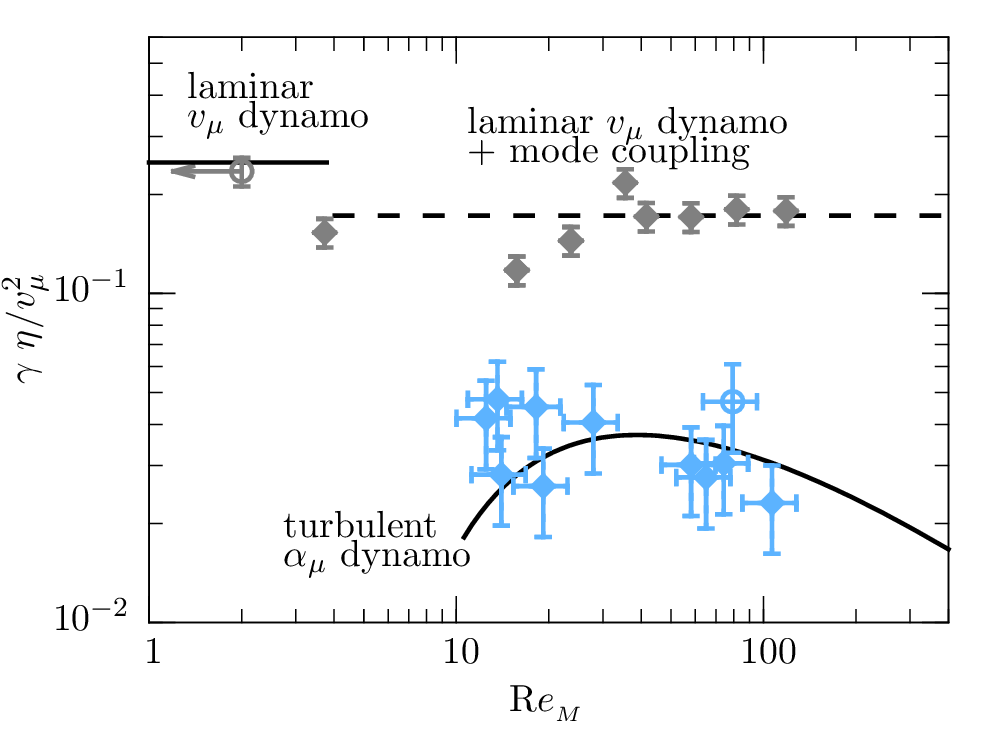}
\caption{
{\bf Externally forced turbulence and chiral magnetically driven turbulence.}
The normalized growth rate $\gamma~\eta/v_\mu^2$ of the magnetic
field as a function of the magnetic Reynolds number $\Rm$.
The gray data points show the
growth rate in the initial, purely kinematic phase of the simulations.
The blue data points show the measured growth rate of the magnetic field
on $k=1$, when the large-scale dynamo occurs.
The diamond-shaped data points represent simulations of forced turbulence, while
the dot-shaped data points refer to the case of
chiral magnetically driven turbulence.
The growth rate observed in the initial laminar phase
for the case of chiral magnetically driven turbulence
is shown at $\Rm=2$, with the left arrow indicating that the actual $\Rm$ is much
lower and out of the plot range at this time; see Figure~\ref{fig_LTkmax}.}
  \label{fig_gamma_Re}
\end{figure}
%%%%%%

In Figure~\ref{fig_gamma_Re} we show the normalized growth
rate $\gamma~\eta/v_\mu^2$ of the magnetic
field as a function of the magnetic Reynolds number $\Rm$.
The gray data points show the growth rate in the initial,
purely kinematic phase of the simulations.
The blue data points show the measured growth rate of the magnetic field
on $k=1$, when the large-scale dynamo occurs.
For comparison of the results with externally forced turbulence
(indicated as diamond-shaped data points), we show
in Figure~\ref{fig_gamma_Re} also the results obtained for the
dynamo in chiral magnetically driven turbulence, which are indicated as dots.

In DNS with externally forced turbulence, we see in all cases a
reduced growth rate due to mode coupling.
Contrary to the case with externally forced turbulence,
in DNS with the chiral magnetically driven turbulence,
we do initially observe the purely laminar dynamo
with the growth rate given by Equation~(\ref{gamma-max}),
because there is no mode coupling in the initial phase
of the magnetic field evolution in this case.
On the other hand, the measured growth rates of the mean-field dynamo
in both cases agree (within the error bars) with the growth rates
obtained from the mean-field theory.

%%%%%%%%%%%%%
% Section 6 %
%%%%%%%%%%%%%

\section{Chiral MHD dynamos in astrophysical relativistic plasmas}
\label{sec_astro}

In this section, the results for the
nonlinear evolution of the chiral chemical potential,
the magnetic field, and the turbulent state of the plasma found in this paper
are applied to astrophysical relativistic plasmas.
We begin by discussing the role of chiral dynamos in the early universe and
identify conditions under which the CME affects the generation and evolution of
cosmic magnetic fields.
Finally, in Section~\ref{sec:PNS}, we examine the importance of the CME
in proto-neutron stars (PNEs).

\subsection{Early Universe}
\label{sec:early-universe}

In spite of many possible mechanisms that can produce magnetic fields in the
early universe
\citep[see, e.g.,][for reviews]{W02,WEtAl12,DN13,Giovannini:2003yn,S16},
understanding the origin of cosmic magnetic fields remains an open problem.
Their generation is often associated with nonequilibrium events in the universe
(e.g., inflation or phase transitions).
A period of particular interest is the electroweak (EW) epoch, characterized
by temperatures of $\unit[10^{15}]{K}$ ($\kB T \sim \unit[100]{GeV}$).
Several important events take place around this time: the electroweak
symmetry gets broken, photons appear while intermediate vector bosons
become massive, and the asymmetry between matter and antimatter appears
in the electroweak baryogenesis scenario \citep{KRS85};
see, for example, the review by \citet{MRM12}.
Magnetic fields of appreciable strength can be generated as a consequence of
these events \citep{V91,Olesen:1992np,Enqvist:1993np,Enqvist:1994dq,
Vachaspati:1994ng,GGV95, 1996PhLB..380..253D,BBM96,V01,Semikoz:2010zua}.
Their typical correlation length $\xi_{\rm M}^{(\rm ew)} \sim (\alphaem T)^{-1}$
corresponds to only a few centimeters today -- much less than the observed correlation
scales of magnetic fields in galaxies or galaxy clusters.
Therefore, in the absence of mechanisms that can increase
the comoving scale of the magnetic field beyond
$\xi_{\rm M}^{(\rm ew)}$, such fields were deemed to be irrelevant
to the problem of cosmic magnetic fields \citep[for discussion,
see, e.g.,][]{Durrer:2003ja,Caprini:2009pr,Saveliev:2012ea,KTBN13}.

The situation may change if (i) the magnetic fields are helical and
(ii) the plasma is turbulent.
In this case, an inverse transfer of magnetic energy may develop,
which leads to a shift of the typical scale of the magnetic field to
progressively larger scales \citep{BEO96, CHB01, BJ04, KTBN13}.
The origin of such turbulence has been unknown.
An often considered paradigm is that a random magnetic field, generated at small
scales, produces turbulent motions via the Lorentz force.
However, continuous energy input is required.
If this is not the case, the magnetic field decays:
$\langle \BB^2 \rangle \sim t^{-2/3}$ as the correlation scale grows
\citep{BM99PhRvL, KTBN13}, so that $\langle \BB^2 \rangle \xi_{\rm M} = \const$.

In the present work, we demonstrated that the presence of a finite chiral
charge in the plasma at the EW epoch is sufficient to satisfy the above
requirements (i) and (ii).
As a result,
\begin{asparaenum}[\it (1)]
\item helical magnetic fields are excited,
\item turbulence with large $\Rm$ is produced, and
\item the comoving correlation scale increases.
\end{asparaenum}
We discuss each of these phases in detail below.

\subsubsection{Generation and evolution of cosmic magnetic fields in the
presence of a chiral chemical potential}
\label{subsec_cosmologybounds}

Although it is not possible to perform numerical simulations with
parameters matching those of the early universe, the results of the
present paper allow us to make qualitative predictions about the fate
of cosmological magnetic fields generated at the EW epoch in the
presence of a chiral chemical potential.

\begin{figure}[!t]
  \centering
  \includegraphics[width=\columnwidth]{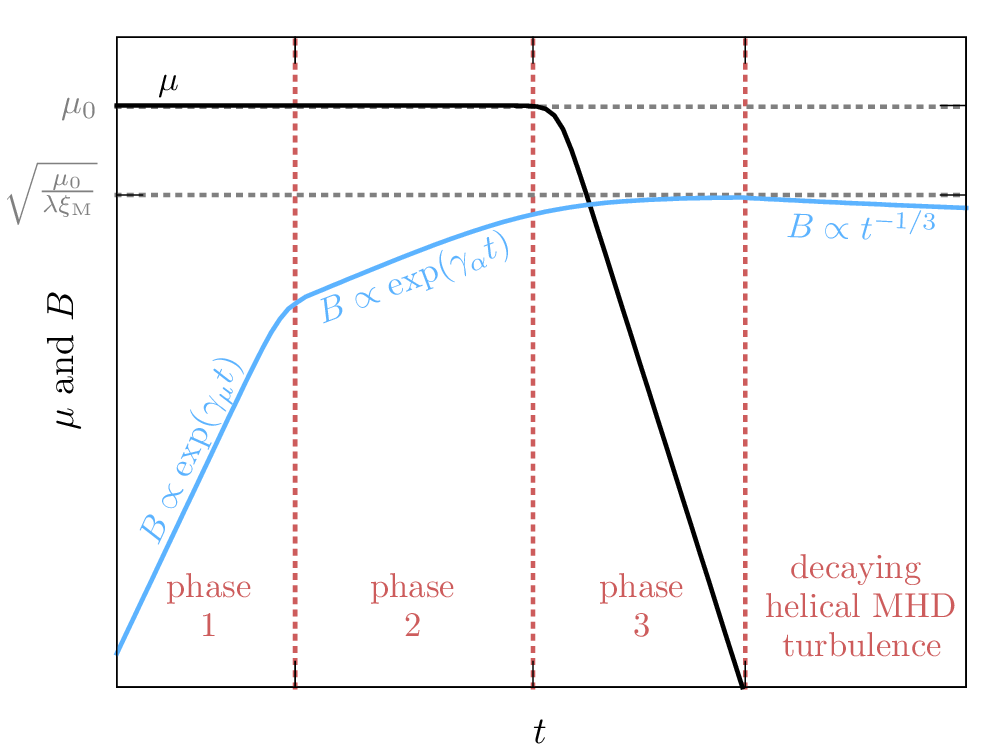}
  \caption{
  \textbf{Chiral MHD dynamos in the early universe.}
Sketch of the different phases of the chiral mean-field dynamo.
From left to right:
small-scale chiral dynamo (phase 1), large-scale turbulent dynamo (phase 2), and
saturation (phase 3).
After saturation of the dynamo, the magnetic magnetic field dissipates.
The upper horizontal dotted line shows the initial value of $\mu$ and the lower
one the ``saturation limit,'' given by Eq.~\protect\eqref{eq:2}.}
  \label{fig:phases}
\end{figure}

All of the main stages of the magnetic field evolution,
summarized in Section~\ref{sec:stages}, can occur in the early universe
(a sketch of the main phases is provided in \Fig{fig:phases}).

Phase 1.
At this initial stage, the small-scale chiral dynamo instability develops at
scales around $\xi_\mu$, where
\begin{equation}
  \xi_\mu \equiv \frac{2}{|\mu_0|} ,
  \label{eq_mu-1}
\end{equation}
and
\begin{equation}
  \label{eq:1}
  \mu_0 \approx 4\alphaem\frac{\mu_5}{\hbar c}\approx
\unit[1.5\times10^{14}]{cm^{-1}}\frac{\mu_5}{\unit[100]{GeV}}.
\end{equation}
The chemical potential $\mu_5$ can be approximated by the thermal energy
$\kB T$ for order-of-magnitude estimates.
In what follows, we provide numerical estimates for $\mu_5 = 100\GeV$ which
corresponds to the typical thermal energy of relativistic particles
at the EW epoch.
The characteristic energy at the quantum chromodynamics phase transition
is $\approx 100\MeV$ where the quark--gluon plasma turns into hadrons.
We stress, however, that the MHD formalism is only valid if the scales
considered are larger than the mean free path given by Equation~(\ref{eq_mfp}).
Comparing the chiral instability scale $k_\mu^{-1}$ with $\ell_\mathrm{mfp}$ 
results in the condition 
$\mu_5 \ll \kB T\, 4\pi^2 \alphaem \ln{((4\pi \alphaem)^{-1/2})}$. 
Strictly speaking, modeling a system that does not fulfill this condition 
requires full kinetic theory as described, for example, in \citet{CPWW13} or
in \citet{AY13}.

The growth rate of an initially weak magnetic field in the linear
stage of the chiral dynamo instability is given by \Eq{gamma-max}:
\begin{equation}
\label{eq:9}
  \gamma^{\rm max}_\mu = \frac{\mu_0^2\eta}4 \approx 2.4\times 10^{19}T_{100}^{-1}\s^{-1}.
\end{equation}
For the value of the magnetic diffusivity $\eta = c^2/(4 \pi \sigma)$
in the early universe, we adopted the conductivity $\sigma$ from
Equation~(1.11) of \cite{ArnoldEtAl2000}.
Numerically,
\begin{equation}
\eta(T)={7.3\times 10^{-4}}\,{\hbar c^2\over\kB T}\approx
{4.3\times10^{-9}}T_{100}^{-1}\cm^2\s^{-1} ,
\label{eq_rDiffvA}
\end{equation}
where $T_{100}=1.2\times10^{15}\K$ (so that $\kB T_{100} =\unit[100]{GeV}$).
As a result, the number of $e$-foldings over one Hubble time $t_H$ is
$$\gamma^{\rm max}_\mu t_H \gg 1,$$ where
\begin{equation}
  \label{eq:tH}
  t_H = H^{-1}(T) \approx 4.8\times10^{-11}\,g_{100}^{-1/2}T_{100}^{-2}\,\s
\end{equation}
(here $g_*$ is the number of relativistic degrees of freedom and $g_{100}=g_*/100$).
We should stress that this picture has been known before and was described
in many previous works \citep{JS97,Frohlich:2000en,Frohlich:2002fg,BFR12}.

We note that a nonzero chiral flipping rate
$\Gamma_\mathrm{f}$ has been discussed in the literature
\citep{CDEO92,BFR12,DS15,BFR15,SiglLeite2016}.
In Section~\ref{sec_flip}, we have found
in numerical simulations that the flipping term affects the evolution
of the magnetic field only
for large values of $f_\mu$, when the flipping term is of the order of or
larger than the $\lambda_\mu$ term in Equation~(\ref{mu-NS});
see also Equation~(\ref{eq_fmu}) and Figure~\ref{fig_ts_flip}.
When adopting the estimate in \citet{BSRKBFRK17} of $f_\mu\approx 1.6\times10^{-7}$,
chirality flipping is not likely to play a significant role for
the laminar $v_\mu^2$ dynamo in the early universe at very high temperatures
of the order of $100 \GeV$.
However, $\Gamma_\mathrm{f}$ depends on the ratio $m_e c^2/(\kB T)$ and
thus suppresses all chiral effects once the universe has cooled down to
$\kB T \approx m_e c^2$ \citep{BFR12}.
At this point, we stress again that the true value of $\mu_0$ is unknown
and has here been set to the thermal energy in Equation~(\ref{eq:1}).
If it turns out that the initial value of the chiral chemical potential
is much smaller than the thermal energy, $f_\mu$ becomes larger, and
the flipping rate can play a more important role already during the initial
phases of the chiral instability in the early universe.
This scenario is not considered in the following discussion.

\begin{figure}[!t]
  \centering
  \includegraphics[width=\columnwidth]{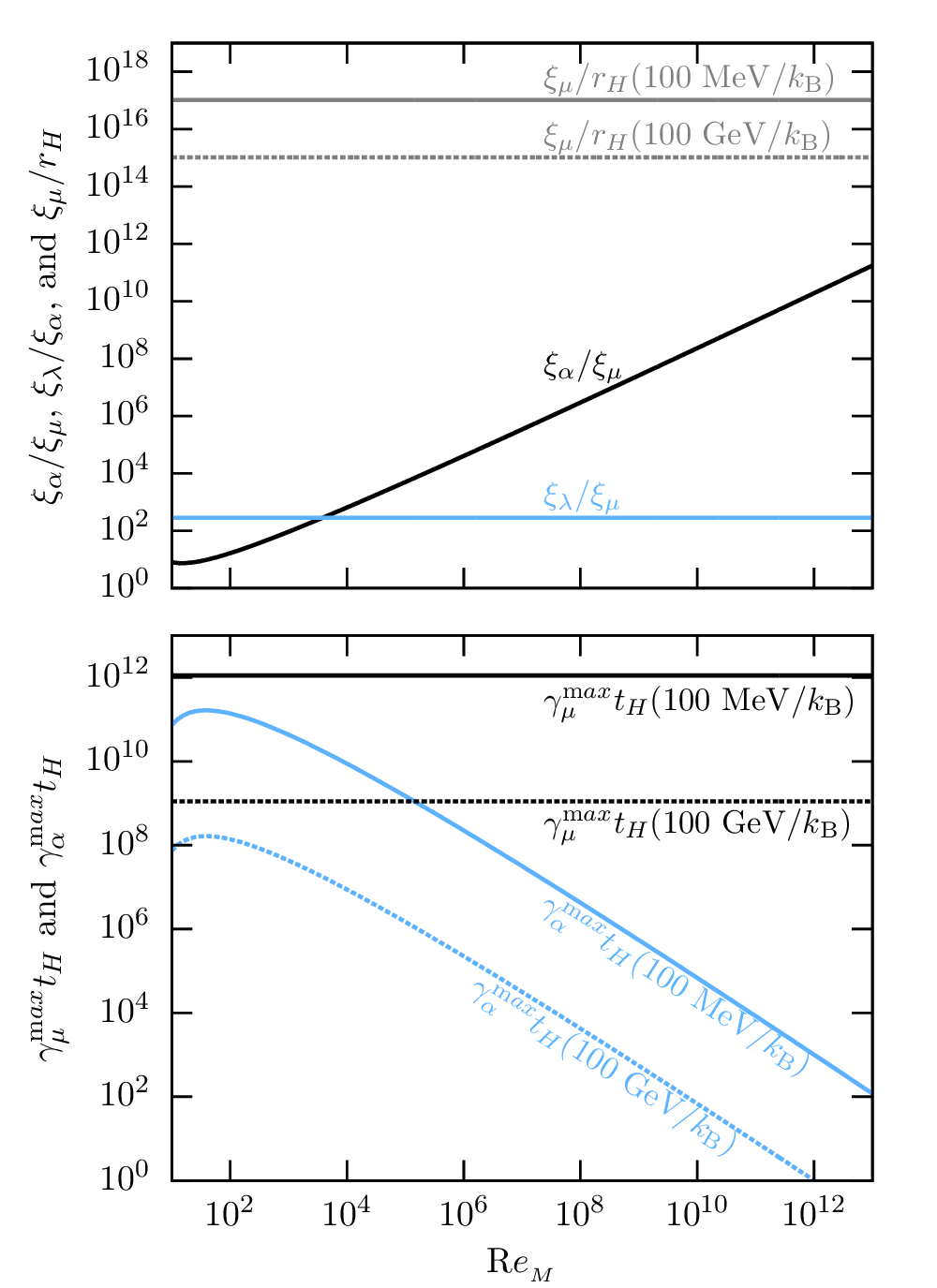}
  \caption[Turbulence-driven instability in the early universe]
  {\textbf{Chiral MHD dynamos in the early universe. }
    The ratios between $\xi_\alpha$ of the turbulence-driven dynamo (Eq.~(\ref{kmax_turb}))
    and scale $\xi_\mu$ (Equation~\protect\eqref{eq_mu-1}), as well as the ratio
    between $\xi_\mu$ and the Hubble radius at different temperatures.
    In the top panel, furthermore, the ratio $\xi_\mu/\xi_\lambda$ is presented.
    Maximum growth rates over the Hubble time for laminar ($\gamma_\mu^{\rm max}$)
    and turbulent ($\gamma_\alpha^{\rm max}$) regimes are shown in the bottom panel.
  }\label{fig:chiral_turbulence}
\end{figure}

In the regime of the laminar $v_\mu^2$ dynamo, one could reach
$\mathcal{O}(10^9)$ $e$-folds over the Hubble time $t_{\rm H}$; see lower panel
of Figure~\ref{fig:chiral_turbulence}.
However, as shown in this work, already after a few hundred $e$-foldings,
the magnetic field starts to excite turbulence via the Lorentz force.
This happens once the magnetic field is no longer force-free.
Once the flow velocities reach the level $v_\mu = \mu_0\eta$, nonlinear
terms are no longer small, small-scale turbulence is produced, and
the next phase begins.

Phase 2.
The subsequent evolution of the magnetic field depends on the strength
of the chiral magnetically excited turbulence.
This has been shown in the mean-field analysis of \citet{REtAl17} and
is confirmed by the present work; see, for example, Figure~\ref{fig_gamma_Re}.
The growth rate and instability scale depend on the magnetic Reynolds
number; see Equations~(\ref{gamma_turb})--(\ref{gammamax_turb}).
The maximum growth rate for $\Rm \gg 1$ is given by
\begin{equation}
  \label{eq:10}
 \gamma^{\rm max}_{\alpha} = \gamma_\mu^{\rm max} \frac{4}3 \frac{(\ln \Rm)^2}{\Rm},
\end{equation}
where $\gamma_\mu^{\rm max}$ is given by \Eq{eq:9}.
For the early universe, it is impossible to determine the exact value of the
magnetic Reynolds number from the numerical simulations, but one
expects $\Rm \gg 1$ and we show in \Fig{fig:chiral_turbulence}
that, in a wide range of magnetic Reynolds numbers,
$1 \ll \Rm \ll 6\times 10^{12}$, the number of $e$-foldings during one
Hubble time is much larger than $1$.
The turbulence efficiently excites magnetic fields at scales much larger than
$\xi_\mu$ (Figure~\ref{fig:chiral_turbulence}, top panel).

Using dimensional analysis and DNS, \citet{BSRKBFRK17} demonstrated
that the resulting spectrum of the magnetic fields behaves as
$E_{\rm M}\propto k^{-2}$ between $k_\mu$ and $k_\lambda$, given by
Equation~(\ref{klambda}).
The wavenumber $k_\lambda$ depends on the nonlinearity parameter $\lambda$,
defined by Equation~(\ref{eq_lambda}), which, in the early universe, is given by
\begin{equation}
  \lambda=3 \hbar c\,\left({8\alphaem\over\kB T}\right)^2\approx1.3\times10^{-17}\,T_{100}^{-2}\,
\cm\erg^{-1}.
  \label{eq_lambda_1}
\end{equation}
We note that this expression is, strictly speaking, only valid when
$\kB T \gg \max(|\mu_L|,|\mu_R|)$ and modifications might be expected 
outside of this regime.
Further, the mean density of the plasma
\begin{equation}
   \meanrho=\frac{\pi^2}{30}\,g_*\frac{(\kB T)^4}{\hbar^3c^5}
      \approx  7.6\times10^{26}g_{100}T_{100}^4\g\cm^{-3}.
\end{equation}
The ratio $\xi_\lambda/\xi_\mu = k_\mu/k_\lambda$ is presented in the top
panel of \Fig{fig:chiral_turbulence}, but we note that the exact numerical
coefficient in the condition $k_\mu/k_\lambda \gg 1$ might depend on $\Rm$.

Phase 3.
The stage of large-scale turbulent dynamo action ends with the
\emph{saturation phase} (see Section~\ref{sec:stages} and
\Fig{fig:phases}).
At this stage, the total chiral charge (determined by the initial
conditions) gets transferred to magnetic helicity.
As shown in \citet{BFR12} (see also~\citet{JS97} for earlier work, as well
as \cite{TVV12} and \cite{Hirono:2015rla} for more discussion), and confirmed by
numerical simulations in \citet{BSRKBFRK17} and in the present work,
the chiral chemical potential $\mu$ follows $k_{\rm M}$ at this stage
and thus decreases with time.
Therefore, most of the chiral charge will be transferred with time into magnetic helicity,
\begin{equation}
  \langle \AAA \cdot \BB\rangle \simeq \xi_{\rm M} \langle \BB^2 \rangle \to \frac{2\mu_0}\lambda ,
  \label{eq:2}
\end{equation}
switching off the CME (the end of Phase 3 in \Fig{fig:phases}).

\subsubsection{Chiral MHD and cosmic magnetic fields}
\label{sec:end_chiral_MHD}

Magnetic fields produced by chiral dynamos are fully helical.
Once the CME has become negligible, the subsequent phase of decaying
helical turbulence begins and the
magnetic energy decreases, while the magnetic correlation length increases
in such a way that the magnetic helicity~\eqref{eq:2} is conserved
for very small magnetic diffusivity \citep{BM99PhRvL, KTBN13}.

Based on \Eq{eq:2}, one can estimate the magnetic helicity \emph{today}; see also \cite{BSRKBFRK17}.
Taking as an estimate for the chiral chemical potential $\mu_5 \sim \kB T$ (this means that the density of the chiral charge is of the order of the
number density of photons), one finds
\begin{multline}
  \label{eq:Br17:3}
  \bra{\BB^2}\xi_{\rm M} \simeq
  \frac{\hbar c}{4\alphaem} \frac{g_{0}}{g_\ast} n_\gamma^{(0)}
  \simeq 6\times 10^{-38}\G^2\Mpc .
\end{multline}
Here, the present number density of photons is $n_\gamma^{(0)} = 411\cm^{-3}$,
and the ratio $g_{0}/g_\ast\approx3.36/106.75$ of the effective
relativistic degrees of freedom today and at the EW epoch appears,
because the photon number density dilutes as $T^3$ while the magnetic helicity
dilutes as $a^{-3}$.
We recall that, to arrive at the numerical value in $\G^2\Mpc$
given in Equation~(\ref{eq:Br17:3}), an additional $4\pi$ factor
was applied to convert to Gaussian units.

Under the assumption that the spectrum of the cosmic magnetic field is sharply
peaked at some scale
$\xi_0$ (as is the case in all of the simulations presented here),
the lower bounds on magnetic fields, inferred
from the nonobservation of GeV cascades from TeV sources
\citep{NV10,Tavecchio:10,Dolag:10} can be directly translated into
a bound on magnetic helicity today. The observational bound scales
as $|\BB| \propto \xi_0^{-1/2}$ for $\xi_0 < 1\Mpc$ \citep{NV10} and
therefore $\bra{\BB^2}\xi_0 = \const > 8\times 10^{-38}\G^2\Mpc$.
The numerical value is obtained using the most conservative bound
$|\BB| \ge 10^{-18}\G$ at $1\Mpc$ (\citealt{DCRFCL11},
see also~\citealt{DN13}).
These observational constraints for intergalactic magnetic fields are
compared to the magnetic field produced in chiral MHD for different values
of the initial chiral chemical potential in Figure~\ref{fig_B_xi__comov}.

The limit given by Equation~(\ref{eq:Br17:3}) is quite general.
It does not rely on chiral MHD or the CME, but simply
reinterprets the bounds of \cite{NV10}, \cite{Tavecchio:10}, \cite{Dolag:10},
and \cite{DCRFCL11} as bounds on magnetic helicity.
Given such an interpretation, we conclude that \emph{if cosmic magnetic fields
are helical and have a cosmological origin, then at some moment in the history
of the universe the density of chiral charge was much larger
than $n_\gamma(T)$}.
This chiral charge can be, for example, in the form of magnetic helicity or of
chiral asymmetry of fermions, or both.
To generate such a charge density, some new physics beyond the Standard Model of
elementary particles is required.
Below we list several possible mechanisms that can generate large initial
chiral charge density:

\begin{figure}[t]
  \centering
  \includegraphics[width=\columnwidth]{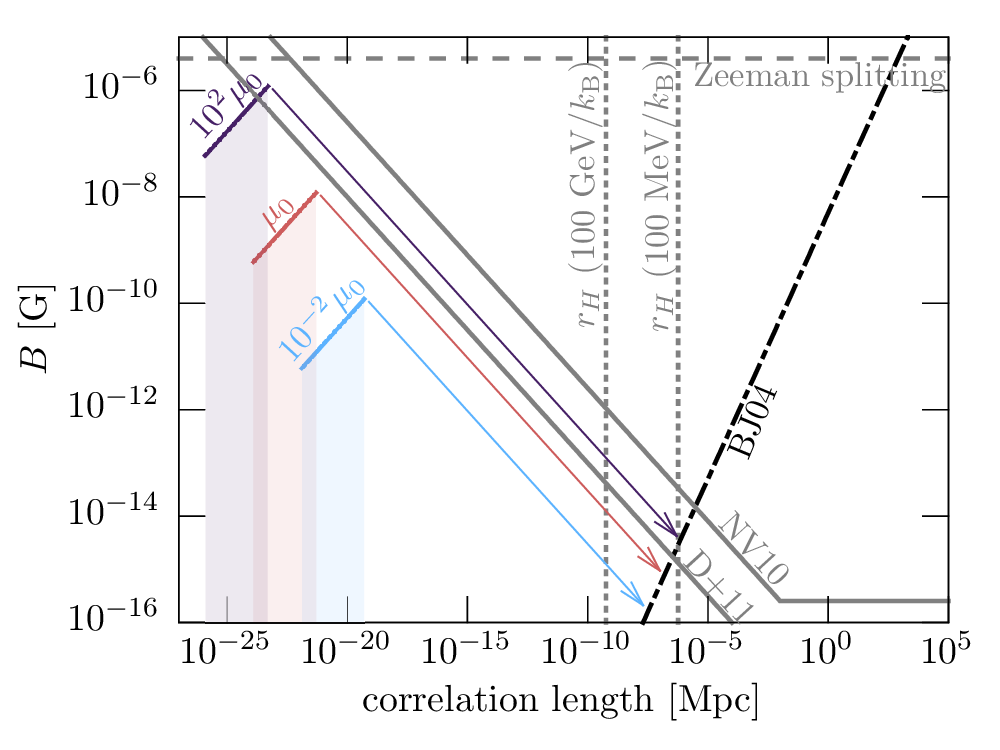}
  \caption{
    {\bf Chiral MHD dynamos in the early universe.}
The magnetic field strength resulting from a chiral dynamo as a function of correlation length in comoving
units and comparison with observational constraints.
The differently colored lines show the chiral magnetically produced magnetic field
strength in the range between the injection length $\mu^{-1}$ and the
saturation length $k_\lambda^{-1}$; see Equations~(\ref{eq_mu-1}) and
(\ref{klambda}), respectively.
The colors indicate different values of the chiral chemical potential:
Red refers to the value of $\mu_0$ given in Equation~(\ref{eq:1}), blue to
$10^{-2}\mu_0$, and purple to $10^{2}\mu_0$.
The dashed gray line is an upper limit on the intergalactic magnetic field from
Zeeman splitting.
Solid gray lines,
refer to the lower limits reported by \citet{NV10} (``NV10'') and \citet{DCRFCL11} (``D+11''), respectively.
The vertical dotted gray lines show the
horizon at $\kB T = 100\GeV$ and $100\MeV$ correspondingly.
The thin colored arrows refer to the nonlinear evolution of magnetic fields in
an inverse cascade in helical turbulence up to the final value as given in
\citet{BJ04}
(line ``BJ04'').
}
  \label{fig_B_xi__comov}
\end{figure}
%%%%%%%

\begin{asparaenum}[\it (1)]
\item The upper bound in \Eq{eq:Br17:3} assumes that only one fermion of the
Standard Model developed a chiral asymmetry $\sim n_\gamma$.
  Many fermionic species are present in the plasma at the electroweak epoch.
  They all can have a left--right asymmetric population of comparable size,
  increasing the total chirality by a factor $\mathcal{O}(10)$, which
  makes the estimate~(\ref{eq:Br17:3}) consistent with the lower bound
  from \citet{DCRFCL11}.
  One should check, of course, whether for more massive fermions the
  chirality flipping rate is much slower than the dynamo growth rate
  determined by Equation~(\ref{eq:9}).
\item The estimate~(\ref{eq:Br17:3}) assumed that left--right asymmetry was
created via thermal processes. Of course, new physics at the EW epoch can result in
nonthermal production of chiral asymmetry (e.g.\ via decays of some long-lived
particles), thus leading to $n_5 \gg n_\gamma$ and so increasing the limit
(\ref{eq:Br17:3}).
\item
  The left--right asymmetry may be produced as a consequence of the decay
  of helical \emph{hypermagnetic} fields prior to the EW epoch.
  Such a scenario, relating hypermagnetic helicity to the
  chiral asymmetry, has been discussed previously, such as in
  \cite{Giovannini:1997eg} and \cite{Semikoz:2012ka}.
  A conservation law similar to that of~(\ref{cons_law}) exists also for
  hypermagnetic fields, and the decay of the latter may cause asymmetric
  populations of left and right states.
\item
In our analysis, we have not taken into account the chiral vortical effect
\citep{Vilenkin:79}.
For nonvanishing chemical potential, it leads to an additional current
along the direction of vorticity \citep[see, e.g.,][]{TVV12}.
\end{asparaenum}

From the point of view of chiral MHD, the value of $\mu_0$ (to which
this bound is proportional) is just an initial condition and therefore
can take arbitrary values.
Once an initial condition with a large value of $\mu_0$ has been
generated, the subsequent evolution (as described above) does not require
any new physics.

Moreover, the coupled evolution of magnetic helicity and chiral chemical
potential is \emph{unavoidable} in the relativistic plasma and should be an
integral part of relativistic MHD (as was discussed in Paper~I).

\subsection{PNSs and the CME}
\label{sec:PNS}

In this section, we explore whether the CME and chiral dynamos can play a
role in the development of strong magnetic fields in neutron stars.
A PNS is a stage of stellar evolution after
the supernova core collapse and before the cold and dense neutron star is
formed \citep[see, e.g.,][]{PRPLM99}.
PNSs are characterized by high temperatures
(typically $\kB T \sim \unit[\mathcal{O}(10)]{MeV}\gg m_e c^2$),
large lepton number density (electrons Fermi energy $\mu_e \sim$ a few
hundreds of MeV), the presence of turbulent flows in the interior, and quickly
changing environments.
Once the formation of a neutron star is completed, its magnetic
field can be extremely large.
Neutron stars that exceed the quantum electrodynamic limit
$B_\mathrm{QED}\equiv m_e^2c^3/(e \hbar)\approx4.4\times10^{13}~\G$ are known as
``magnetars'' \citep[see, e.g.,][for recent reviews]{MPM15,TZ15,KB17}.
The origin of such strong magnetic fields remains unknown, although
many explanations have been proposed; see, for example, \citet{DT92}, 
\cite{AWML03}, and \cite{FW06}.

The role of the CME in the physics of (proto)neutron stars and
their contribution to the generation of strong magnetic fields have been
discussed in a number of works
\citep{Charbonneau:2009ax,Ohnishi:2014uea,Dvornikov:2015lea,DS15,
GKR15,Dvornikov:2016cmz,SiglLeite2016,Yamamoto:2015gzz}.

\subsubsection{Chiral MHD in PNSs}
\label{sec:chiral-mhd-PNS}

During the formation of a PNS, electrons and protons are converted into neutrons,
leaving behind left-handed neutrinos.
This is known as the Urca process
\citep[$e + p \to n + \nu_e$;][]{Haensel:95}.
If the chirality-flipping timescale, determined by the electron's
mass, is longer than the instability scale, the net chiral
asymmetry in the PNS can lead to the generation
of magnetic fields. This scenario has been discussed previously
\citep{Ohnishi:2014uea,SiglLeite2016,GKR15}.
The chiral turbulent dynamos discussed in this work can be relevant for
the physics of PNSs and can affect our conclusions
about the importance of the CME.
However, to make a detailed quantitative analysis, a number of factors
should be taken into account:
\begin{asparaenum}[\it (1)]
\item The rate of the Urca process is strongly temperature dependent
  \citep{LPPH91,Haensel:95}.
  The temperatures inside PNSs are only known with large uncertainties, and
  the cooling occurs on a scale of seconds \citep[see, e.g.,][]{PRPLM99},
  making estimates of the Urca rates uncertain by orders of magnitude.
\item The chirality flipping rate that aims to restore the depleted population of
  left-chiral electrons is also expected to be temperature dependent
  \citep[see, e.g.,][]{GKR15,SiglLeite2016}.
\item The neutrinos produced via the Urca process are trapped in the
  interior of a PNS and can release the chiral asymmetry back into the plasma
  via the $n + \nu_e \to e + p$ process.
  Therefore, only when the star becomes transparent to neutrinos
  (as the temperature drops to a few MeV) does the creation of chiral asymmetry
  can become significant.
\end{asparaenum}

Modeling the details of PNS cooling and neutrino propagation is
beyond the scope of this paper.
Below we perform the estimates that demonstrate that chiral MHD can
significantly change the picture of the evolution of a PNS.

\subsubsection{Estimates of the relevant parameters}

An upper limit of the chiral chemical potential can be estimated by
assuming that
$n_\mathrm{L}=0$ and $n_\mathrm{R}=n_e$ (all left-chiral electrons have been
converted into neutrinos, and the rate of chirality flipping is much slower than
other relevant processes).
This leads to the estimate $\mu_5 \simeq \mu_e$ and correspondingly
\begin{equation}
  \mu_{\rm max} = 4\alphaem \frac{\mu_e}{\hbar c} \approx \unit[4\times 10^{11} ]{cm}^{-1}\left(\frac{\mu_e}{\unit[250]{MeV}}\right),
  \label{eq_muupper_NS}
\end{equation}
where we have used a typical value of the electron's Fermi energy $\mu_e$
\citep{PRPLM99}.
For an ultrarelativistic \emph{degenerate} electron gas (i.e., when
$\mu_e \gg \kB T \gg m_e c^2$), the relation between the number density
of electrons, $n_e$, and their Fermi energy, $\mu_e$, is
\begin{equation}
  \label{eq:5}
  \mu_e = \hbar c(2\pi^2 n_e)^{1/3}\approx 250\MeV \left(\frac{n_e}{\unit[10^{38}]{cm^{-3}}}\right)^{1/3}.
\end{equation}
The interior of neutron stars is a conducting medium whose
conductivity is estimated to be \citep{BPP69,Kelly1973}:
\begin{eqnarray}
  \sigma(T) &=& \sqrt 3\left(\frac4\pi\right)^{3/2}\frac{\hbar^4 c^2}{e\,m_p^{3/2}}\frac{n_e^{3/2}}{\kB^2 T^2}\\
  &\approx& 1\times  10^{27}\left(\frac{1~\mathrm{MeV}}{\kB T}\right)^{2}\left(\frac{n_e}{\unit[10^{38}]{cm^{-3}}}\right)^{3/2}\s^{-1}
\label{eq_sigmaNS}
\end{eqnarray}
(there is actually a difference in the numerical coefficient $\mathcal{O}(1)$
between the results of \citet{BPP69} and \citet{Kelly1973}).
Using Equation~(\ref{eq_sigmaNS}), we find the magnetic diffusion coefficient to be
\begin{equation}
  \label{eq:3}
  \eta(T)
  \approx 7\times 10^{-8}\cm^2\s^{-1} \left(\frac{\kB T}{1~\mathrm{MeV}}\right)^{2}\left(\frac{\unit[10^{38}]{cm^{-3}}}{n_e}\right)^{3/2}.
\end{equation}
Therefore, we can determine the
the maximum growth rate of the small-scale
chiral instability (\ref{gamma-max}) as
\begin{equation}
  \label{eq:4}
  \gamma^{\rm max}_\mu = \frac{\mu_{\rm max}^2\eta}4 \approx 2\times 10^{15}\s^{-1} \left(\frac{\mu_e}{\unit[250]{MeV}}\right)^2 \left(\frac{\kB T}{1~\mathrm{MeV}}\right)^{2}.
\end{equation}
We see that over a characteristic time $\tau_\mathrm{cool}\sim 1 \s$
(the typical cooling time), the magnetic field would increase by many
$e$-foldings.
In fact, using a flipping rate of $\Gamma_\mathrm{f}=10^{14} \s$, as
suggested in \citet{GKR15} for $\mu_\mathrm{e}= 100 \MeV$ and $\kB T=30 \MeV$,
we find that $f_\mu$ ranges from
$\approx 9\times 10^{-3}$ down to $\approx 9\times 10^{-7}$ for the range
between $\kB T = 1 \MeV$ and $\kB T = 100 \MeV$.
Hence the evolution of the chemical potential
and the chiral dynamo
is weakly affected by flipping reactions.

\begin{figure}[!t]
  \centering
  \includegraphics[width=\columnwidth]{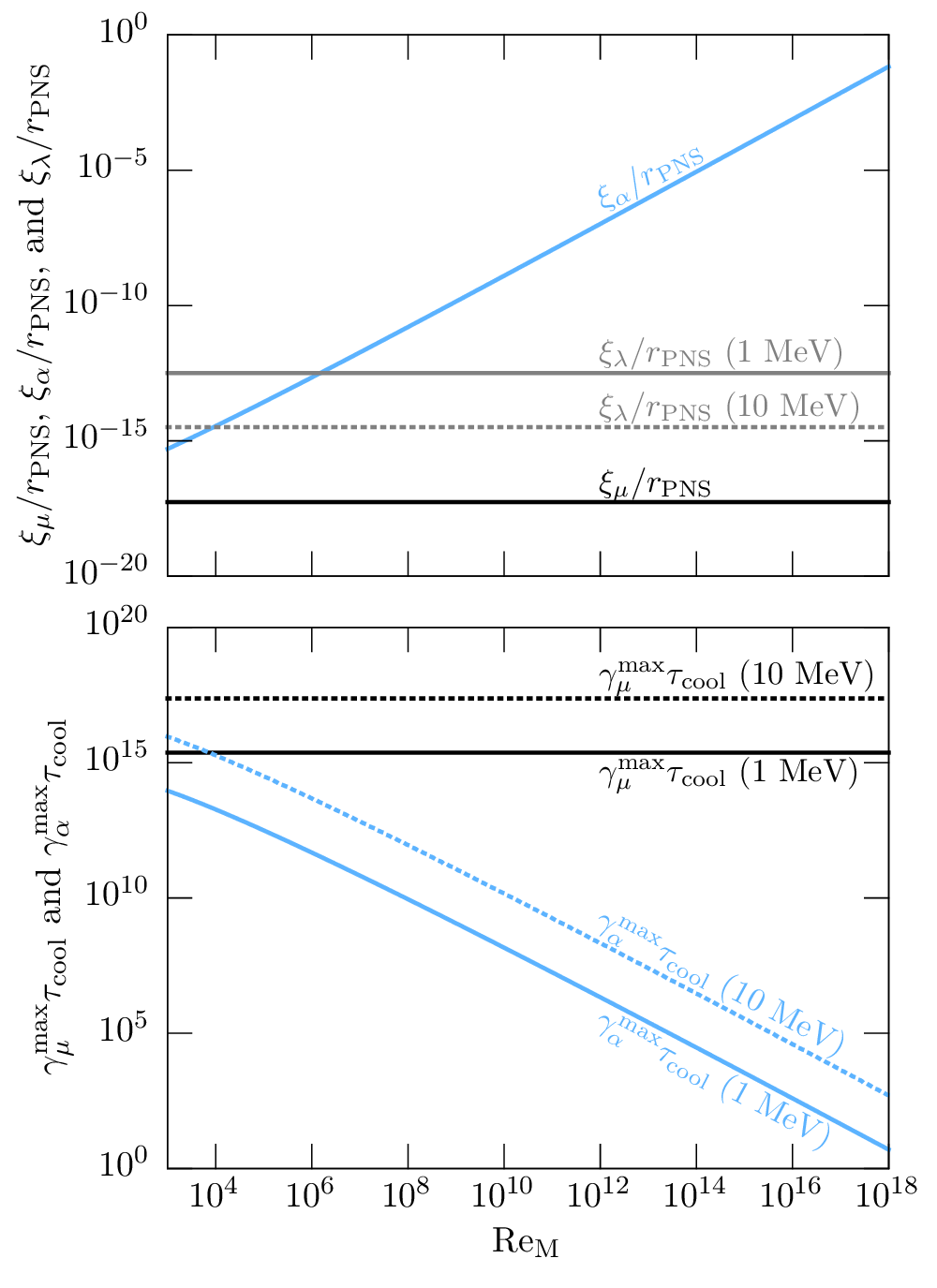}
  \caption[Laminar and turbulent scales in PNS]{
  %\textbf{Chiral MHD dynamos in in PNS.}
%AB: proofs
%JS: removed one "in"
%  \textbf{Chiral MHD dynamos in in PNSs.}
  \textbf{Chiral MHD dynamos in PNSs.}
Laminar and turbulent growth rate multiplied by the
cooling timescale (top panel) and the characteristic scales of chiral MHD
normalized by the typical radius of the PNS $r_{\rm NS} \sim 10\km$
(bottom panel).
The estimates are presented as a function of $\Rm$.
The initial value of the chiral chemical potential is assumed at the
level~(\protect\ref{eq_muupper_NS}) and the we use $\mu_e =\unit[250]{MeV}$.
Since the conductivity is temperature dependent, the ratios including $\eta$ are
presented for both $\kB T = 1~\mathrm{MeV}$ and $\kB T = 10~\mathrm{MeV}$.
  }
  \label{fig:turbulence_PNS}
\end{figure}

As in Section~\ref{subsec_cosmologybounds}, the phase of the small-scale
instability ends when turbulence is excited.
It should be stressed, however, that unlike the early universe, the
interiors of PNSs are expected to be turbulent with high $\Rm$ even in
the absence of chiral effects
(with $\Rm$ as large as $10^{17}$); see \cite{TD93}.
Therefore, the system may find itself in the forced turbulence regime of Section~\ref{sec:forced-chiral-dynamos}.
Figure~\ref{fig:turbulence_PNS} shows that in a wide range of magnetic
Reynolds numbers, one can have many $e$-foldings over a typical timescale of
the PNS and that the scale of the magnetic field can reach macroscopic size.

\subsubsection{Estimate of magnetic field strengths}

A dedicated analysis, taking into account temperature and density evolution of
the PNS as well as its turbulent regimes, is needed to make detailed predictions.
Here we will make the estimates of the strength of the magnetic field,
similar to Section~\ref{sec:early-universe} above.
To this end, we use the conservation law~(\ref{cons_law}), assuming
$\mu_0 = \mu_{\rm max}$.
In the PNS case, the plasma is degenerate, and therefore the relation between
$n_5$ and $\mu_5$ is given by
\begin{equation}
  n_5 = \frac{\mu_5}{3\pi^2}(3\mu_e^2 + \pi^2 T^2)
  \label{eq:7}
\end{equation}
(in the limit $\mu_5 \ll T$).
{
As a result, the chiral feedback parameter
$\lambda$ is
\begin{equation}
  \label{lambda_PNS}
  \lambda_\pns = \frac{\hbar c\pi^2}{2}\left(\frac{8\alphaem}{\mu_e}\right)^2,
\end{equation}
which determines the wavenumber $k_\lambda$; see Equation~(\ref{klambda}).
The corresponding length scale $\xi_\lambda = k_\lambda^{-1}$ is presented in
the top panel of Figure~\ref{fig:turbulence_PNS}, where we assume a
mean density of the PNS of $\meanrho_\pns = 2.8\times10^{14}~\mathrm{g}\mathrm{cm}^{-3}$.

Using \Eqs{eq_muupper_NS}{lambda_PNS}, we find
\begin{eqnarray}
  \label{eq:8}
  (B^2\xi)_{\max}
   &=&\frac{4\pi\mu_{\rm max}}{\lambda_\pns} = \frac{\mu_e^3}{2(\hbar c)^2\alphaem}\\
  &\approx& 1.4\times 10^{24}\G^2\cm\left(\frac{\mu_e}{\unit[250]{MeV}}\right)^3.
\end{eqnarray}
Assuming for the maximum correlation scale $\xi_{\rm PNS} \sim 1\cm$
(see \Fig{fig:turbulence_PNS}), we find that
magnetic field strength is of the order of
\begin{equation}
  \label{eq:6}
  B_{\rm max} \approx 1.2\times 10^{12}\G \left(\frac{\mu_e}{\unit[250]{MeV}}\right)^{3/2}
  \left(\frac {1\cm}{\xi_{\rm M}}\right)^{1/2} .
\end{equation}
Notice that the estimate~(\ref{eq:8}) is independent of $T$
(but depends strongly on the assumed value of $\mu_e$).

Our estimates have demonstrated that the chiral MHD could be capable of generating
strong small-scale magnetic fields. \emph{Therefore, chiral effects should be
included in the modeling of evolution of PNSs.}}

% %%%%%%%%%%%%%
% % Section 7 %
% %%%%%%%%%%%%%

\section{Conclusions}
\label{sec_concl}

In this work, we have presented results from numerical simulations of chiral MHD
that include the temporal and spatial evolution of magnetic fields,
plasma motions, and the chiral chemical potential.
The latter, characterizing the asymmetry between left- and right-handed fermions,
gives rise to the CME, which
results in the excitation of a small-scale chiral dynamo instability.

Our numerical simulations are performed for the system of
chiral MHD equations~(\ref{ind-DNS})--(\ref{mu-DNS}) that was derived in
Paper~I.
This system of equations is valid for plasmas with high electric conductivity,
that is, in the limit of high and moderately high Reynolds numbers.
Chiral flipping reactions are neglected in most of the simulations.
In the majority of the runs, the initial conditions are a very weak magnetic seed
field and a high chiral chemical potential.
Both initially force-free systems and systems with external forcing of
turbulence are considered.
With our numerical simulations, we confirm
various theoretical predictions of the chiral laminar
and turbulent large-scale dynamos discussed in Paper~I.

Our findings from DNS can be summarized as follows:
\begin{itemize}
\item{
The evolution of magnetic fields studied here in DNS agrees with
the predictions made in Paper~I for all types of laminar dynamos.
In particular, the scalings of
the maximum growth rate of the chiral dynamo instability
$\gamma_\mu\propto v_\mu^2$ for the
$v_\mu^2$ dynamo (see Figure~\ref{fig__Gamma_eta}) and
$\gamma_\mu\propto (S v_\mu)^{2/3}$
for the $v_\mu$--shear dynamo (see Figure~\ref{Gamma_Us_Beltrami}) have been
confirmed.
Additionally, the transitional regime of an $v_\mu^2$--shear dynamo, where the
contributions from the $v_\mu^2$- and shear terms
are comparable, agrees with theoretical predictions, as can be seen in
Figure~\ref{fig_Gamma_Us_aShear}.
In our DNS, the scale-dependent amplification of the magnetic field
in the laminar chiral dynamo is observed in the energy spectra; see, for example,
Figure~\ref{fig_LTspec} where the maximum
growth rate of the $v_\mu^2$ dynamo instability is attained
at wavenumber $k_\mu=\mu_0/2$.
}
\item{
The conservation law~(\ref{CL}) for total chirality
implies a maximum magnetic field strength
of the order of $B_\mathrm{sat}\approx(\mu_0 \xi_\mathrm{M}/\lambda)^{1/2}$.
This dependence of $B_\mathrm{sat}$ on the chiral nonlinearity parameter
$\lambda$ has been confirmed numerically and is presented in
Figure~\ref{fig_Bsat_lambda}.
}
\item{
The CME can drive turbulence efficiently via the Lorentz force,
which has been demonstrated in our numerical simulations
through the measured growth rate of the turbulent velocity, which is
larger by approximately a factor of two than that of the magnetic field;
see, for example, the middle panel of Figure~\ref{fig_LTts}.
}
\item{
In the presence of small-scale turbulence, the large-scale dynamo
operates due to the chiral $\alpha_\mu$ effect,
which is not related to the kinetic helicity;
see Equation~(\ref{alphamu}).
In the limit of large magnetic Reynolds numbers,
the maximum growth rate of the large-scale dynamo instability
is reduced by a factor of
$(4/3)(\ln\Rm)^2/\Rm$ as
compared to the laminar case; see Equation~(\ref{gammamax_turbRm}).
The dynamo growth rate is close to this prediction
of mean-field chiral MHD for both
chiral magnetically produced turbulence and
for externally driven turbulence; see
see Figure~\ref{fig_gamma_Re}.
}
\item{
Using DNS, we found a new scenario of the magnetic field evolution
consisting of three phases
(see also the schematic overview in Figure~\ref{fig:phases}):
\begin{asparaenum}[\it (1)]
\item small-scale chiral dynamo instability;
\item production of small-scale turbulence, inverse transfer of magnetic energy,
and generation of a large-scale magnetic field by the chiral $\alpha_\mu$ effect;
\item
saturation of the large-scale chiral dynamo
by a decrease of the CME
controlled by the conservation law for
the total chirality:
$\lambda \, \bra{{\bm A} {\bm \cdot} \BB}/2 + \bra{\mu} = \mu_0$.
\end{asparaenum}
The previously discussed scenario of magnetic field
evolution caused by the CME \citep{BFR12}
did not include the second phase.}
\end{itemize}

While the results summarized above have been obtained
in simulations of well-resolved periodic domains,
astrophysical parameters are beyond the regime accessible to DNS.
Hence we can only estimate the effects of the chiral anomaly in relativistic
astrophysical plasmas, like in the early universe or in neutron stars.
The main conclusions from the astrophysical applications are the following:
\begin{itemize}
\item{The chiral MHD scenario found in DNS may help to explain the origin of
the magnetic field observed in the interstellar
medium. The chiral dynamo instability produces helical magnetic
fields initially at
small spatial scales and simultaneously drives turbulence,
which generates a magnetic field on large scales.
After the chiral chemical potential has been transformed into magnetic
helicity during the dynamo saturation phase,
the magnetic field cascades to
larger spatial scales according to the phenomenology of decaying MHD
turbulence.
We have estimated the values of $\mu_0$ and $\lambda$ for the early universe.
These parameters determine the time and spatial scales associated with
the chiral dynamo instability
(see Figure~\ref{fig:chiral_turbulence}) and the maximum magnetic
helicity (see Equation~\ref{eq:Br17:3}).
Our estimates for magnetic fields produced by chiral dynamos in the
early universe are consistent with the observational lower limits
found by \citet{DCRFCL11} (see Figure~\ref{fig_B_xi__comov})
if we assume that the initial chiral chemical potential is of the order of
the thermal energy density.
}
\item{In PNSs, chiral dynamos
operating in the first tens of seconds after the supernova explosion
can produce magnetic fields of approximately $10^{12}\G$ at a magnetic
correlation length of $1\cm$; see Equation~(\ref{eq:6}).
However, we stress that many questions remain open, especially regarding the
generation of a chiral asymmetry and the role of the chiral flipping term
in PNEs.
}
\end{itemize}

Finally, we stress again that the parameters and the initial conditions,
including the initial chiral asymmetry, are unknown in the astrophysical 
systems discussed in this paper. 
Hence, the purpose of our applications should be classified as a study
of the conditions under which the CME plays a significant 
role in the evolution of a plasma of relativistic charged fermions. 
With the regimes accessible to our simulations not being truly realistic in the 
context of the physics of the early universe and 
in neutron stars, our applications have a rather exploratory nature. 
In this sense, our results from DNS can be used to answer
the question in which area of plasma physics -- the physics of the early
universe, the physics of neutron stars, or the physics of heavy ion
collisions---the CME is important and can modify the
evolution of magnetic fields.

\begin{acknowledgements}
We thank the anonymous referee for constructive criticism that
improved our manuscript.
We acknowledge support from Nordita, which is funded by the
Nordic Council of Ministers, the Swedish Research Council, and the two host
universities, the Royal Institute of Technology (KTH) and
Stockholm University.
This project has received funding from the
European Union's Horizon 2020 research and
innovation program under the Marie Sk{\l}odowska-Curie grant
No.\ 665667.
We also acknowledge the University of Colorado's support through
the George Ellery Hale visiting faculty appointment.
Support through the NSF Astrophysics and Astronomy Grant Program (grant 1615100),
the Research Council of Norway (FRINATEK grant 231444),
and the European Research Council (grant number 694896) are
gratefully acknowledged. Simulations presented in this work have been performed
with computing resources
provided by the Swedish National Allocations Committee at the Center for
Parallel Computers at the Royal Institute of Technology in Stockholm.
\end{acknowledgements}

\end{document}